\documentclass[traditabstract]{aa} 
\pdfoutput=1
\usepackage[varg]{txfonts}
\usepackage{graphicx,rotating,amssymb}
\usepackage{booktabs}
\usepackage{subfig}
\usepackage{float,capt-of}
\usepackage{natbib}
\usepackage{multirow}
\usepackage[english]{babel}
\usepackage{morefloats}

\begin{document}
   \title{Massive Molecular Outflows and Evidence for AGN Feedback \\
   from CO Observations}

   \author{C. Cicone
          \inst{1, 2} 
          \and
           R. Maiolino  \inst{1, 2}
	\and
         E. Sturm \inst{3}
         \and
         J. Graci\'{a}-Carpio  \inst{3}
         \and
         C. Feruglio \inst{4} 
        \and
        	R. Neri \inst{4}
         \and 
          S. Aalto \inst{5}
           \and
         R. Davies \inst{3}
	 \and
         F. Fiore \inst{6}
	\and
         J. Fischer \inst{7}
         \and
         S. Garc\'{i}a-Burillo \inst{8}
          \and 
         E. Gonz\'{a}lez-Alfonso \inst{9}
         \and
         S. Hailey-Dunsheath \inst{10}
         \and
         E. Piconcelli \inst{6}
         \and
         S. Veilleux \inst{11}              
	}
      \institute{Cavendish Laboratory, University of Cambridge 19 J. J. Thomson Avenue, Cambridge CB3 0HE, UK\\ 
      	\email{c.cicone@mrao.cam.ac.uk}
	  \and Kavli Institute for Cosmology, University of Cambridge, Madingley Road, Cambridge CB3 0HA, UK\\ 
	  \and Max Planck Institute f\"{u}r Extraterrestrische Physik (MPE), Giessenbachstra\ss e 1, 85748, Garching, Germany\\ 
	  \and Institute de Radioastronomie Millimetrique (IRAM), 300 Rue de la Piscine, F-38406 St. Martin d'Heres, Grenoble, France\\ 
	  \and Department of Earth and Space Sciences, Onsala Observatory, Chalmers University of Technology, 43992 Onsala, Sweden\\ 
	  \and Osservatorio Astronomico di Roma (INAF), via Frascati 33, I-00040 Monteporzio Catone, Italy \\ 
	  \and Naval Research Laboratory, Remote Sensing Division, 4555 Overlook Ave SW, Washington, DC 20375, USA \\
	  \and Observatorio Astron\'{o}mico Nacional (OAN), Observatorio de Madrid, Alfonso XII 3, 28014 Madrid, Spain \\ 
	  \and Universidad de Alcal\'{a} de Henares, Departamento de F\'{i}sica y Matem\'{a}ticas, Campus Universitario, 28871 Alcal\'{a} de Henares, Madrid, Spain\\ 
	  \and California Institute of Technology, Mail Code 301-17, 1200 E. California Blvd., Pasadena, CA 91125, USA \\
	  \and Department of Astronomy, University of Maryland, College Park, MD 20742, USA \\
	      }

   \date{Received 09 August 2013 / Accepted: 07 November 2013 }

\abstract{We study the properties of massive, galactic-scale outflows of molecular gas and investigate their
impact on galaxy evolution. We present new IRAM PdBI CO(1-0) observations of local ULIRGs and QSO hosts: clear
signature of massive and energetic molecular outflows, extending on kpc scales, is found in the CO(1-0) kinematics 
of four out of seven sources, with measured outflow rates of several $\rm 100~M_{\odot}~yr^{-1}$. We combine these new 
observations with data from the literature, and explore the 
nature and origin of massive molecular outflows within an extended sample of 19 local galaxies.
We find that starburst-dominated galaxies have an outflow rate comparable to their SFR, 
or even higher by a factor of $\sim$2-4, implying that starbursts can indeed be effective 
in removing cold gas from galaxies. Nevertheless, our results suggest that the presence of an AGN can 
boost the outflow rate by a large factor, which is found to increase with the $\rm L_{AGN}/L_{bol}$ ratio.
The gas depletion time-scales due to molecular outflows are anti-correlated with the presence and luminosity
of an AGN in these galaxies, and range from a few hundred million years in starburst galaxies, down to just a few
 million years in galaxies hosting powerful AGNs. In quasar hosts the depletion time-scales
due to the outflow are much shorter than the depletion time-scales due to star formation. We estimate the
outflow kinetic power and find that, for galaxies hosting powerful AGNs, it corresponds to about
5\% of the AGN luminosity, as expected by models of AGN feedback. Moreover,
we find that momentum rates of about ${\rm 20~L_{AGN}/c}$ 
are common among the AGN-dominated sources in our sample.
For ``pure'' starburst galaxies our data tentatively support models in which outflows are mostly 
momentum-driven by the radiation pressure from young stars onto dusty clouds.
Overall, our results indicate that, although starbursts are effective in powering massive molecular outflows, the
presence of an AGN may strongly enhance such outflows and, therefore, have 
a profound feedback effect on the evolution of galaxies, by efficiently removing fuel for star formation, 
hence quenching star formation.}

  \keywords{Galaxies: active  -- Galaxies: evolution -- quasars: general -- Radio lines:ISM -- ISM: molecules}
 
   \maketitle

%

\section{Introduction}

The recent discovery of massive, highly-energetic and kpc-scale molecular outflows in local AGN 
hosts and ultra-luminous infrared galaxies (ULIRGs) 
\citep{Feruglio+10, Fischer+10, Sturm+11, Alatalo+11, Aalto+12a, Cicone+12, Dasyra+Combes12, Nesvadba+10, Nesvadba+11,
Veilleux+13, Spoon+13, Gonzalez-Alfonso+13}  has provided 
important evidence for negative feedback on star formation in action in galaxies 
(see also the review by \citealt{Fabian12}). 
The major breakthrough achieved by these studies
is the confirmation that, whatever is the driving mechanism (AGN or star formation itself), these outflows
affect the phase of the interstellar medium (ISM) out of which stars forms, i.e. the cold molecular gas.
Therefore, such feedback mechanism can have a significant impact on the evolution of 
the host galaxy, by regulating, through outflows, the amount of cold and dense gas available in the ISM for
star formation and black hole accretion. 
In particular, negative feedback from quasars
is believed to prevent massive galaxies from overgrowing (hence explaining the steep decline of the stellar
mass function at high masses), to account for the ``red-and-dead'' properties of massive ellipticals
\citep{Cowie+96,Baldry1+04,Perez-Gonzalez+08}, and to drive the correlations between supermassive 
black holes and bulge properties that are observed in the local Universe 
(e.g. \citealt{Magorrian+98, Marconi+Hunt03, Ferrarese+Ford05, Kormendy+Ho13}).

Indeed, most theoretical models propose that the super massive black hole, during 
its bright active phase (``quasar-mode" feedback), characterised by a high accretion rate, can expel the ISM out of the
host galaxy, and eventually clears the galaxy of its cold gas reservoir, although the details of the mechanism responsible
for coupling the quasar radiation with the galactic ISM are still debated
(e.g. \citealt{Silk+Rees98, Granato+04, Lapi+05, DiMatteo+05,
Springel+05, Croton+06, Hopkins+08, Menci+08, Narayanan+08, Fabian+09, King10, Zubovas+12, Faucher-Giguere+12}).

While various techniques have been used to identify other phases of outflows in active galaxies, both locally
(e.g. \citealt{Rupke+Veilleux11, Rupke+Veilleux13a, Muller-Sanchez+11, Mullaney+13, Emonts+05, Morganti+03, Morganti+05, Rodriguez-Zaurin+13})
and at high redshift (e.g. \citealt{Harrison+12, Cano-Diaz+12, Diamond-Stanic+12, Bradshaw+13, Alexander+10, Maiolino+12}),
the interferometric mapping of the high velocity component of CO millimeter emission has proved to be very effective.
Indeed this method, as mentioned, allows us to directly trace outflows of the molecular phase out of which stars form, which
in most galaxies also represents the bulk of the ISM; moreover the spectro-imaging information delivered by 
millimeter interferometers allows us to determine the size and geometry of the outflow, hence to accurately estimate
the outflow rate and energetics.

In this work we aim to address some open questions, such as: 
are massive molecular outflows effective in quenching star formation in the host galaxy? 
Is the central AGN the dominant power source and, consequently, are AGNs controlling the growth and evolution of galaxies? 
Are current AGN-feedback models adequate to explain observations?

In this paper we present new IRAM PdBI CO(1-0) observations for a sample of 7 local active galaxies, aimed at identifying the
presence of molecular outflows. Our new data are combined with previous results from our and other teams,
with the goal of characterising the physical properties, origin and nature of massive molecular outflows, and their
impact on galaxy evolution.

A $H_0$=70.4 km s$^{-1}$ Mpc$^{-1}$, $\Omega_M=0.27$, $\Omega_{\Lambda}=0.73$ 
cosmology is adopted throughout this work.

\section{Sample Selection and IRAM PdBI Observations}

We observed seven local active luminous and ultra-luminous infrared galaxies
in their CO(1-0) molecular transition with the IRAM Plateau de Bure Interferometer.
Since these are still exploratory observations, the sample has a mixed composition.
Out of the seven sources, five are extracted from the SHINING project, which is a
Herschel--PACS spectroscopic guaranteed time key program, targeting 
starbursts, Seyfert galaxies, and (ultra) luminous infrared galaxies \citep{Fischer+10, Sturm+11, Veilleux+13}.
Four of these ULIRGs, namely Mrk~273, IRAS~F08572+3915, 
IRAS~F10565+2448 and IRAS~23365+3604, show evidence, in their far-IR spectrum,
for P-Cygni profiles of the OH transitions at 79 and 119$\mu$m, which unambiguously trace
molecular outflows \citep{Sturm+11, Veilleux+13}. In the other SHINING source, 
Mrk~876, as well as in the quasar I~Zw~1, also 
observed with PACS in a cycle 1 open-time program, OH 
is detected only in emission \citep{Veilleux+13}.
These two sources are incorporated in our sample to investigate whether
evidence of outflow is observed in CO emission, hence to test whether the detection
of OH blue-shifted absorption is a pre-requisite for the presence of molecular outflows.
Finally, IRAS~F23060+0505 is one of the
most powerful type 2 quasars, and, by observing it, we aim 
to investigate the presence of molecular outflows
in this class of very powerful, but heavily embedded objects.
The list of targets is given in
Table~\ref{table:obs}.

\begin{table*}
 \centering
 \scriptsize
 \begin{minipage}{150mm}
  \caption{Description of the IRAM PdBI CO(1-0) Observations}
  \label{table:obs}
\begin{tabular}{@{}llllllcccc@{}}
\hline
\hline
Object 				&	RA0		&	DEC0		&  Date(s) 		& Obs. Freq. 	& Min, Max 		& Conf.		& On Source  	& FoV                   	& Synt. Beam \\
					&	(J2000)	&	(J2000)		&				&	(GHz)	&  Baseline (m)		& 			& Time (hrs)	&  (arcsec)	  	&   (arcsec)   \\
\hline
IRAS F08572+3915		& 09:00:25.37	&	39:03:53.68	& May-Oct 2011	& 108.930  	& (15.0, 176.0) 		&	C+D		& 20			& 46.3 $\times$ 46.3	& 3.1 $\times$ 2.7\\
IRAS F10565+2448		& 10:59:18.14	&	24:32:34.42	& Jun-Oct 2011 	& 110.509		& (16.5, 175.2)		&	C+D		&  10			& 45.6 $\times$ 45.6	& 3.5 $\times$ 2.9			\\
IRAS 23365+3604		& 23:39:01.23	&	36:21:09.30	& Jun-Oct 2011		& 108.299		& (17.0, 175.8)		&	C+D		&  10		         & 46.5 $\times$ 46.5 & 3.1 $\times$ 2.6 				\\
Mrk 273				& 13:44:42.16 	& 	55:53:13.21	& Jun-Jul 2012		& 111.076		& (17.5, 102.5)		&	D		& 12			& 45.4 $\times$ 45.4 & 5.0 $\times$ 4.0 \\
IRAS F23060+0505		& 23:08:33.93	&	05:21:29.90	& May 2010		&   98.270		& (15.0, 94.5)		&	D		& 3			& 51.3 $\times$ 51.3	& 5.3 $\times$ 4.5		\\
Mrk 876				& 16:13:57.21	&	65:43:10.60	& May 2010		& 102.100		& (16.6, 93.2)		&	D		& 6			& 49.4 $\times$ 49.4	& 6.0 $\times$ 3.9		\\
I Zw 1				& 00:53:34.94	&	12:41:36.20	& May 2010		& 108.629		& (17.8, 95.6)		&	D		& 8			& 46.4 $\times$ 46.4	& 4.7 $\times$ 3.7 						\\
\hline
\end{tabular}
\end{minipage}
\end{table*}
 
Observations of the CO(1-0) rotational transition (${\rm \nu_{rest}}$ = 115.271 GHz) were
performed with the IRAM Plateau de Bure Interferometer (PdBI) 
between May, 2010 and July, 2012.  A technical description of the observational parameters
is provided in Table \ref{table:obs}.
In the compact (D) array configuration, only
5 of the 15m antennas were used, while observations in C configuration
were carried out with the complete array (6 antennas).
Data calibration and analysis were performed by using the CLIC and MAPPING
softwares within the GILDAS package.
The flux calibration accuracy at these frequencies is about 10\% \citep{PdBI_cookbook}.
We exploit the wide-band (WideX) correlator offered by the PdBI, which provides
a spectral resolution of 1.95 MHz (corresponding, at an average observed frequency of $\sim$110 GHz, 
to channel widths of 
$\sim$ 5.3 km s$^{-1}$) over its full bandwidth of 3.6 GHz. 
For display purposes, we 
bin the spectra using a re-binning factor that varies, from source to source, from 5 to 20 channels.

\section{Results}

\subsection{Criterion for Molecular Outflow Detection}

The adopted criterion to claim the detection of a molecular outflow is that at least two of the following 
conditions must be satisfied:
\begin{enumerate}
\item CO(1-0) emission with velocities higher than 500 km s$^{-1}$ is detected in the 
interferometric continuum-subtracted maps with a significance of at least {\rm 5$\sigma$};
\item Broad CO(1-0) wings with velocities higher than 300 km s$^{-1}$ are detected 
in the interferometric continuum-subtracted maps with a significance of at least {\rm 5$\sigma$}
{\it and} are found to deviate from the rotational
pattern (for example in the position-velocity diagram);
\item A molecular outflow has been already detected through P-Cygni profiles
of molecular transitions (e.g. OH, H$_2$O).
\end{enumerate}
This criterion selects only those galaxies for which unambiguous evidence
for molecular outflows is found; sources which
only meet conditions 1 or 2 may also be hosting a powerful molecular outflow,
but additional observational evidence is needed in order to exclude that the
high velocity CO emission is tracing other mechanisms, such as gas rotation, infalls
or also disturbed gas kinematics due to an ongoing galaxy merging.

According to the criterion outlined above, we find evidence for massive molecular outflows
in the CO(1-0) data for four out of seven galaxies of our sample, namely: IRAS~F08572+3915, 
IRAS~F10565+2448, IRAS~23365+3604 and Mrk~273, which also 
exhibit prominent OH (79 and 119$\mu$m)
P-Cygni profiles in their Herschel--PACS spectra \citep{Sturm+11, Veilleux+13}. This proves that, once that
the presence of a molecular outflow has been assessed by the detection
of blue-shifted OH absorption, interferometric
CO observations are a very successful complementary technique to
study more in detail the characteristics of the molecular outflow, allowing us to
directly estimate the mass and size of the outflow and hence derive other important properties 
(e.g. outflow rates, kinetic energy, momentum rates etc.).
In the powerful type 2 quasar IRAS F23060+0505 we detect, at high significance,
broad CO(1-0) wings with ${\rm |v|>300~km~s^{-1}}$, which deviate
from the rotation pattern in the PV diagram. IRAS F23060+0505 therefore meets 
one of the conditions mentioned above and is very likely to host
a massive molecular outflow.  However, since gas at ${\rm |v|>500~km~s^{-1}}$ is only marginally
detected, and since IRAS F23060+0505 was not observed by Herschel, this source
does not fully qualify as a molecular outflow ``detection''.
The remaining two AGN-host galaxies of our sample, Mrk~876 and I~Zw~1, do not 
fully meet any of the previously mentioned conditions for the detection of
molecular outflows, although in the case of Mrk~876 we do observe some marginal evidence
for high velocity and spatially extended CO(1-0) emission. 
In the absence of additional observational evidence and of better quality data, we
use our CO(1-0) observations to estimate, for IRAS F23060+0505, Mrk~876 and I~Zw~1,
upper limits on the mass of molecular outflowing gas, on the outflow rates and other
outflow properties.

\subsection{Deriving Physical Properties of Molecular Outflows}

Our approach to the analysis of the broad wings of the CO(1-0) emission lines
in the seven sources is similar to that adopted for studying
the massive outflows detected in the CO(1-0) and CO(2-1) transitions 
in Mrk 231 \citep{Feruglio+10,Cicone+12} and in the [CII] 158$\mu$m line in 
the high redshift quasar SDSS J1148+5152 \citep{Maiolino+12}.

Our method for estimating the integrated flux and the spatial
extension of the CO(1-0) wings relies
almost exclusively on the analysis of the {\it uv} visibility data,
therefore producing results which are independent of aperture
effects and of the cleaning process. 
More specifically, we first estimate the 
3mm continuum emission from the line-free frequency
ranges in our wide-band (WideX) observations, and then subtract it directly from the {\it uv} data.
We then average the frequency channels corresponding to the CO(1-0)
wings, therefore producing {\it uv} tables for both the blue
and the red-shifted CO wing, which are used to estimate the 
integrated flux and the size of the CO wings.
To infer the flux (both for the core of the line and for the wings), we bin the visibility vs {\it uv} radius plots (form here 
simply referred to as ``{\it uv}'' plots) in baseline steps of 20m, and use the
visibility amplitude of the first point of this plot, which
approximates the total flux (i.e. the flux that would be measured at a zero antenna separation).
To infer the spatial extent of the wings, we fit the {\it uv} plot
with a circular Gaussian model, and use the resulting FWHM
as an estimate of the size (diameter) of the CO wings emission.
We stress that this technique, based on the {\it uv} data rather
than on the interferometric deconvolved maps, allows us
to estimate the size of structures which are comparable to the 
synthesised beam, on condition that the signal from the source
is detected at least at a ${\rm 8\sigma}$ significance. 
Moreover, the advantage of working directly in the $uv$ plane
for estimating the fluxes and the sizes, is that at this level no alterations are made on the
flux distribution as consequence of the ``cleaning'' process.
The fluxes and sizes inferred for the broad CO wings of our
seven sources are listed in Table \ref{table:COwings}. For I~Zw~1, where CO(1-0) broad wings
are not detected(\footnote {We note that  throughout this paper, the equivalence
``broad CO wings = molecular outflow'' is not valid, because
of the criterion that we define for the molecular outflow detection (Section 3.1).
More specifically, although IRAS~F23060+0505, Mrk~876, and
I~Zw~1 are all treated as non detections {\it for the molecular
outflow}, IRAS~F23060+0505 and Mrk~876 differ from I~Zw~1 in that
broad {\it CO wings} are (marginally) detected, and we can estimate 
their flux and size from our CO(1-0) observations.}), we provide upper limits.

We use the quantities listed in Table \ref{table:COwings} to derive
physically relevant outflow properties: outflow molecular
gas masses, mass-loss rates, molecular gas depletion time-scales due to
the outflow, kinetic powers and momentum rates of the molecular 
outflows, as detailed in the following.
Outflow rates and energetics are estimated by adopting the same method as for
Mrk 231 \citep{Feruglio+10}, NGC 6240 \citep{Feruglio+13a, Feruglio+13b}
and SDSS J1148+5152 \citep{Maiolino+12}.
We refer to those papers for an exhaustive description 
of our outflow model and its implications.
Here, we only briefly summarise the method.

The excitation study of the massive molecular outflow in the ULIRG Mrk 231 indicated
that the bulk of the molecular gas in the galaxy
and the outflowing gas have roughly similar excitation \citep[][Feruglio et al. in prep.]{Cicone+12}.
A direct implication
of this result is that, to evaluate
the molecular outflow mass from the CO luminosity of the broad wings, 
a better choice would be to use the standard
ULIRG CO-to-H$_2$  conversion factor, rather than
the lower, more conservative value assumed in \cite{Feruglio+10}. A similar result has been found
in M82, where a detailed modelling of the conversion factor of the gas in the outflow
has been performed by \cite{Weiss+01} and \cite{Walter+02}. As a consequence,
for all molecular outflows treated in this paper, we 
adopt $\alpha_{\rm CO(1-0)}$= 0.8 {$\rm M_{\sun} (K\,km\,s^{-1}\,pc^2)^{-1}$}.	

As already mentioned, for the galaxies of our PdBI sample, as well
as for Mrk 231 and NGC 6240, we 
assume the same spherical, or multi-conical, geometry as in
\cite{Maiolino+12}. As discussed in the latter paper, if the outflowing clouds
are assumed to uniformly populate the spherical (or multi-conical) region affected by the outflow,
then the outflow rate is given by the
relation:
\begin{equation} 
{\rm \dot{M}_{OF} = 3v\,(M_{OF}/R_{OF})}
\label{eq_outfl}
\end{equation} 
The formula above provides an appropriate description if the outflow is continuously refilled with clouds
ejected from the galactic gaseous disc, as expected in feedback scenarios. However, if the outflow
is instead associated with a single explosive event, in which the clouds are ejected, then the outflow
rate is more properly described by the gas mass of the clouds divided by
the dynamical time required by the clouds to reach the current location, resulting in an outflow rate
which is one third of the one given by Eq.~\ref{eq_outfl}. 
Higher angular resolution data would be required to discriminate between these two
possible scenarios, and in particular to clearly distinguish between a uniform
distribution of clouds within outflow cones (former scenario, described by Eq.~\ref{eq_outfl}),
or a shell-like geometry (latter scenario).
The data available so far
for our sources (including the well-studied Mrk 231), are more consistent with a uniform outflow, rather than with the shell-like
explosive scenario, hence in this work we adopt the prescription given by Eq.~\ref{eq_outfl}. However, the reader should
be aware that if future data favour an explosive-like (i.e. shell-like) scenario, then the figures for outflow rates,
kinetic powers and momentum rates may need to be revised downward. For some objects taken from the literature and discussed
in Section 4 this seems indeed to be the case (and this is properly taken into account
for the determination of the outflow rate for these sources).
In Eq.~\ref{eq_outfl} we conservatively use the average velocity in the range used to integrate
the broad CO wings (rather than the maximum velocity, both listed in Table \ref{table:res2}).
We note that, if the outflowing gas intersects our line of sight
(which is always true, as probed by the Herschel--PACS
detection of OH P-Cygni profiles),
Eq. ~\ref{eq_outfl} would provide a fairly accurate estimate
of the outflow rate also in the case of a bi-conical wind geometry.
The kinetic power and momentum rate of the outflow are 
simply calculated as ${\rm1/2\,v^2\,\dot{M}_{H_2,\,OF}}$ and
 ${\rm v\,\dot{M}_{H_2,\,OF}}$, respectively.
 
All the outflow properties described above 
are listed in Table \ref{table:res2} for the extended
sample of galaxies, consisting of our seven sources observed
with the PdBI, and of an heterogeneous sample of 12 local
galaxies hosting molecular outflows, drawn from the literature 
(further explanation in Section 4).

For IRAS F23060+0505 and Mrk~876, we use the values reported
 in Table \ref{table:COwings} for flux and size of the CO wings
to infer upper limits on the outflow properties.
In the case of I~Zw~1, the upper limits on the outflow parameters are 
calculated by using the upper limit that we estimate for the
flux in the CO wings, listed in Table \ref{table:COwings}, and 
hypothetical values for the
outflow extension (radius) and velocity of 0.5 kpc and 500 km s$^{-1}$, respectively.

\subsection{Analysis of the IRAM PdBI CO(1-0) Observations}

In the following we present our IRAM PdBI CO(1-0) observations 
of the seven local AGN-host galaxies listed in Table \ref{table:obs}, mainly focusing
on the evidence for broad CO(1-0) wings and their properties.
For additional information about the individual sources and 
further discussion of our CO(1-0) observations we refer to
Appendix A.


\begin{figure*}[!]
\centering
{\includegraphics[width=.7\textwidth, angle=0]{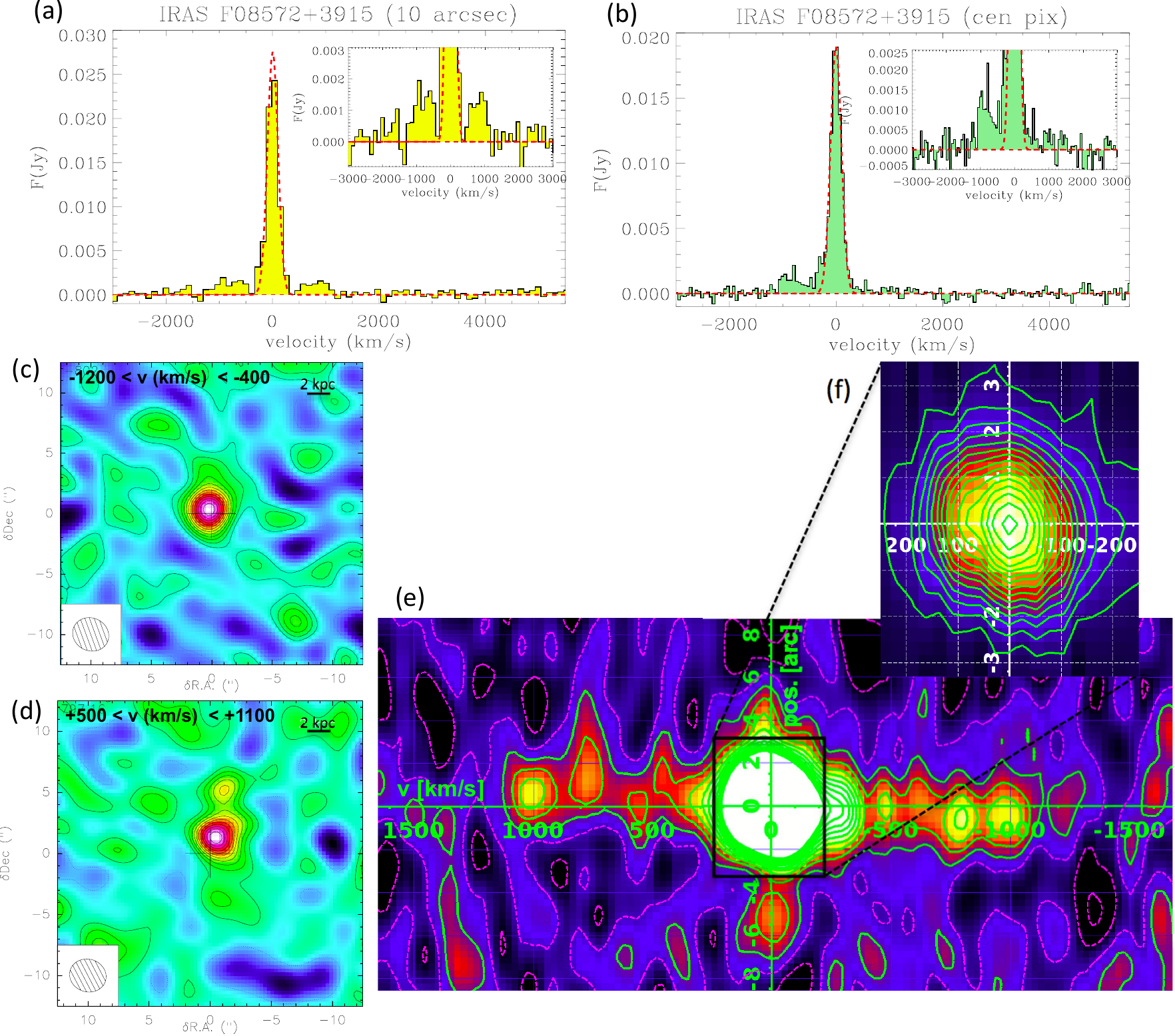}}\\
\caption{Continuum-subtracted IRAM-PdBI spectra, maps and position-velocity
diagram of the CO(1-0) emission line of IRAS F08572+3915. Panels (a) and (b) show 
the spectra extracted from a circular aperture with diameter
of 10 arcsec and from the central pixel (i.e. the centroid of the integrated CO emission), respectively. 
For display purposes, the spectra are re-binned in channels of 107 km s$^{-1}$ (a) and 54 km s$^{-1}$ (b).
In both panels the narrow core of the line is fitted with a single
Gaussian function (red dashed line). Note the clear detection of broad wings in the spectra.
(c-d) Map of the CO(1-0) emission integrated in the blue and red wings. 
Contours correspond to 1$\sigma$ (1$\sigma$ rms level of the two maps is
0.1 mJy beam$^{-1}$).
The physical scale at the redshift of the source is 1.122 kpc arcsec$^{-1}$.
The cross indicates the peak of the radio (VLBI) emission, which corresponds to the
NW nucleus of the merging pair IRAS F08572+3915.
(e-f) Position-velocity diagram along the major axis of the molecular disk rotation (as traced by the 
narrow core of the CO(1-0) line). The inset (f) highlights the rotation pattern
traced by the narrow component of the CO(1-0) line.
Contours are in steps of 1$\sigma$ (up to 10$\sigma$) in panel (e) and 5$\sigma$ in panel (f);
negative contours are in steps of 1$\sigma$ (magenta dashed lines).
Each pixel corresponds to 0.9 arcsec $\times$ 90 km s$^{-1}$ in diagram (e),
and to 0.3 arcsec $\times$ 30 km s$^{-1}$ in the inset (f).
Note the presence of high velocity gas, which does not follow the rotation
pattern and extends to velocities as high as 1200 km s$^{-1}$. This traces the same
molecular outflow detected in OH by Herschel--PACS.
} \label{fig:f0857} \end{figure*}

\begin{figure*}[!]
\centering
{\includegraphics[width=.7\textwidth, angle=0]{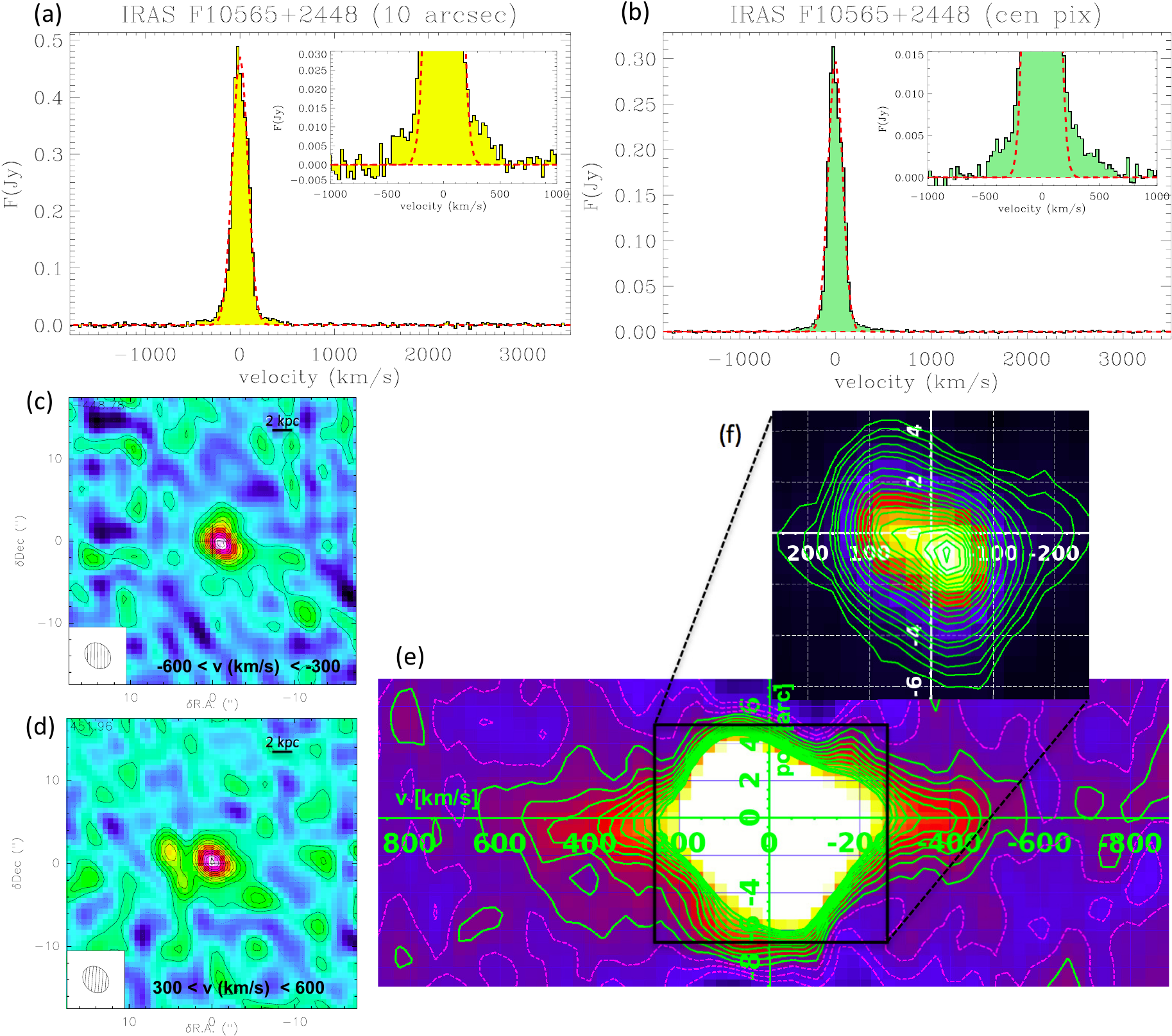}}\\
\caption{Continuum-subtracted IRAM-PdBI spectra, maps and position-velocity diagram 
of the CO(1-0) emission line
of IRAS F10565+2448. Panels (a) and (b) show the spectra extracted from a circular aperture with diameter
of 10 arcsec and from the central pixel, respectively. 
For display purposes, the spectra are re-binned in channels of 27 km s$^{-1}$.
In both panels the narrow core of the line is fitted with a single
Gaussian function (red dashed line). Note the appearance, in both spectra, of broad CO wings up to 
$\pm$600 km s$^{-1}$. 
(c-d) Cleaned map of the emission integrated in the blue and red-shifted CO(1-0) wings. 
Contours correspond to 1$\sigma$ (1$\sigma$ rms level of the two maps is 0.2 mJy beam$^{-1}$).
The cross indicates the peak of the radio (VLBI) emission.
The physical scale at the redshift of the source is 0.846 kpc arcsec$^{-1}$.
(e-f) Position-velocity diagram along the major axis of rotation. The inset (f) highlights the rotation pattern
traced by the narrow component of the CO(1-0) line.
 Contours are in steps of 1$\sigma$ 
(up to 10$\sigma$) in panel (e), and 5$\sigma$ in panel (f) (starting from 10$\sigma$ in panel (f));
negative contours are in steps of 1$\sigma$ (magenta dashed lines).
Each pixel corresponds to 1.3 arcsec $\times$ 53 km s$^{-1}$ in diagram (e),
and to 0.6 arcsec $\times$ 26 km s$^{-1}$ in the inset (f).
Note the clear signature of gas at velocities ${\rm |v| > 300~km~s^{-1}}$, which does not follow the rotation
curve and is therefore ascribed to the molecular outflow detected by Herschel--PACS.
} \label{fig:f1056} \end{figure*}

\begin{figure*}[!]
\centering
{\includegraphics[width=.7\textwidth, angle=0]{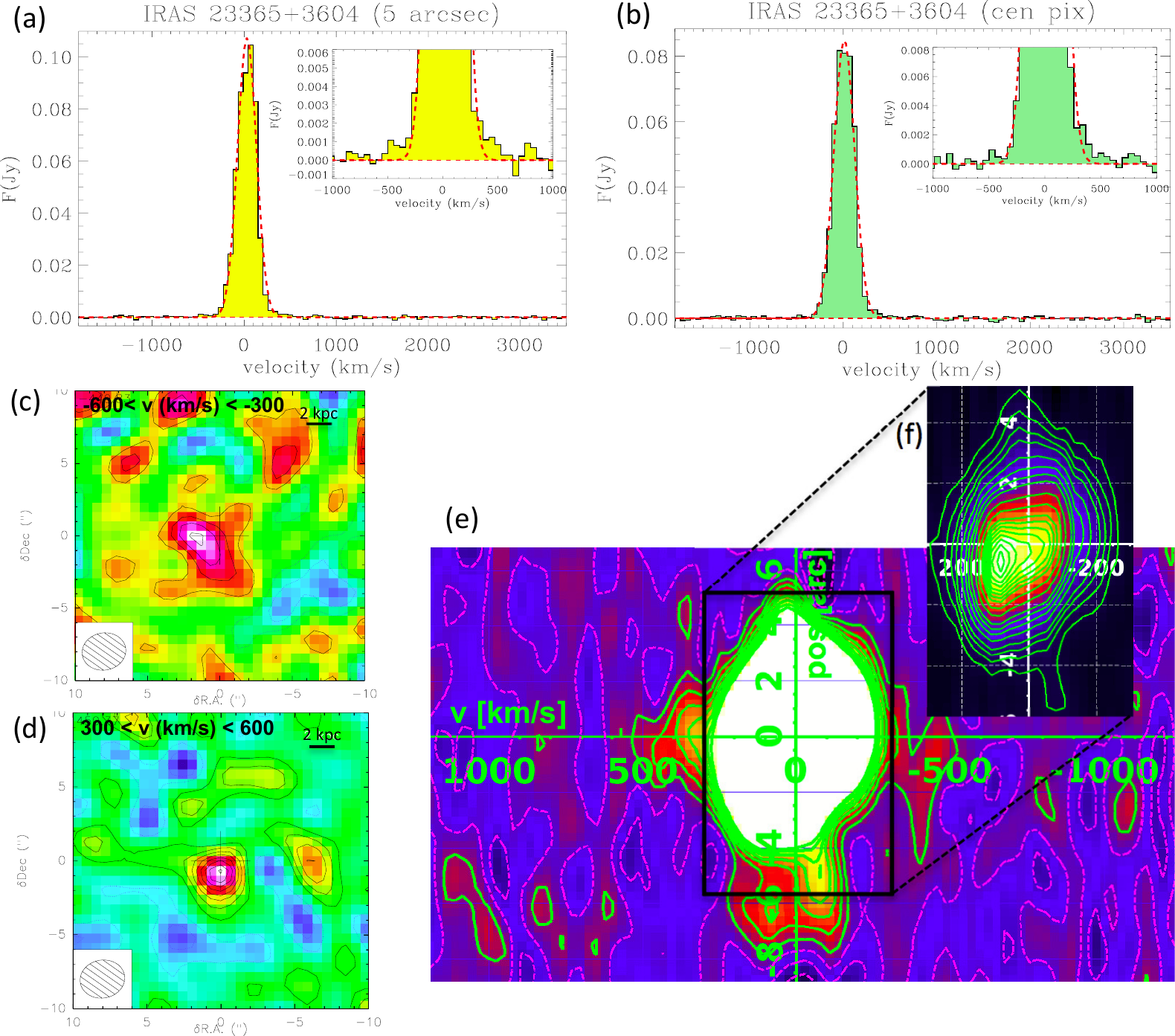}}\\
\caption{Continuum-subtracted IRAM-PdBI spectra, maps and position-velocity diagram 
of the CO(1-0) emission line
of IRAS 23365+3604. Panels (a) and (b) show the spectra extracted from a circular aperture with diameter
of 5 arcsec and from the central pixel, respectively. 
For display purposes, the spectra are re-binned in channels of 54 km s$^{-1}$.
In both panels the narrow core of the line is fitted with a single
Gaussian function (red dashed line).
(c-d) Cleaned map of the emission integrated in the blue and red-shifted CO(1-0) wings. 
Contours correspond to 1$\sigma$ (1$\sigma$ rms level of the two maps is 0.2 mJy beam$^{-1}$).
The physical scale at the redshift of the source is 1.232 kpc arcsec$^{-1}$.
(e) Position-velocity diagram along the major axis of rotation. 
The inset (f) highlights the rotation pattern traced by the narrow component of the CO(1-0) line; this
also suggests the presence of a counter-rotating disk visible at large radii (i.e. up to about 8 arcsec from
the nucleus, which correspond to $\sim$10 kpc).
Contours are in steps of 1$\sigma$ (up to 7$\sigma$) in panel (e), and 5$\sigma$ in the inset (f);
negative contours are in steps of 1$\sigma$ (magenta dashed lines).
Each pixel corresponds to 0.6 arcsec $\times$ 54 km s$^{-1}$ in diagram (e),
and to 0.3 arcsec $\times$ 27 km s$^{-1}$ in the inset (f).
Note the signature of CO(1-0) emission at velocities higher than $\pm$300 km s$^{-1}$, which does
not follow the rotation pattern and is therefore interpreted as associated with the molecular outflow
detected by Herschel--PACS.
} \label{fig:i2336} \end{figure*}


\subsubsection*{IRAS F08572+3915}

The continuum-subtracted
CO(1-0) emission line profile of IRAS F08572+3915 is shown in Figure \ref{fig:f0857}, in which we report the PdBI spectra
extracted from an aperture of diameter $\simeq$ 10 arcsec (a) and from the central beam
(b), which is actually simply measured from the central pixel (since the maps are
produced in units of mJy/beam).
In this source we estimated the continuum emission over the velocity ranges 
v$\in$(-3700, -2500) km s$^{-1}$
and v$\in$ (2500, 4000) km s$^{-1}$.
We measure a total integrated line flux of ${\rm S_{CO, TOT}}$ = (10.80 $\pm$ 0.70) Jy km s$^{-1}$
within $\pm$ 2000 km s$^{-1}$ from the systemic velocity, which is consistent
with previous single dish and interferometric observations
\citep{Evans+02, Solomon+97, Papadopoulos+12a}.

The large bandwidth of WideX allows us to detect, for the first time in this source, spectacular 
broad wings of the CO(1-0) emission line, tracing an outflow with velocities of up to 1200 km s$^{-1}$.
The two spectral profiles presented in Figure \ref{fig:f0857}(a-b)  show interesting differences. 
In the spectrum extracted from the 10 arcsec aperture, the blue and the red wing appear almost 
perfectly symmetric with respect to the line core, and their emission peaks at velocities of about 
$\pm$900 km s$^{-1}$ from the systemic.
Moreover, the narrow core can be fitted quite well with a single Gaussian function.
Conversely, in the spectrum extracted from the central pixel, the fit to the narrow core 
shows a clear excess of emission at v$\in$ (-200, -400) km~s$^{-1}$, which is not discernible in the
10 arcsec aperture spectrum, probably because of the lower signal-to-noise.
Nonetheless, the most obvious difference between the two spectra is the appearance of 
the high velocity wings: in the central pixel spectrum, while the blue-shifted wing appears 
very prominent, the red-shifted emission at ${\rm v\geq 500~km s^{-1}}$ is not recovered, 
which would suggest that this component is mostly extended. 
We note that some very recent follow-up observations of this source, obtained by us in PdBI A-configuration, 
resolve out the red wing, hence confirming that this must trace a very extended outflow.

The cleaned maps of the blue and red-shifted wings are presented in panels
(c-d) of Fig. \ref{fig:f0857}. Both the wings are detected at a $>10\sigma$ significance 
in the maps, which is also the significance
of detections of wings at ${\rm |v| > 500~km~s^{-1}}$(\footnote{ By restricting
the velocity range of integration for the blue wing to velocities ${\rm v<-500~km~s^{-1}}$
we obtain, similarly, a S/N=12 detection.}).
The molecular outflow is detected in correspondence of the north-west galaxy of the merging system,  
confirming findings of previous studies.
The peak of the red-shifted emission is slightly offset 
($\sim$ 1.4 arcsec) with respect to the pointing and phase centre (indicated by the cross).
Moreover, the red wing map exhibits a structure extended to the north, which
further confirms the hypothesis of a very extended red-shifted emission.
The fluxes measured in the blue and red-shifted wings of the CO(1-0) emission line
are reported in Table \ref{table:COwings}.
The analysis of the {\it uv} data gives an average size (FWHM) of
1.56 $\pm$ 0.33 kpc for the massive molecular outflow in IRAS F08572+3915 (see Appendix A for the details).
The velocity map of the central core emission (shown in the Appendix A)
shows a regular rotation pattern, with a major axis oriented east-west, similarly to the narrow
ionised gas component \citep{Rupke+Veilleux13a}. 
The position-velocity diagram along the rotation major axis is presented in panels (e-f) of Fig. \ref{fig:f0857}. 
The inset (f) shows the central rotation pattern, which is instead burned out
in panel (e) with the adopted colour cuts.
The signature of the extreme molecular outflow is clearly seen through 
the high velocity gas (along the horizontal axis of the plot) that does not follow the rotation curve of the galaxy. 

Summarising, the analysis of our PdBI CO(1-0) observations of the Sy2-ULIRG IRAS F08572+3915 convincingly
reveals the presence of a powerful and massive molecular outflow, which extends on both sides
of the galaxy up to velocities of $\sim$1200 km s$^{-1}$ and on physical scales of $\sim$1 kpc, carrying
a mass of molecular gas of about ${\rm 4\times 10^8~M_{\odot}}$, resulting in an outflow mass-loss rate
of $\sim$ 1200 ${\rm M_{\odot}~yr^{-1}}$.


\subsubsection*{IRAS F10565+2448}

The previous CO(1-0) observations of the Sy2-ULIRG IRAS F10565+2448, due to their narrow bandwidth,
could not detect the broad component superposed on the narrow core of the line. This broad
component is clearly visible in our new PdBI spectra shown in Figure \ref{fig:f1056}(a-b).
The broad wings in this source are detected up to about v = $\pm$ 600 km s$^{-1}$.
The continuum emission was estimated in the velocity ranges 
v$\in$(-3600, -1200) km s$^{-1}$
and v$\in$ (1200, 4000) km s$^{-1}$. We stress that the choice of a different
continuum window with respect to IRAS F08572+3915 and to the other sources, does not
introduce artificial bias on the shape and velocity extension of the broad wings. 
Indeed for each source we first inspect the appearance and shape of the broad wings in the 
non continuum-subtracted spectrum. 

We note that the spectrum extracted from a circular aperture with a diameter of 10 arcsec (Fig \ref{fig:f1056}a) and the
one extracted from the central pixel (Fig \ref{fig:f1056}b) show similar line profiles.
We measure a total integrated flux of ${\rm S_{CO, TOT}}$ = (108.0 $\pm$ 1.0) Jy~km~s$^{-1}$, which
is significantly ($>$ 50 \%) larger than the flux recovered by \cite{Downes+Solomon98} and 
$\sim$17\% higher than the value given by \cite{Chung+09} (see also \cite{Papadopoulos+12a}).
We report in Fig. \ref{fig:f1056}(c-d) the blue and red wing
maps obtained by integrating the CO(1-0) emission in the velocity ranges 
(-600, -300)~km~s$^{-1}$ and (300, 600)~km~s$^{-1}$.
CO wings at ${\rm |v| > 300~km~s^{-1}}$ are detected with a S/N = 12, and the position-velocity
diagram along the major axis of rotation (panels (e-f) of Fig. \ref{fig:f1056}), proves that this high velocity gas does deviate
from the central rotation pattern. Because Herschel--PACS observations revealed, in this source, 
prominent OH P-Cygni profiles with blue-shifted velocities of up to 950~km~s$^{-1}$ \citep{Veilleux+13}, the high significance 
detection of CO wings with ${\rm|v|>300~km~s^{-1}}$ deviating from the rotational pattern 
is sufficient, according to our criterion (Section 3.1), to claim
the detection of a massive molecular outflow. We note that very high velocity (${\rm|v|>500~km~s^{-1}}$) CO(1-0) emission
is detected in this galaxy at a significance level of 4$\sigma$.

Our observations resolve the outflow traced by the broad wings of the CO(1-0) line,
and the fit to the {\it uv} data results in an outflow radius of 1.1 kpc (see Appendix A for details).
 We note that this is probably a conservative estimate of the outflow extension, since
the maps of the wings in Fig. \ref{fig:f1056}(c-d) suggest the presence 
of even more extended structures, although at a low significance level (3$\sigma$).
The molecular outflow discovered in IRAS F10565+2448 appears to be
less powerful than in the other Sy-ULIRGs Mrk 231 and IRAS F08572+3915: indeed it has lower
velocity and, carrying a mass of molecular gas of ${\rm \sim 2\times 10^8~M_{\odot}}$,
has a mass-loss rate of ``only'' 300 ${\rm M_{\odot}~yr^{-1}}$.


\subsubsection*{IRAS 23365+3604}

Previous IRAM PdBI observations of the CO(1-0) transition in IRAS 23365+3604 suggest the presence of a 
compact rotating ring or disk
\citep{Downes+Solomon98}, responsible for the observed narrow line profile with a FWZI $\sim$ 500 km s$^{-1}$.
Our new observations recover a total CO flux of ${\rm S_{CO, TOT}}$ = (48.90 $\pm$ 0.20) Jy km s$^{-1}$
(Table \ref{table:COwings}), which is 60\% larger than the IRAM PdBI flux measured by \cite{Downes+Solomon98}, but 
consistent with the IRAM 30m flux reported by the same authors (see also \cite{Papadopoulos+12a}).

In Fig. \ref{fig:i2336} we present the continuum-subtracted CO(1-0) spectrum, extracted 
from an aperture with a diameter of 5 arcsec (a) and 
from the central pixel (b). The continuum emission was evaluated in the velocity ranges
v$\in$(-3700, -1200) km s$^{-1}$ and v$\in$ (1200, 4000) km s$^{-1}$.
The 5 arcsec aperture spectrum shows wings of the CO(1-0) emission line which
extend to about ${\rm \sim 600~km~s^{-1}}$ from the systemic velocity, as well as some
low significance emission at higher velocities. The CO(1-0) wings are instead less
prominent in the spectrum extracted from the central pixel.
The cleaned maps of the blue and red-shifted CO(1-0) wings (Fig. \ref{fig:i2336}(c-d)), integrated within 
the velocity ranges (-600, -300)~km~s$^{-1}$ and (300, 600)~km~s$^{-1}$, show that these are
detected at 4$\sigma$ and 8$\sigma$ significance, respectively. By combining their emission, we reach
a signal-to-noise of 9 (the combined map and {\it uv} plot are shown in the Appendix, Fig. \ref{fig:wings_vc26}).
The position-velocity diagram along the rotation major axis (Fig.~\ref{fig:i2336}(e-f)) reveals that such molecular 
gas at ${\rm |v|> 300~km~s^{-1}}$ does not follow the rotation pattern traced by the central core emission:
this suggests that this gas is tracing the same molecular outflow discovered by Herschel--PACS \citep{Veilleux+13}.

It is interesting to note that
the position-velocity diagram exhibits the signature of a second
disk or ring, counter-rotating with respect to the bulk of the molecular gas, which was not
detected by \cite{Downes+Solomon98}. This may provide further support to the hypothesis 
that IRAS 23365+3604 is a later merger.

We estimate the molecular outflow in IRAS 23365+3604 to carry a mass of molecular
gas of ${\rm \sim 1.5\times 10^8~M_{\odot}}$ and to have an approximate
radius of 1.2 kpc (obtained by fitting directly the $uv$ data), resulting in an outflow mass-loss rate of 170 ${\rm M_{\odot}~yr^{-1}}$, significantly
smaller than in the other ULIRGs.


\begin{figure*}[!]
\centering
{\includegraphics[width=.7\textwidth, angle=0]{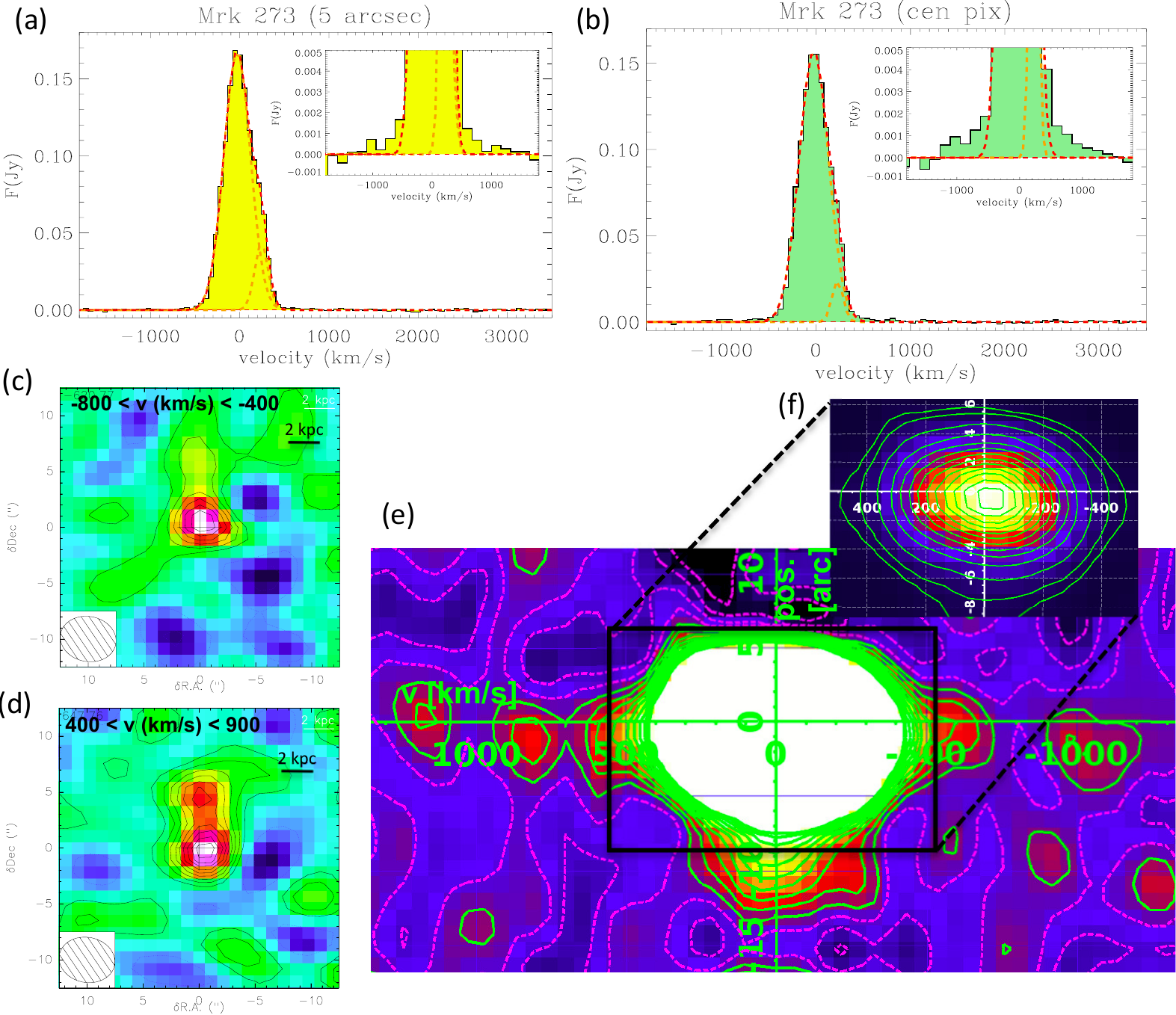}}\\
\caption{Continuum-subtracted IRAM-PdBI spectra, maps and position-velocity diagram of the CO(1-0) emission line
of Mrk 273. Panels (a) and (b) show the spectra extracted from a circular aperture with diameter
of 5 arcsec and from the central pixel, respectively. 
For display purposes, the spectra are re-binned in channels of 53 km s$^{-1}$ 
(and in channels of 160 km s$^{-1}$ in the insets).
In both panels (a) and (b) the narrow core of the line is fitted with two
Gaussian functions (red dashed lines, further comments in the text and in the Appendix).
Note the appearance, in both spectra, of broad CO wings, and in particular with velocities
higher than 500 km s$^{-1}$.
(c-d) Cleaned map of the emission integrated in the blue and red-shifted CO(1-0) wings.
Contours correspond to 1$\sigma$ (1$\sigma$ rms noise of the two maps is 0.2 mJy beam$^{-1}$). 
The physical scale at the redshift of the source is 0.746 kpc arcsec$^{-1}$.
(e-f) Position-velocity diagram along the major axis of rotation. Note that
our D-configuration observations do not completely resolve the disk rotation in this source.
Contours are in steps of 1$\sigma$ (up to 10$\sigma$) in panel (e), and 5$\sigma$ in the inset (f);
negative contours are in steps of 1$\sigma$ (magenta dashed lines).
Each pixel corresponds to 2.2 arcsec $\times$ 120 km s$^{-1}$ in diagram (e),
and to 1.1 arcsec $\times$ 54 km s$^{-1}$ in the inset (f).
Gas at velocities higher than 400 km s$^{-1}$ does not follow the
rotation pattern and is most likely associated to the molecular outflow detected by Herschel--PACS.
} \label{fig:mrk273} \end{figure*}

\subsubsection*{Mrk 273}

The continuum-subtracted CO(1-0) spectra of Mrk 273, extracted from a circular aperture
of 5 arcsec diameter and from the central pixel, 
are shown in Figure \ref{fig:mrk273}(a-b).
The continuum emission was estimated in the velocity ranges v$\in$(-3000, -2000) km s$^{-1}$
and v$\in$ (2000, 3700) km s$^{-1}$.
The total CO(1-0) line flux that we measure within $\pm$ 2000 km s$^{-1}$ is
${\rm S_{CO, TOT}}$ =(90.60 $\pm$ 0.50) Jy km s$^{-1}$ (Table \ref{table:COwings}),
consistent with the flux recovered by previous single dish observations 
\citep{SSS91, Solomon+97, Albrecht+07, Papadopoulos+12a}, and about 15\% larger 
than the value obtained by \cite{Downes+Solomon98} with the PdBI. 
The CO(1-0) line spectra show a broad emission core,
which has already been detected by previous observations, mostly tracing 
molecular gas in a nuclear rotating disk (see Appendix for details).
We fit this spectral core component with the sum of two Gaussian functions (red dashed line in Fig. \ref{fig:mrk273}a-b).
In addition to this component, 
our new observations, obtained with a much larger bandwidth than previous
data, reveal extra CO(1-0) emission at high blue and red-shifted velocities. 
Such broad blue and red CO wings are detected at a significance of 6$\sigma$ 
and 9$\sigma$, respectively,
as shown by their integrated maps in panels (c) and (d). 
The rotation in the central concentration of molecular gas is not clearly resolved by our D-configuration (i.e.
low spatial resolution) observations (Fig. \ref{fig:mrk273}f). However, PdBI
observations of the CO(2-1) transition presented by
\citealt{Downes+Solomon98} resolve the kinematics of the nuclear
disk of Mrk~273, revealing a velocity centroid that changes from -300 to 200 km s$^{-1}$.
The PV diagram in panels (e-f) shows that the CO(1-0) emission at ${\rm |v| > 400~km~s^{-1}}$ 
does not follow the central rotation pattern, and is therefore ascribable to the molecular outflow
detected by Herschel--PACS \citep{Veilleux+13}.
We note that even if restricting the velocity range of integration to velocities
${\rm |v| > 500~km~s^{-1}}$, the broad CO(1-0) wings of Mrk~273 are still
both detected at a significance of 5$\sigma$.

In principle, the need for two Gaussians to reproduce the core of the CO(1-0) emission is
not an anomaly (see also the case of Mrk 231, \citealt{Cicone+12}), but
it can reflect the kinematics of the rotating gas in the central molecular disk.
However, we cannot rule out the possibility that the asymmetry on the red side of the 
line core, which is fitted by the smaller Gaussian centred at about 250 km s$^{-1}$, is also 
a signature of outflow. Indeed we detect in our channel maps (not shown)
a blob elongated towards the North at velocities ${\rm v > 150~km~s^{-1}}$,
further confirming our hypothesis that the red-shifted peak of CO(1-0) emission, manifestly
visible in the spectra in Fig. \ref{fig:mrk273}(a-b), is tracing the
low-velocity component of a prominent red-shifted outflow. Such outflow is mostly
extended to the North up to velocities of 500 km s$^{-1}$, while it is more compact 
at larger velocities.

For the molecular outflow in Mrk~273, we estimate an average 
radius of about 600 pc (see Appendix
for details) and a mass of ${\rm \sim 1.7\times 10^8~M_{\odot}}$, yielding
a mass-loss rate of 600 ${\rm M_{\odot}~yr^{-1}}$.


\begin{figure*}[!]
\centering
{\includegraphics[width=.7\textwidth, angle=0]{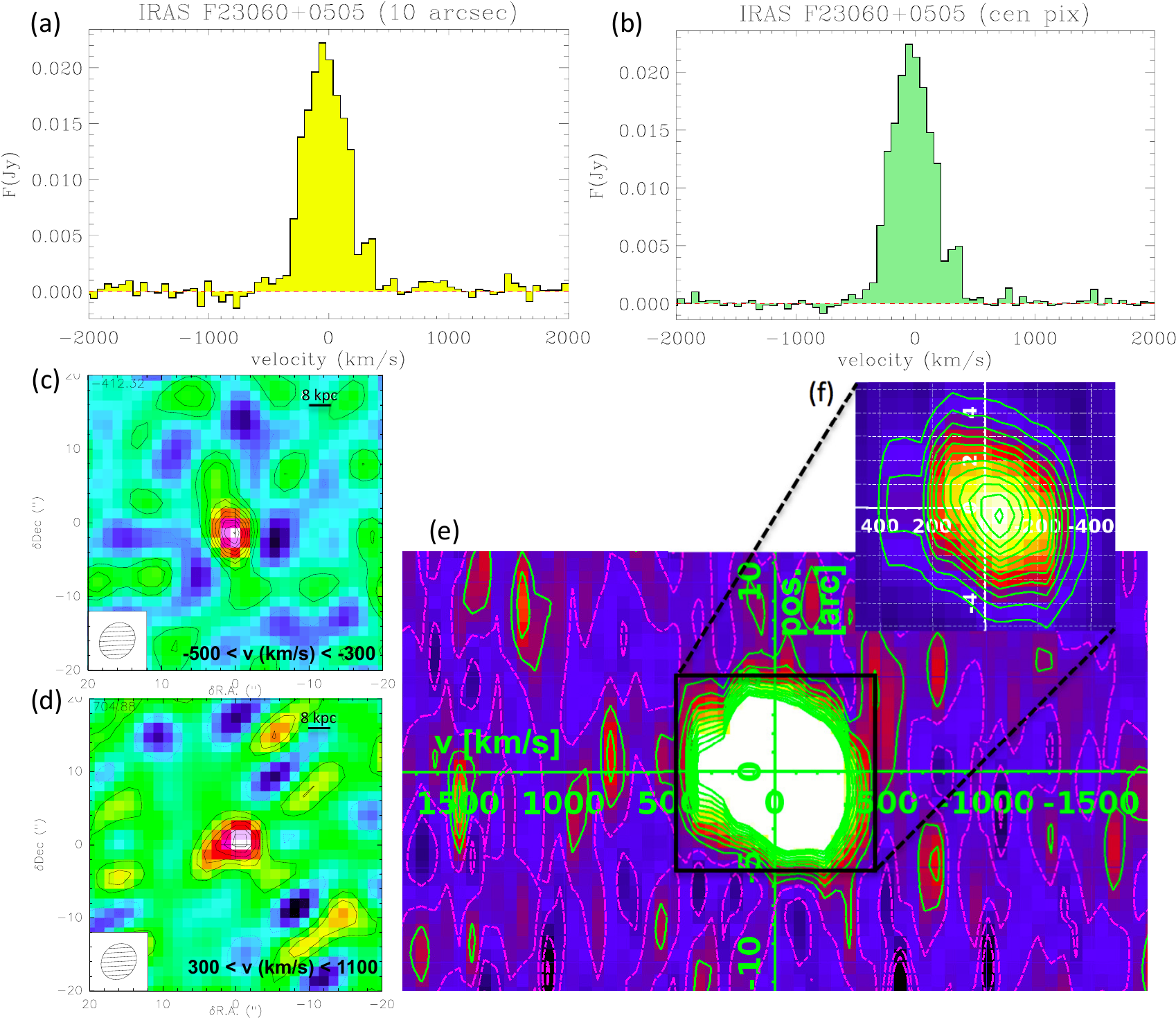}}\\
\caption{Continuum-subtracted IRAM-PdBI spectra, maps and position-velocity diagram 
of the CO(1-0) emission line in IRAS F23060+0505.
Panels (a) and (b) show the spectra extracted from a circular aperture with diameter
of 10 arcsec and from the central pixel, respectively.
For display purposes, the spectra are re-binned in channels of 60 km s$^{-1}$.
(c-d) Cleaned maps of the emission integrated in the blue and red-shifted CO(1-0) wings.
Contours correspond to 1$\sigma$ (1$\sigma$ rms noise level is 0.4 mJy beam$^{-1}$ in the blue 
wing and 0.2 mJy beam$^{-1}$ in the red wing). 
The physical scale at the redshift of the source is 2.934 kpc arcsec$^{-1}$.
(e-f) Position-velocity diagram along the major axis of rotation. Panel (f) highlights the
rotational kinematics traced by the narrow core of the CO(1-0) emission.
Contours are in steps of 1$\sigma$ (up to 10$\sigma$) in panel (e) and of 5$\sigma$ in the inset (f).
Negative contours are in steps of 1$\sigma$ (magenta dashed lines).
 Each pixel corresponds to 0.7 arcsec $\times$ 60 km s$^{-1}$ in both diagrams (e) and (f).
Note that the CO(1-0) wings at ${\rm |v|> 300~km~s^{-1}}$ deviate from the rotation pattern,
and are likely tracing an outflow of molecular gas, similarly to the one discovered in the other sources.}
 \label{fig:i2306} \end{figure*}

\begin{figure*}[!]
\centering
{\includegraphics[width=.7\textwidth, angle=0]{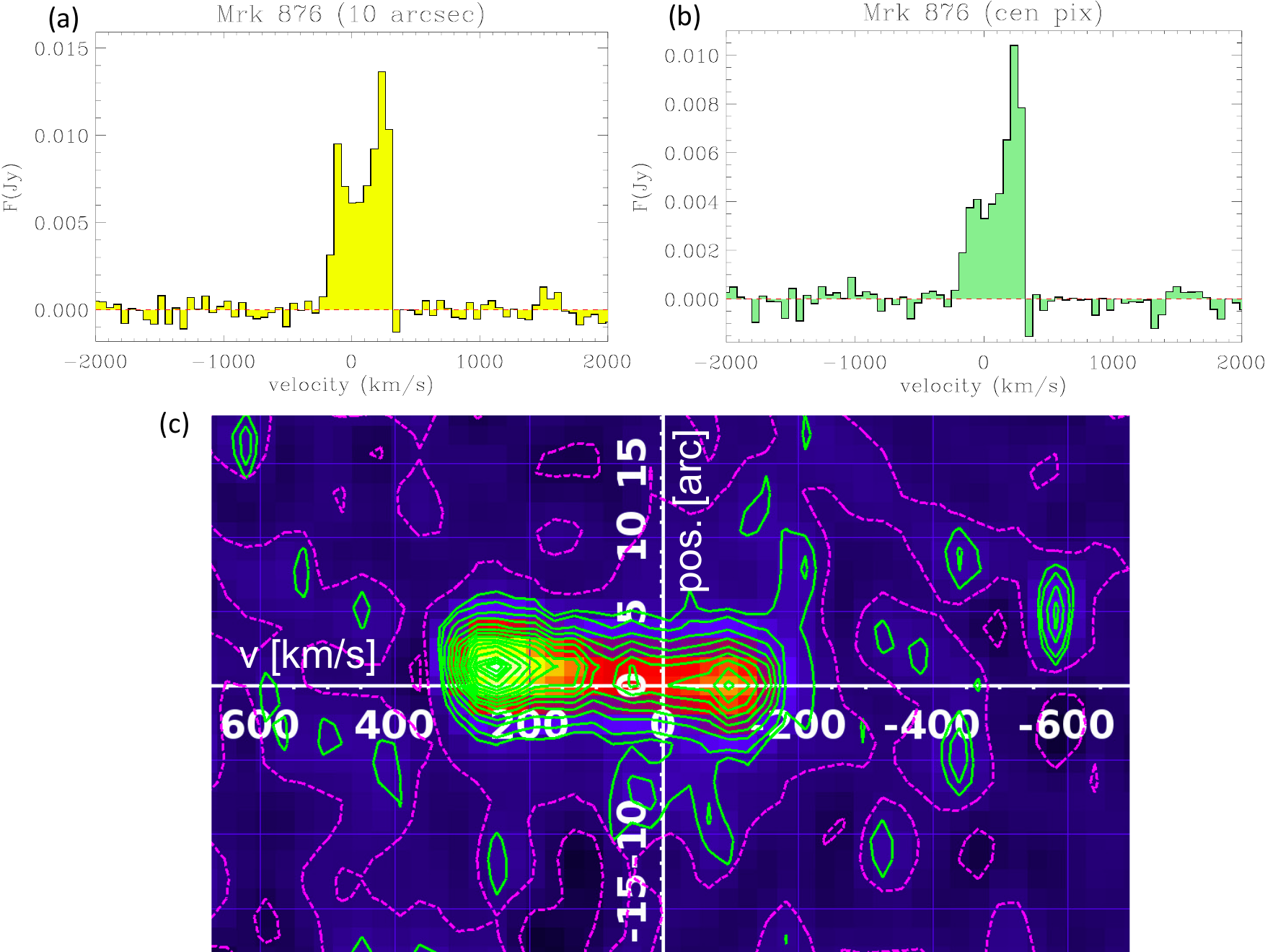}}\\
\caption{Continuum-subtracted IRAM-PdBI spectra and position-velocity diagram 
of the CO(1-0) emission line
of Mrk 876. Panels (a) and (b) show the spectra extracted from a circular aperture with diameter
of 10 arcsec and from the central pixel, respectively. 
For display purposes, the spectra are re-binned in channels of 57 km s$^{-1}$.
No clear evidence for broad wings is observed.
(c) Position-velocity diagram along the major axis of the CO(1-0) rotation. Contours are in steps of 1$\sigma$;
negative contours are in steps of 1$\sigma$ (magenta dashed lines).
 Each pixel in diagram (c) corresponds to 1.3 arcsec $\times$ 29 km s$^{-1}$.
The physical scale at the redshift of the source is 2.295 kpc arcsec$^{-1}$.
} \label{fig:mrk876} \end{figure*}

\begin{figure*}[!]
\centering
{\includegraphics[width=.7\textwidth, angle=0]{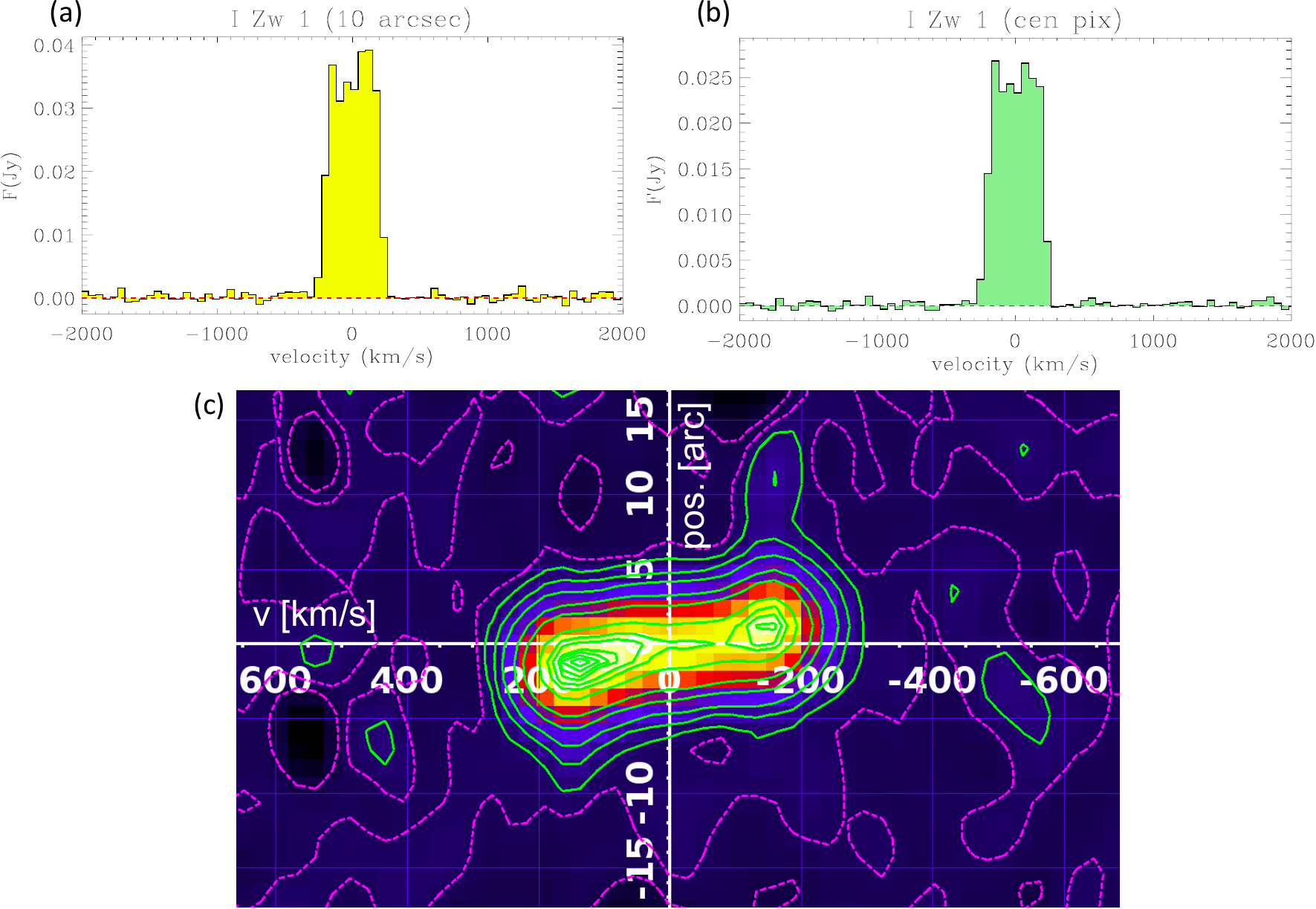}}\\
\caption{Continuum-subtracted IRAM-PdBI spectra and position-velocity diagram of the CO(1-0) emission line
of I~Zw~1. Panels (a) and (b) show the spectra extracted from a circular aperture with diameter
of 10 arcsec and from the central pixel, respectively. 
For display purposes, the spectra are re-binned in channels of 54 km s$^{-1}$.
No evidence for broad wings is observed.
(c) Position-velocity diagram along the major axis of rotation. 
Contours 
correspond to: 1$\sigma$, 3$\sigma$, 6$\sigma$, 12$\sigma$, 18$\sigma$, 30$\sigma$,
36$\sigma$, 39$\sigma$, 42$\sigma$, 48$\sigma$; negative contours are in steps of 1$\sigma$ (magenta dashed lines).
Each pixel in diagram (c) corresponds to 1.2 arcsec $\times$ 27 km s$^{-1}$.
The physical scale at the redshift of the source is 1.175 kpc arcsec$^{-1}$.
} \label{fig:izw1} \end{figure*}


\subsubsection*{IRAS F23060+0505}

To our knowledge, this is the first published observation of the CO(1-0) molecular transition in the
powerful type 2 QSO and ULIRG IRAS F23060+0505. 
We measure a total integrated CO flux of ${\rm S_{CO, TOT}}$ = (15.30 $\pm$ 0.20) Jy km s$^{-1}$.
The continuum-subtracted CO(1-0) spectra, extracted from an aperture with diameter
of 10 arcsec and from the central pixel, are shown in panels (a-b) of Figure \ref{fig:i2306}.
The spectrum extracted from a 10 arcsec aperture shows some indication of broad CO
emission, which extends up to at least ${\rm v \sim 1000~km~s^{-1}}$ in the red-shifted side. Such evidence
of high velocity emission is much more marginal in the central pixel spectrum. An interesting emission feature
at red-shifted velocities of ${\rm v\in (200, 400)~km~s^{-1}}$ is clearly distinguishable in both
spectra.

The cleaned maps of the broad CO wings, integrated within velocities of
$\rm -500 <v<-300$ km s$^{-1}$ and $\rm 300 <v<1100$ km s$^{-1}$, are presented in
panels (c-d) of Fig. \ref{fig:i2306}: these show that the blue and red wings are 
detected at 9$\sigma$ and 6$\sigma$ significance. Moreover, the position-velocity diagram
in Fig. \ref{fig:i2306}(e-f) reveals that such emission at ${\rm |v| > 300~km~s^{-1}}$ is clearly
deviating from the central rotational pattern. This is particularly evident for the 
red-shifted CO emission: the PV diagram confirms that the feature appearing in the spectra
at ${\rm v\in (200, 400)~km~s^{-1}}$ does not trace rotating gas, but it is likely tracing
either an outflow or an inflow of molecular gas. 
The blue-shifted gas at ${\rm v < -300~km~s^{-1}}$ is mostly elongated to the north
of the galaxy, as shown by the map in panel (c) of Fig. \ref{fig:i2306}.
We also note the detection, in the  PV diagram (as well as in the 10 arcsec aperture
spectrum), of very high velocity  (${\rm v > 500~km~s^{-1}}$)
red-shifted CO emission, although at a low significance level (i.e. S/N$<$5).
As already mentioned in Section 3.1, according to our criterion, IRAS F23060+0505 does not fully qualify as
a reliable ``outflow'' detection, since high velocity gas is only detected at low significance
and we lack independent confirmation from Herschel data of the presence of a molecular outflow in this object.
Additional observations are certainly required to understand the nature
of the high velocity CO emission in this powerful QSO.

We use the flux and the spatial extent of the broad CO wings (estimated to be as large
as 4 kpc in radius, see Appendix for further details) to estimate upper limits
on the outflow mass and mass-loss rate in this source, which may be as large
as ${\rm \sim 4\times 10^9~M_{\odot}}$ and 
1500 ${\rm M_{\odot}~yr^{-1}}$, respectively.


\subsubsection*{Mrk 876 (PG 1613+658)}

Previous interferometric and single dish observations of the 
unobscured QSO and LIRG Mrk 876 detected 
the CO(1-0) emission at high signal-to-noise \citep{Evans+01, Evans+06}, and estimated a total CO integrated flux
that is completely consistent with our PdBI measure of ${\rm S_{CO, TOT}}$ = (8.5 $\pm$ 1.5) 
Jy km s$^{-1}$. 
The continuum-subtracted PdBI CO(1-0) spectrum, extracted from an aperture with
diameter of 10 arcsec and from the central pixel, is presented in Fig. \ref{fig:mrk876}(a-b).
The signal-to-noise of our spectrum is low due to the short integration (6 hours of on source time 
with the PdBI, see Table \ref{table:obs}). Both the aperture and the central pixel spectra
do not show any clear evidence of broad CO wings.
We marginally detect some high velocity blue and red-shifted CO emission in the maps,
obtained by averaging the frequency channels corresponding to velocities of 
(-500, -300) km s$^{-1}$ and (400, 1700) km s$^{-1}$ (not shown). The significance
of this detection is however low. Moreover, in the position-velocity diagram (in Fig. \ref{fig:mrk876}c) there is no
apparent trace of the broad CO wings that we detect in the maps.
For this reason, and because Herschel--PACS observations of the OH transition in this source
did not find any evidence for a blue-shifted OH absorption (which would be indicative of a molecular
outflows), Mrk~876 does not satisfy our criterion for claiming the detection of a massive molecular outflow.  

By combining the blue and red-shifted CO(1-0) emission, we obtain
enough S/N (i.e. S/N = 10, Fig. \ref{fig:wings_uc2a}) to evaluate the physical extent of the
broad wings through the analysis of the {\it uv} data (see Appendix for details). Similarly to
IRAS F23060+0505, we use the flux and radius of 3.6 kpc of these putative CO wings to estimate the
upper limits of ${\rm \sim 3\times 10^9~M_{\odot}}$
on the mass of outflowing gas and of 1800 ${\rm M_{\odot}~yr^{-1}}$ on the outflow rate in Mrk~876.
We stress that additional interferometric observations are required to 
investigate this source in details, and to draw more definite conclusions
about the presence and the properties of a possible molecular outflow.

Finally, we note that the position-velocity diagram in panel (c) of Fig. \ref{fig:mrk876}
 shows evidence for a double peak (also clearly discernible in the spectrum
 in panel (a)), which may be tracing a ring-like distribution of molecular gas.


\subsubsection*{I Zw 1 (PG 0050+124)}

No broad CO(1-0) emission is detected in the spectra and maps of the LIRG and luminous
QSO host galaxy I~Zw~1.
A comparison between our PdBI flux, ${\rm S_{CO, TOT}}$ = (23.5 $\pm$ 1.0) 
Jy km s$^{-1}$, and previous single-dish observations of this galaxy \citep{Evans+06, Papadopoulos+12a}, shows that we
may be missing over 20\% of more extended emission, and therefore the presence
of a faint and broad CO(1-0) component, cannot be ruled out with our data. 
 We note that this object does not show OH absorption either: in contrast, 
OH is detected in pure emission, red-shifted by $\sim$ 200 km s$^{-1}$ \citep{Veilleux+13}.
We present in Fig. \ref{fig:izw1}(a-b) the continuum-subtracted CO(1-0) spectrum of I~Zw~1, which is
overall narrow, with a FWZI of 500 km s$^{-1}$. The position-velocity diagram obtained along
the major axis of the CO(1-0) rotation (panel (c) of Fig. \ref{fig:izw1}) shows a clearly
double-peaked emission, suggesting that I~Zw~1 may host a ring-like distribution of molecular 
rotating gas in its nucleus.

For this source, we estimate upper limits for the flux in the broad CO(1-0) component 
as follows. We average the frequency channels corresponding to 
the velocity ranges of v$\in$(-650, -250) km s$^{-1}$ and v$\in$(500, 750) km s$^{-1}$, and produce
the cleaned maps, which however do not reveal any emission above the noise
level. Our upper limits for the fluxes in the blue and red CO wings correspond to 
to the rms noise of the two maps (i.e. 0.13 mJy beam$^{-1}$ and 0.19 mJy beam$^{-1}$) multiplied by a 
factor of 3, and integrated over 400 and 250 km s$^{-1}$, respectively.
We use these upper limit fluxes and an hypothetical CO wings radius of 500 pc (i.e. roughly comparable to 
the extent of the narrow core emission in I~Zw~1, see Appendix for more details), to estimate
upper limits to the outflowing molecular mass and outflow rate in this source.


\begin{table*}
\footnotesize
 \centering
 \begin{minipage}{150mm}
  \caption{Integrated Fluxes and Spatial Extensions of the CO(1-0) Broad Wings}
  \label{table:COwings}
\begin{tabular}{@{}lcccccc@{}}
\hline
\hline
Object 				&	Total Line		 			&  \multicolumn{2}{c}{Blue Wing}  					& \multicolumn{2}{c}{Red Wing} 		& Wings Size			\\
\hline	
					&	${\rm S_{CO}}$				& Vel. Range 			& ${\rm S_{CO}}$			&Vel. Range		& ${\rm S_{CO}}$	& FWHM	 				\\
				 	&	(Jy km s$^{-1}$)			& (km s$^{-1}$)			& (Jy km s$^{-1}$)			& (km s$^{-1}$)		& (Jy km s$^{-1}$)	& (kpc)						\\	
\hline
 IRAS F08572+3915	 	& 	(10.80  $\pm$ 0.70)			& (-1200, -400)			&  (1.69 $\pm$ 0.42)		& (500, 1100)		& (1.22 $\pm$ 0.36) &  (1.64 $\pm$ 0.34)	 					\\
IRAS F10565+2448		& 	(108.0 $\pm$ 1.0)			& (-600, -300)			&  (1.72 $\pm$ 0.54)		& (300, 600)		& (1.50 $\pm$ 0.24)	&  (2.19 $\pm$ 0.22)						\\
 IRAS 23365+3604	 	&	(48.90 $\pm$ 0.20)			& (-600, -300)			& (0.40 $\pm$ 0.12) 		&  (300, 600)		& (0.46 $\pm$ 0.12) &  (2.45 $\pm$ 0.70)			\\
Mrk 273				&	(90.60 $\pm$ 0.50)			& (-800, -400)			& (1.04 $\pm$ 0.50)		& (400, 900) 		& (2.07 $\pm$ 0.52)  &  (1.10 $\pm$ 0.50)				\\	
IRAS F23060+0505		&	(15.30 $\pm$ 0.20)			& (-500, -300)			& (0.75 $\pm$ 0.18)	         & (300, 1100)		& (1.51 $\pm$ 0.40)	&  (8.1 $\pm$ 2.9)		\\
Mrk 876				&	(8.5  $\pm$ 1.5)			& (-500, -300)			& (0.65 $\pm$ 0.63) 			& (400, 1700)		& (3.1 $\pm$ 1.5) 		& (7.1 $\pm$ 1.4) 				\\
I Zw 1				&	(23.5 $\pm$ 1.0)			& (-650, -250)			& $\leq$ 0.16       			& (500, 750)		& $\leq$ 0.14			& Non detection 			\\
\hline
\end{tabular}
\end{minipage}
\end{table*}


\section{Extended Sample}
\subsection{Description}

In the following analysis we combine our 7 sources observed with the PdBI with a heterogeneous sample of 12 local galaxies
in which outflows of molecular gas have been constrained by mapping the
two lowest-J CO rotational transitions (i.e. the \textit{J=}1-0 and \textit{J=}2-1).
The complete list of sources is given in Table \ref{table:res1}. In the following, we briefly discuss the properties of the
sample of sources drawn from the literature.

The Sy1-ULIRG Mrk~231 can be naturally incorporated in our analysis, since its powerful molecular outflow
was studied using the same methods that we employ for the new sources presented in this paper.
We adopt the CO integrated fluxes in the blue and red-shifted wings and the outflow
size (estimated from the \textit{J=}1-0 transition) that are listed in \cite{Cicone+12}, where the PdBI 
data exploited by \cite{Feruglio+10} were combined with more recent observations.

More careful consideration is needed for the other cases. 
\cite{Alatalo+11} discovered a molecular outflow in the early-type LINER-host galaxy NGC 1266, 
using a combination of single dish (IRAM 30m) and interferometric (SMA, CARMA)
observations of the CO~\textit{J=}1-0, \textit{J=}2-1 and \textit{J=}3-2 transitions. To calculate the 
outflow mass, we exploit the flux of the CO(1-0) broad component
provided by these authors. 
By assuming the bipolar outflow geometry
proposed by \cite{Alatalo+11}, the outflow size (radius) is about 0.450 kpc, which refers
to the average width of the blue and red-shifted CO(2-1) emission as measured with the SMA.
We note that, by using the \textit{J=}2-1 transition, we may be slightly underestimating the outflow size
\citep{Cicone+12}.

The molecular outflow in the starburst galaxy M82 has been investigated
by \cite{Walter+02} using detailed CO(1-0) maps obtained with the OVRO interferometer.
We adopt the CO integrated flux of ${\rm S_{CO}}$ = 7240 Jy km s$^{-1}$, estimated
by \cite{Walter+02} for the outflow north and south of the molecular disk.
The wind geometry is almost spherical, with a radius of about 0.8 kpc.

The lenticular galaxy NGC 1377 and the LINER NGC 1266 share some similarities. SMA CO(2-1) observations of
NGC 1377 revealed the presence of a molecular outflow, traced by broad wings of the emission line profile
\citep{Aalto+12b}. In this case, both the flux of the broad component, determined by fitting the
spectrum with two Gaussians (narrow plus broad), and the  outflow radius of 0.2 kpc (under the hypothesis
of a biconical outflow), were estimated using the CO(2-1) transition. The same caveats as for NGC 1266
are valid, with the addition that, in this object, the outflow mass is also obtained using the CO(2-1) line
instead of the CO(1-0) line. To estimate the total CO(1-0) flux in the outflow we assume
thermalised optically thick CO emission, i.e. ${\rm L'_{CO(1-0)}=L'_{CO(2-1)}}$ \citep{Solomon+05}.

The complex morphology of the CO(1-0) emission in the AGN-dominated ULIRG and merger NGC 6240
has been investigated in detail using the compact and the most extended configurations of the PdBI
\citep{Feruglio+13a, Feruglio+13b}.  \cite{Feruglio+13b} clearly show that the blue-shifted CO(1-0) wing
traces molecular gas outflowing from the southern nucleus. Conversely, most of the red-shifted emission
arises from the region between the two nuclei and, although its origin is strongly debated, 
it is believed to be closely related to the merging process. 
Given the very high velocities observed in this red-shifted component, it is tempting
to presume that, among the other mechanisms proposed and discussed in \cite{Feruglio+13b} and references therein, 
there may be an outflow contribution, 
but the current data do not allow us to quantify it. We therefore follow \cite{Feruglio+13b} 
and estimate the outflow rate from the blue wing of the CO(1-0) line, in the velocity range
v$\in$(-500, -200) km s$^{-1}$, which has an integrated flux of 17.8 Jy km s$^{-1}$
and a total size of 1.3 kpc. We note that in \cite{Feruglio+13b} a conservative CO-to-H$_2$
conversion factor of 0.5 {$\rm M_{\sun} (K\,km\,s^{-1}\,pc^2)^{-1}$} has been used, while
here we adopt the same value of 0.8 as in the other ULIRGs.

\cite{Sakamoto+06b} discovered a wind of molecular gas in the starburst merger
NGC 3256, by using SMA observations of the CO(2-1) emission line. They integrate the
blue and red-shifted high-velocity emission in the CO(2-1) spectrum, within the velocity
ranges (-195, -165) km s$^{-1}$ and (215, 425) km s$^{-1}$. The resulting total
CO(2-1) flux in the broad wings is 28 Jy km s$^{-1}$, and we use this flux to estimate the
mass of the molecular outflowing gas. We also adopt the same biconical geometry of the outflow
as \cite{Sakamoto+06b}, with an outflow extension of 0.5 kpc.

The starburst galaxy NGC 3628 hosts a sub-kpc scale molecular outflow to the north
of the galactic disk traced by CO(1-0) emission \citep{Tsai+12}.
This is spatially coincident with the northern ejection point of a large scale ($\sim$10 kpc)
plasma outflow, which instead extends both to the north and to the south of the disk, as evidenced by
soft X-ray Chandra observations. \cite{Tsai+12} infer a CO(1-0) flux of  82 Jy km s$^{-1}$ for the molecular outflow, by using detailed 
interferometric (NMA) maps. They suggest that this gas is expanding isotropically to the northern side of the galactic disk,
with an average velocity of 50 km s$^{-1}$, up to a radius of 0.4 kpc.

The CO(2-1) emission in the prototypical starburst galaxy NGC 253, imaged with the SMA by \cite{Sakamoto+06a}, displays 
two molecular gas features at a distance
of 0.2 and 0.7 kpc from the galactic centre, which have been
interpreted as expanding super-bubbles. \cite{Sakamoto+06a} estimate their
molecular gas masses and diameters of $\sim$100 pc from the CO(2-1) channel maps, and assume cylindrical symmetry and
non-linear expansion (i.e. R(t) $\propto$ t$^{1/2}$) to model the outflow. We adopt the same assumptions as
these authors, but we apply a correction to the mass estimates to account for the use of a different
CO-to-H$_2$ conversion factor (see notes in Table \ref{table:res2}).
Very recently, \cite{Bolatto+13} have studied with unprecedented detail the molecular outflow of NGC 253, using 
new ALMA observations of the CO(1-0) transition. The outflow rate inferred from these new data of ${\rm \dot{M}_{OF}\sim 3-9~M_{\odot}~yr^{-1}}$, is
in excellent agreement with our estimate based on the work by \cite{Sakamoto+06a} (Table  \ref{table:res2}).

The integrated CO(1-0) PdBI map of the LINER and starburst galaxy NGC 6764
shows evidence for a molecular outflow extending to the north of the galactic disk \citep{Leon+07}. 
To estimate the outflow rate, we use the CO flux of 1.4 Jy km s$^{-1}$ provided by \cite{Leon+07} and
we assume that this is emitted by gas uniformly distributed in a conical volume with radius of
0.6 kpc, outflowing at a velocity of 170 km s$^{-1}$.

Multi-transitional CO interferometric observations have shown the presence of a massive molecular
outflow in the Sy2-LIRG NGC~1068, revealed by broad wings of the molecular emission lines \citep{Krips+11}.
The blue and the red-shifted high velocity CO components are co-spatial, and they reach velocity shifts
of up to 250 km s$^{-1}$, with respect to the systemic. \cite{Krips+11} suggest that up to
30\% of the observed PdBI CO(2-1) emission is ascribable to gas blowing radially
outward from the galaxy disk, and we use this estimate to evaluate the mass of the outflowing gas. 
We model the outflow using a biconical geometry with radius of 
1 arcsec (0.1 kpc) and an average velocity of 150 km s$^{-1}$.

In the early type galaxy IC~5063, which
hosts a radio-loud Sy2 nucleus, \cite{Morganti+13} detect with APEX a prominent blue-shifted wing of the CO(2-1)
line, tracing gas with velocities exceeding
those of the rotating gas in the disk. Interestingly, such velocities are consistent
with those found in blue-shifted HI absorption, also detected in IC~5063, tracing an outflow along the
line of sight of the radio jet.
Since the APEX observations do not provide any information about the
spatial extension of the CO(2-1) emission, the outflow rate estimation by \cite{Morganti+13} relies
on the (likely but not yet proven) assumption that the molecular and neutral winds
are co-spatial. Therefore, following these authors, we model the blue-shifted
CO(2-1) wing as a conical outflow with radius of 0.5 kpc and velocity of
400 km s$^{-1}$. Similarly to the other cases, we assume ${\rm L'_{CO(1-0)}=L'_{CO(2-1)}}$.

\cite{Tsai+09} discovered various diffuse molecular features extended above and below the molecular disk of the
nearby star-bursting galaxy NGC 2146, by using very deep NMA CO(1-0) observations. One of 
these structures, which can be clearly seen in the PV diagram and in the integrated intensity 
and velocity maps, traces a molecular outflow with velocities of up to 200 km s$^{-1}$, extending to a 
distance of about 25 arcsec from the galactic centre (corresponding to 1.6 kpc with the cosmology 
adopted in this paper). The CO(1-0) integrated flux in this component is 150 Jy km s$^{-1}$.

\subsection{Outflow Properties}

As for the seven sources of our PdBI sample, to estimate the total mass of outflowing molecular gas
in the 12 sources belonging to the literature sample, we adopt a unique 
CO-to-H$_2$ conversion factor, i.e. the value commonly used for ULIRG-like
gas conditions (${\rm \alpha_{ CO(1-0)}= 0.8~M_{\odot}~(K~km~s^{-1}pc^2)^{-1}}$, see also
explanation in Section 3.2).
However, for the non-ULIRG galaxies belonging to this sample, we also report, in Table \ref{table:res2},
the outflow mass, mass-loss rates and energetics as obtained by using the same $\alpha_{\rm CO(1-0)}$ adopted
by the authors of the respective papers for estimating the total amount of molecular gas in that specific host
galaxy (see column 6 of Table \ref{table:res1} and notes in Table \ref{table:res2}).
When only the CO(2-1) transition is available for estimating
the outflow mass, as in the case of NGC 1377 and NGC 3256, 
we assume that the CO emission is thermalised and optically thick,
hence the line luminosity ${\rm L'_{CO}}$ is independent of
the transition $\rm J$ \citep{Solomon+05}.

In the 12 sources from the literature, we derive the molecular outflow properties by assuming
the same geometry as in the corresponding papers.  These works usually provide
the dynamical time-scale ${\rm \tau_{dyn}}$ of the wind, and 
calculate the outflow mass-loss rate using the relation: 
${\rm \dot{M}_{OF} \simeq M_{OF}/\tau_{dyn}}$. As already explained
in Sect. 3.2, this method is appropriate in a shell-like explosive scenario, where
one or more clouds are clearly resolved and their distance from the centre can be measured. 
However, if the outflowing gas appears to be uniformly distributed within a
spherical or multi-conical volume, our prescription is
more correct; hence, in the latter case, we re-calculate the outflow rate with our method.

\begin{table*}
 \centering
 \scriptsize
  \caption{Properties of the Extended Sample of Galaxies}
  \label{table:res1}
\begin{tabular}{@{}llcccccccc@{}}
\hline
\hline
Object 			&	type 		& z$_{\rm CO}$ &  	SFR					& log(${\rm P_{kin,SF}}$)	  	& $\alpha_{\rm CO(1-0)}$			 & ${\rm log(M_{H_2, TOT})}^\dag$	&   ${\rm log(L_{AGN})}$	&	${\rm L_{AGN}/L_{Bol}}$	& Refs.	 \\	
				&			&			&	[M$_{\odot}$ yr$^{-1}$]	&	[erg s$^{-1}$]		& [$\rm M_{\sun}/(K\,km\,s^{-1}\,pc^2)$]	& [M$_{\odot}$]			& 	[erg s$^{-1}$]			&						&      						\\
(1)				&	(2)		& (3)			&	(4)					&		(5)		  	& 	(6)							&	(7)				& (8)					&	(9)						&   (10)		\\
\hline
IRAS F08572+3915	&	Sy 2		& 0.05821	 	& 	20					&	43.15			&			0.8					&	9.18 				&	45.72					&	0.860			&	1, b			\\
IRAS F10565+2448	&	Sy 2		& 0.04311		&	95					&	43.82			&			0.8					&	9.90				&	44.81				         &	0.170			&	1, c			\\
IRAS 23365+3604	&	LINER	& 0.06438		&	137					&	43.98			&			0.8					&	9.93				&	44.67					&       0.072			&	1, b				\\	
Mrk 273			&	Sy 2		& 0.03777		&	139					&	43.99			&			0.8					&	9.70				&	44.73					&	0.080			&	1, a				\\
IRAS F23060+0505	&	Sy 2		& 0.17300		&	75				      	&	43.72			&			0.8					&	10.39			&	46.06					&	0.780			&	1, b		\\
Mrk 876			&	Sy 1		& 0.12900		&	6.5					&	42.66			&			0.8					&	9.84				&	45.84					&	0.930			&	1, d					\\
I Zw 1			&	Sy 1		& 0.06114		&	36					&	43.40			&			0.8					&	9.56				&	45.37					&	0.520			&	1, d				\\
\hline
Mrk 231			&  Sy 1		& 0.04217		&	234					&	44.21			&			0.8					&	9.73				&	45.72					&	0.340			&	2, b				\\
NGC 1266		&  LINER		& 0.00719		&	1.6					&	42.05			&			4.4					&	9.23				&	43.31				         &	0.250			&	3, f			\\
M 82				&   HII		& 0.00068		&	10					&	42.85			&			1.2					&	8.64				&	$\leq$41.54				&	$\leq$0.00090		&	4, i				\\
NGC 1377		& LINER		& 0.00578		&	0.9					&	41.80			&			4.4					&       8.44				&	42.93					&	0.200			&	5, f		\\
NGC 6240 		& Sy 2		& 0.02448		&	16					&	43.05			&			0.8					&	9.86				&	45.38					&	0.780			&	6, e	\\
NGC 3256		& HII			& 0.00926		&	36					&	43.40			&			1.2					&	9.68				&	$\leq$41.97				&	$\leq$0.00070		&	7, j				\\
NGC 3628		& HII			& 0.00280		&	1.8					&	42.10			&			1.2					&	9.53				&	$\leq$40.79				&	$\leq$0.00090		&	8, k			\\
NGC 253 			& HII			& 0.00081		&	3					&	42.32			&			1.2					&       8.15				&	$\leq$40.66				&	$\leq$0.00040		&	9, j			\\
NGC 6764		& LINER		& 0.00807	         &	2.6					&	42.26			&			1.2					&	8.90				&	42.23					&  	0.017		   	 & 10, g    	\\
NGC 1068		& Sy 2		& 0.00379		&      18					&	43.10	 		&			0.8					&      9.11				&      43.94  					&      0.097			& 11, h 				\\
IC 5063			& Sy 2		& 0.01100          &     0.6					&	41.62		        &			4.4					&	8.85				&	44.30					&	0.900			& 	12, l		 \\
NGC 2146		& HII			& 0.00298		& 	12					&	42.92			&			1.2					&	8.94				&	$\leq$41.09				& $\leq$0.00030		&  13, m	   \\
\hline			 
\end{tabular}

\begin{flushleft}
\small
The following information is listed for each galaxy in each of the columns: (1) name; (2) optical classification; (3) redshift based on the CO
spectrum; (4) star formation rate; (5) kinetic energy associated with the supernovae; (6) CO-to-H$_2$ conversion factor adopted to infer M$_{\rm H_{2}}$
from the CO(1-0) luminosity; (7) total mass of molecular gas; (8) AGN  bolometric luminosity; (9)  ratio between the bolometric luminosity of the AGN and the total bolometric luminosity of the galaxy; 
(10) references for the CO fluxes used to estimate M$_{\rm H_{2}}$: 1 - this work, 2 - \cite{Cicone+12}, 3 - \cite{Alatalo+11}, 4 - \cite{Walter+02}, 5 - \cite{Aalto+12b}, 6 - \cite{Feruglio+13a, Feruglio+13b}, 7 - \cite{Sakamoto+06b}, 8 - \cite{Tsai+12}, 
9 - \cite{Mauersberger+96}, 10 - \cite{Sanders+Mirabel85}, 11 - \cite{Maiolino+97}, 12 - \cite{Wiklind+95}, 13 - \cite{Tsai+09}; references
for $\rm L_{AGN}$: a - \cite{Nardini+09}, b - \cite{Nardini+10},
c - \cite{Veilleux+09a}, d - \cite{Piconcelli+05}, e - Risaliti (Priv. Comm.), f - \cite{Moustakas+Kennicutt06}, g - \cite{Croston+08}, h - \cite{Prieto+10}, i - \cite{Verrecchia+07},
j - \cite{Ranalli+03}, k - \cite{Gonzalez-Martin+09}, l - \cite{Koyama+92}, m - \cite{DellaCeca+99}, \cite{Inui+05}. \\
Notes: $^\dag$ In these cases of NGC 1377 and NGC 3256, we estimate
the total molecular gas mass by using the CO(2-1) transition (instead of the CO(1-0)), by assuming that the CO gas is thermalised and optically
thick, i.e. that ${\rm L'_{CO(1-0)}=L'_{CO(2-1)}}$.
\end{flushleft}

\end{table*}

\begin{table*}
 \centering
 \scriptsize
  \caption{Outflow Properties of the Extended Sample of Galaxies}
  \label{table:res2}
\begin{tabular}{@{}lccccccccc@{}}
\hline
\hline
Object 						& ${\rm log(M_{H_2, OF})}$$^{\dag}$ 	&        ${\rm \dot{M}_{H_2, OF}}$$^{\dag}$ 	& 	${\rm R_{OF}}$		&   ${\rm v_{OF,avg}}$ & 	${\rm v_{OF, max}}$ 		& 	log(${\rm \tau_{dep}})$$^{\dag}$ 	& log(${\rm P_{kin, OF}}$)$^{\dag}$ 	&	(${\rm \dot{M}_{H_2, OF}v)/(L_{AGN}/c)}$$^{\dag}$ 		&	Refs. 	\\	
							&		[M$_{\odot}$]			&	[M$_{\odot}$ yr$^{-1}$]		  			& 	[kpc]				& [km s$^{-1}$]
							&	[km s$^{-1}$]			&	[yr]							&	[erg s$^{-1}$]   				&										& 		\\
(1)							&	(2)						&	(3)						 			& 	(4)				&	(5)				& (6)					&		(7)						&    (8)						&		(9)								&	(10)	\\
\hline
IRAS F08572+3915				&	8.61					&	1210							&      0.82			&	800			& 1200				&	 6.10								&	 44.39					&			 35						&1\\
IRAS F10565+2448				&	 8.37					&	300							&	 1.10			&	450			& 600				&	 7.42								&	43.28					&			40					&1\\
IRAS 23365+3604				&	8.17					&	 170						&	1.23		&	450		&  600				&	 7.70				 				&   	43.04					& 			31	  					&1\\	
Mrk 273						&	8.24						&	600								&	0.55 				&	620		 	& 900				&	6.92									&	43.86					&			130							&1\\
IRAS F23060+0505				&	$\leq$ 9.56				&	 $\leq$ 1500					&      $\leq$ 4.05		&	(550)		&  (1100)		&	 $\geq$ 7.21						& 	$\leq$ 44.16			&    			  $\leq$14 					& 1\\
Mrk 876						&	$\leq$ 9.48				&	$\leq$ 1830						&	$\leq$ 3.55		&	(700)			& (1700)				&	$\geq$6.57							& 	$\leq$ 44.45				&   			$\leq$35  						& 1\\
I Zw 1						&	$\leq$ 7.67				&	$\leq$ 140							&	(0.50)		         &	(500)			&(750)			         &	$\geq$7.41							&    $\leq$ 43.04				&  			$\leq$6						&1\\
\hline
Mrk 231						&	8.47						&	1050								& 	0.60				&	700			& 1000				&	6.71									&	44.21					& 			26							& 2	\\
NGC 1266					&       7.93	- 8.66				&      33 - 180							&	0.45				&	177			& 362				&	7.71 - 6.97							&	41.51 - 42.25				&  			54 - 300						& 3		\\
M 82							&      	8.08 - 8.25				&     12 - 18							&	0.80				&      100			& 230				&	7.56 - 7.38							&	40.58 - 40.75				& 			$\geq$650 - $\geq$980			& 4		\\
NGC 1377					&      7.29 - 8.03				&     14 - 76							&	0.20				&	110			& 140				&	7.29 - 6.56							&	40.73 - 41.46				&   			34 - 190					        & 5			\\
NGC 6240 					&	8.61						& 	800								&	0.65				&	400			& 500				&	6.96									&	43.61					&			25 						        & 6			\\
NGC 3256				         &	7.34 - 7.51				&	11 - 16							&	0.50				&	250			& 425				&	8.64 - 8.48							&	41.34 - 41.50				&			$\geq$560 - $\geq$810			& 7		\\
NGC 3628					&	7.36 - 7.54				&	4.5 - 6.7							& 	0.40				&	50			& 110				&	8.88 - 8.70							&	39.55 - 39.72				&			$\geq$690 - $\geq$1030			&	8		\\
NGC 253 						&       6.32 - 6.50				&      4.2 - 6.3							&	0.20				&	50			& 100				&	7.53 - 7.35							&	39.52 - 39.70				&			$\geq$870 - $\geq$1300 			&	9		\\
NGC 6764					&	6.52 - 6.69				&	3.1 - 4.7							&	0.60				&	170			& 280				&	8.41 - 8.23							&	40.45 - 40.63				&			59 - 89 						&	10	\\
NGC 1068					&       7.26						&	84								&	0.10				&	150			& 250				&	7.19									&    	41.77					&			27 							&	11		\\
IC 5063						&    7.37 - 8.10					&     23 - 127							& 	0.50				&	300			& 450				&	7.48 - 6.75							&	41.82 -  42.56				&			7 - 36						&	12	\\
NGC 2146					&   7.68 - 7.86					&    14 - 22 							&     1.55				&     150			& 200				&  	7.78 - 7.61							&	41.00 - 41.18				&			$\geq$3300 - $\geq$4900 	 	& 	13	\\
\hline			
\end{tabular}

\begin{flushleft}
\small
The following information is listed for each galaxy in each of the columns: (1) name;
(2) mass of the molecular gas in the outflow; (3) molecular outflow mass-loss
rate; (4) radius of the outflow extension; (5) average velocity of the outflow; (6) maximum velocity of the outflow;
(7) depletion time-scale associated with the outflow; (8) kinetic power of the outflow; (9) outflow momentum rate relative to $\rm L_{AGN}/c$;
(10) references for the molecular outflow measurements:
 1 - this work, 2 - \cite{Cicone+12}, 3 - \cite{Alatalo+11}, 4 - \cite{Walter+02}, 5 - \cite{Aalto+12b}, 6 - \cite{Feruglio+13b}, 
 7 - \cite{Sakamoto+06b}, 8 - \cite{Tsai+12}, 9 - \cite{Sakamoto+06a} and \cite{Bolatto+13}, 
10 - \cite{Leon+07}, 11 - \cite{Krips+11}, 12 - \cite{Morganti+13}, 13 - \cite{Tsai+09}\\
Notes:
$^{\dag}$ For non-ULIRG galaxies we report two values of the outflow mass, outflow rate, depletion time, kinetic power and momentum rate of the outflow (columns 2, 3, 7, 8, 9), obtained by adopting in one case the 
conservative CO-to-H$_2$ conversion factor $\alpha_{\rm CO(1-0)}$=0.8, which was found appropriate for Mrk231, and in the other case the
same $\alpha_{\rm CO(1-0)}$ (column 6 in table \ref{table:res1} ) as for estimating the total molecular gas mass in the same
host galaxy (column 7 in table \ref{table:res1}). 
The two values of $\alpha_{\rm CO}$ yield respectively to the lower and the 
higher estimates of ${\rm M_{H_2, OF}}$, ${\rm \dot{M}_{H_2}}$ and ${\rm P_{kin}}$, and conversely for ${\rm \tau_{dep}}$ (see also
discussion in the text).
\end{flushleft}
\end{table*}


\section{Ancillary Information}

In this section we discuss how the ancillary information listed in Table 3 was derived.
The star formation rates in column 4 are inferred from the total IR (8-1000 $\mu$m) luminosity by assuming the relation in 
\cite{Kennicutt98} and a Chabrier IMF (see also \cite{Sturm+11}): ${\rm SFR = (1-a)\cdot 10^{-10}\,L_{IR}}$,
using the factor ${\rm a\equiv L_{AGN}/L_{Bol}}$ reported in column 9. From these SFRs, we 
calculate the kinetic power injected by supernovae (column 5), by using the relationship
provided by \cite{Veilleux+05}: ${\rm P_{kin,SF} (erg\,s^{-1}) = 7\cdot 10^{41}\,SFR(M_{\odot}\,yr^{-1})}$ 
(see also \citealt{Maiolino+12}).

The total mass of molecular gas in column 7 is corrected for He and is derived from the CO integrated flux, using,
for each source, the corresponding CO-to-H$_2$ conversion factor in column 6.
The value of the CO-to-H$_2$ conversion factor in different environments and its dependance on the galaxy properties is highly debated,
and it may constitute a major source of uncertainty in our estimates of the molecular
gas masses. However, such discussion goes beyond the scope of this paper. Here we simply assume
 $\alpha_{\rm CO(1-0)}$= 0.8 in ULIRGs \citep{Downes+Solomon98},
1.2 in M82 and M82-like galaxies \citep{Weiss+01}, 4.36 (rounded to 4.4 in Table 3) in the other galaxies (Milky Way value, \citealt{Genzel+12}).

Obtaining a reliable estimate of the AGN luminosity and of the AGN contribution to the total bolometric luminosity of a galaxy
is in general a difficult task, but it is even more complicated in the case of the heavily obscured objects included in our sample. 
For the AGN-host (U)LIRGs, we mostly rely on the analysis of their infrared properties as revealed by Spitzer-IRS observations; in particular, for 
IRAS F08572+3915,  IRAS 23365+3604, Mrk 273, IRAS F23060+0505 and Mrk 231, we adopt the
${\rm L_{AGN}/L_{Bol}}$ ratios inferred using a spectral decomposition method, which relies on the use of AGN and starburst templates,
in the rest-frame 5-8 $\mu$m wavelength range, where the starbursts and AGNs have quite distinct spectral properties \citep{Nardini+09,Nardini+10}. 
In the case of IRAS F10565+2448, we assume the average AGN bolometric fraction measured by \cite{Veilleux+09a}, using a combination
of continuum, emission and absorption line diagnostics based on MIR and FIR Spitzer observations of this galaxy.
We then evaluate for these ULIRGs the AGN bolometric luminosity by assuming ${\rm L_{Bol} = 1.15\,L_{IR}}$ \citep{Veilleux+09a}.
In the case of I~Zw~1, Mrk 876, NGC 6240, NGC 6764 and
IC 5063, we estimate ${\rm L_{AGN}}$ from their absorption-corrected hard X-ray luminosity ${\rm L_{X}(2-10\,keV)}$ using
the relation in \cite{Marconi+04}.
For the other two LINERs, NGC 1266 and NGC 1377, we
calculate ${\rm L_{AGN}}$ using the [OIII]$\lambda$5007 luminosity (reported by \citealt{Moustakas+Kennicutt06}) as a tracer of AGN activity, following \cite{Heckman+04}.
For the extremely obscured Seyfert 2 NGC 1068, we adopt directly the AGN luminosity calculated by \cite{Prieto+10}.
For the starburst galaxies M82, NGC 3256, NGC 3628, NGC 253 and NGC 2146 , we assume  ${\rm L_{Bol} \simeq L_{IR}}$ and
set as an upper limit for the ${\rm L_{AGN}}$ the value estimated from the unabsorbed hard X-ray luminosity ${\rm L_{X}(2-10\,keV)}$,
using the relation provided by \cite{Marconi+04}.


\section{Discussion}

Recent works based on the analysis of Herschel--PACS FIR observations of local ULIRGs and
AGN hosts have shown that molecular outflows are rather common in this class of objects 
\citep{Veilleux+13, Spoon+13}. In our study, thanks to the
additional information provided by the interferometric CO observations, we find that molecular outflows
can be very massive and energetic, with outflow rates of several $\rm 100~M_{\odot}~yr^{-1}$, 
and that the outflowing gas extends on kiloparsec scales, hence
affecting the host galaxy.

There is a good correspondence between the detection of molecular outflows through the OH P-Cygni profile
and the detection of the same wind through the CO wings. 
In particular, we note that in the case of IRAS 08572+3915 (as well as for Mrk 231 and for the starburst galaxy NGC 253),
the outflow mass-loss rate estimate obtained by \cite{Sturm+11} by modelling the OH P-Cygni profile is in good agreement
with our results. 
For the other three sources in which OH P-Cygni profiles were detected by Herschel
and which we followed up with the PdBI in their CO(1-0) transition (i.e. IRAS 10565+2448, IRAS 23365+3604 and Mrk 273), 
we find a good correspondence between the outflow velocities traced by the CO broad wings
and the velocity of the OH blue-shifted absorption reported by \cite{Veilleux+13}.
Our PdBI observations of Mrk 876 and I Zw 1 only allow us to set upper limits on the 
molecular outflow properties in these sources. These two galaxies 
share some similarities: both are dominated by a powerful unobscured QSO, 
and both exhibit pure OH emission profiles in the Herschel--PACS data
presented by \cite{Veilleux+13}, with no evidence for blue-shifted absorption.
However, the OH profile of Mrk 876 is the broadest among the sources
with OH detected exclusively in emission: this suggests non-gravitational motion of molecular gas.
The poor signal-to-noise of Herschel--PACS observations of Mrk 876 and I Zw 1 does not allow us
to draw any conclusion about the possibility of very high-velocity (i.e. ${\rm |v| > 1000~km~s^{-1}}$) 
molecular outflows.

In the following we discuss the relationships between the properties of the molecular outflows and the physical
properties  of the AGN and of the host galaxies, to shed light on the origin and nature of the molecular outflows, and their role
in the evolution of galaxies.
A general caveat of this study is that our sample is certainly not representative of the local population of galaxies.
It is incomplete
and subject to biases. Indeed, the bulk of the SHINING subsample (4 targets out of 5) has been selected to show
evidence of molecular outflows according to the OH P-Cygni profiles. 
 As a consequence, only IRAS~F23060+0505, Mrk~876 and I~Zw~1 were observed with the PdBI 
without a previous knowledge of the presence of a molecular outflow. In IRAS~F23060+0505 we detect
molecular gas at velocities larger than 300~km~s$^{-1}$ whose kinematics deviates from the central rotational
pattern (traced by the narrow core of the CO line): however, the lack of detection of high velocity CO emission
at high significance prevents us from drawing firm conclusions about the outflowing origin of this gas.
As mentioned before, in Mrk~876 and I~Zw~1 we do not find any significant evidence for 
the presence of a massive molecular outflow, although in Mrk~876 we tentatively detect some high velocity 
CO emission. Summarising, none of our three ``unbiased'' sources constitutes a reliable molecular outflow detection.
On the other hand, the sample collected from the literature
only contains sources for which a molecular outflow has been constrained: this unavoidably biases the sample
towards galaxies for which a molecular outflow has been detected, since in most cases in which an outflow has {\it not}
been detected (or overlooked) the authors generally do not provide upper limits. 
Yet, even so, the biases affecting our sources selection do not prevent us from analysing the outflow properties
within this sample, which is nevertheless an interesting and quite heterogeneous collection of local galaxies,
spanning a wide range of physical properties, especially in terms of star formation rate and AGN luminosity.

All of the plots shown in this section are obtained by assuming a CO-to-H$_2$ conversion
factor $\alpha_{\rm CO(1-0)}$=0.8 for the molecular gas in the outflow (see also Table \ref{table:res2}
and the related discussion in Section 3.2, for the PdBI sample, and Section
4.2, for the literature sample); nevertheless, the use of the alternative conversion factors (whose
resulting outflow rates and energetics are also reported in Table \ref{table:res2})
does not affect the observed trends significantly. 
The errors on the outflow radius (i.e. $\pm$ 0.1 dex, on average), 
outflow rate and depletion time-scale ($\pm$ 0.3 dex), 
 outflow kinetic power ($\pm$ 0.5 dex) and momentum rate ($\pm$ 0.45 dex), are calculated
for our 4 PdBI sources with detected molecular outflows, by using the uncertainties 
on ${\rm S_{CO}}$ and ${\rm R_{OF}}$ obtained from the {\it uv} data (reported in Table \ref{table:COwings}),
and by assuming a conservative
average outflow velocity error of $\pm$50\%, which accounts for uncertainties associated with projection effects. 
We assume these error estimates to be representative also of the sample extracted from the literature, 
for which obtaining reliable error-bars is quite difficult, given the incompleteness of the information provided in the
literature. These error-bars do not take into account the uncertainty on the CO-to-H$_2$ conversion factor;  however, we
note that including it would simply increase the upper error-bars, 
since we are using, for all the sources, the most conservative value of $\alpha_{\rm CO(1-0)}$ (in the molecular
outflow).

\subsection{The Relation Between Outflow Rate, SFR and AGN Luminosity.}

\begin{figure}[h!]
\centering
   \includegraphics[width=.8\columnwidth,angle=0]{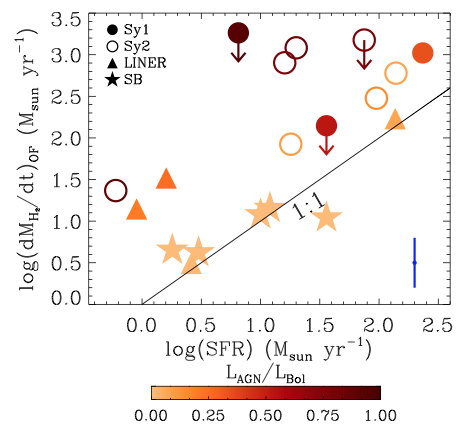}\\
   \includegraphics[width=.8\columnwidth,angle=0]{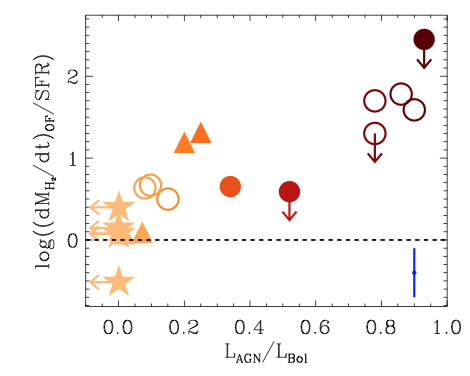}\\
   \caption{{\it Top:} Outflow mass-loss rate as a function of the star formation rate for the extended sample of galaxies, whose general
    properties and outflow characteristics are summarised in Tables \ref{table:res1} and \ref{table:res2}. Filled and open 
    circles represent respectively unobscured and obscured AGNs, LINERs are plotted as upward triangles and ``pure'' starburst 
    galaxies as stars (see legend at the top-left corner of the plot). 
    Symbols are colour-coded according to the fraction of bolometric luminosity attributed to the AGN 
    (${\rm L_{AGN}/L_{Bol}}$). 
     Outflow rates are derived by assuming, in the outflow, the conservative CO-to-H$_2$ 
     conversion factor $\alpha_{\rm CO(1-0)}$=0.8 for all 
    of the sources. The black dashed line represents the 1:1 correlation between SFR and outflow mass-loss rate.
    {\it Bottom:} This plot indicates a positive correlation between the outflow mass loading factor (${\rm \dot{M}_{H_2, OF}}$/SFR) and ${\rm
    L_{AGN}/L_{Bol}}$
    that emerges from the diagram in the top panel.
    {\it Notes:} At the bottom-right of each panel we show an error-bar representative for the whole sample, which does not take into account the uncertainty on the
    CO-to-H$_2$ conversion factor (see text in Section 6 for a more detailed explanation), and corresponds to an average error of $\pm$0.3 dex.}
   \label{fig:plot1}
\end{figure}

\begin{figure}[h!]
\centering
   \includegraphics[width=.8\columnwidth,angle=0]{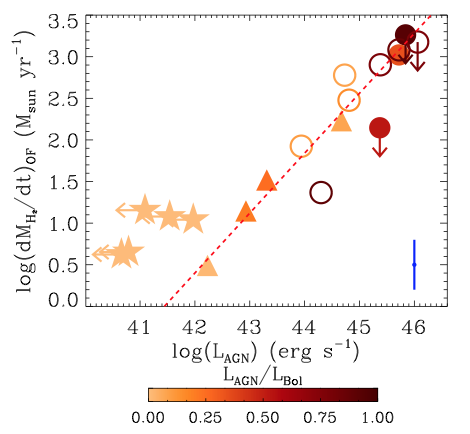}\\
   \caption{Outflow mass-loss rate as a function of the AGN bolometric luminosity, for the extended sample of galaxies 
   (Tables \ref{table:res1} and \ref{table:res2}). Symbols and colour-coding are as in Figure \ref{fig:plot1}. 
   Error bars as in Fig. \ref{fig:plot1}. 
   For the ``pure''
   starburst galaxies we set an upper limit to the AGN contribution by hypothesising that all of their hard X-ray (2-10 keV)
   unobscured luminosity is produced by an AGN (further explanation in Section 5).
   The red dashed relationship results from a linear fit to the AGN host-galaxies, 
   in which the upper limits have been excluded.}
   \label{fig:plot2}
\end{figure}

The top panel of Figure \ref{fig:plot1} shows the outflow rate $\rm \dot{M}_{H2,OF}$
as a function of the star formation rate, by
colour-coding the different symbols according to the fraction of the total galaxy bolometric luminosity
that is ascribed to the AGN (i.e. ${\rm L_{AGN}/L_{Bol}}$). Solid and open circles indicate galaxies optically
classified as type 1 (unobscured) and type 2 (obscured) AGNs, respectively; triangles represent LINER-type galaxies 
and stars indicate galaxies classified as ``pure'' starbursts (HII). 
In this diagram, the pure starbursts and the starburst-dominated
galaxies lie close to the relation ${\rm SFR = \dot{M}_{H_2, OF}}$, 
i.e. outflow mass loading factor ${\rm \eta=\dot{M}_{OF}/SFR \sim 1}$. 
Conversely, objects with higher AGN fraction depart from this sequence, and the deviation appears to increase
with ${\rm L_{AGN}/L_{Bol}}$. This result would suggest that the presence of an AGN can 
boost the outflow rate by a large factor, which approaches two orders of magnitude in the most powerful QSOs
 of our sample.

However, we note that also some of the SB-dominated sources have outflow rates which depart slightly from a
1:1 correlation by having outflow mass loading factors $\eta \sim 1-4$, indicating that they are also in a 
star formation--quenching regime. Although models of starburst-driven outflows can predict mass loading factors of up to $\sim$2-3, such
high values of the outflow mass loading factor in SB-dominated sources may indicate that even in these case an 
AGN can contribute in driving the outflow,
even if the AGN
accounts only for a minor fraction of the total bolometric luminosity of the source. Alternatively, an AGN
may have been present, but switched off recently.
Indeed, recent models show that AGN-driven massive molecular outflows can persist
for $\sim10^8$ yrs after the central nucleus has turned off (\citealt{King+11, Zubovas+12}).

The tentative positive correlation between the mass loading factor ${\rm \dot{M}_{H_2, OF}}$/SFR and the AGN fraction 
of the bolometric luminosity of the galaxies is better illustrated by the the bottom panel of
Figure \ref{fig:plot1}.

The relevance of the AGN in driving outflows is also highlighted in the plot of  the
outflow rate as a function of AGN luminosity reported in Fig. \ref{fig:plot2}.
For the galaxies of our sample which host an AGN,
the AGN luminosity correlates with the outflow rate.
Interestingly, ``pure'' starburst galaxies are all outliers in this correlation, possibly indicating
that the feedback mechanism in action in these objects is substantially different from the AGN-host galaxies,
even from those AGNs whose bolometric luminosity is dominated by the starburst.
We fit the correlation for the AGN-host galaxies, by 
excluding the upper limits, and we obtain:

\begin{equation}
{\rm log(\dot{M}_{H_2, OF})=(-29.8 \pm 3.7) + (0.720 \pm 0.083)\,log(L_{AGN})},
\end{equation}

resulting in the red dashed line in Figure \ref{fig:plot2}.
This is the first time that a correlation between the molecular outflow rate and
the AGN luminosity is found from CO data, and we interpret it as a direct evidence for these massive molecular
outflows being mostly powered by the AGN and, therefore, for AGN negative feedback in action
in the AGN-host galaxies included in our study. However, I~Zw1 and IC~5063 are remarkable counter examples,
deviating from the relation. Given the biases and incompleteness of our sample, these two objects may be the
tip of the iceberg of a population of AGNs with low outflow rates. 
In this scenario, the correlation in Fig. \ref{fig:plot2} would be just the upper envelope of the entire
distribution of AGN host-galaxies in the outflow rate vs AGN luminosity diagram.
This possibility should be investigated
with extensive CO surveys, targeting a complete sample of galaxies. Alternatively, I~Zw1 and IC~5063 
may represent a transient phase in which the molecular outflow has been swept away
or the molecules photo-dissociated and ionised, and additional molecular clouds from the disk have
still to be accelerated.

\subsection{Gas Depletion Time-Scales}

\begin{figure}[h!]
\centering
\includegraphics[width=.8\columnwidth,angle=0]{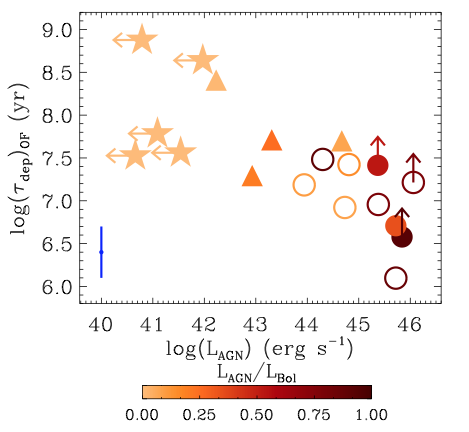}\\
   \caption{Molecular gas depletion time-scale due to the molecular outflow
   	(${\rm \tau_{dep, OF} \equiv M_{H_2,TOT}/\dot{M}_{H_2, OF}}$) versus AGN bolometric luminosity. 
	Symbols and colour-coding as in Fig. \ref{fig:plot1}. 
  	The uncertainties ($\pm$ 0.3 dex) on the depletion time-scales 
  	(a representative error-bar is shown at the bottom-left of the plot) are dominated by the errors on the 
  	outflow rates.}
   \label{fig:plot3}
\end{figure}

\begin{figure}[h!]
\centering
   \includegraphics[width=.8\columnwidth,angle=0]{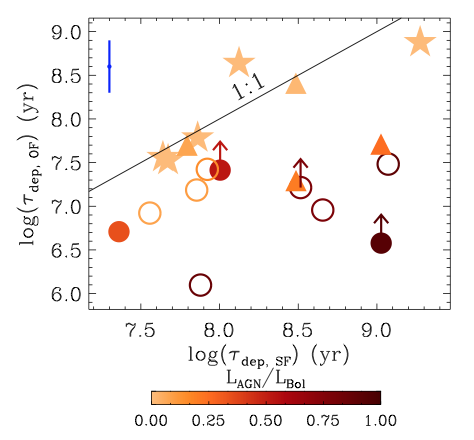}\\
   \caption{Comparison between the depletion time-scale due to gas removal through molecular outflows with the
   depletion time-scale due to gas consumption by star formation. Symbols and colour-coding as in Fig. \ref{fig:plot1}.
   Error-bars as in Fig. \ref{fig:plot3}.}
   \label{fig:plot3bis}
\end{figure}

Fig. \ref{fig:plot3} shows the molecular gas depletion time-scale (due to outflows) as a function of the AGN luminosity.
The gas depletion time-scale associated with the outflow is defined as the time required by
the outflow (assuming that it continues at the same rate) to completely remove the whole
gas content in the host galaxy, i.e.
${\rm \tau_{dep, OF} \equiv M_{H_2,TOT}/\dot{M}_{H_2, OF}}$.
In Fig. \ref{fig:plot3} the gas depletion time-scales are significantly reduced by the presence of a luminous AGN.  
We observe indeed an anti-correlation between
these two quantities, as already observed by \cite{Sturm+11}, implying that the objects hosting more powerful AGNs
are depleted of their molecular gas content on shorter time-scales. 
Starburst galaxies have outflow depletion time-scales up to several
hundred million years, while in galaxies hosting powerful quasars the depletion time-scales
can be as short as a few million years.
We note however that these are time-{\it scales}, which would correspond to the real
depletion times only if the outflow continues at this rate. Moreover, as already 
stressed at the beginning of this section, our sample is strongly biased toward
galaxies and quasars which {\it do host} massive molecular outflows, and therefore these results may
not hold for the {\it general} population of local galaxies.

Fig. \ref{fig:plot3bis} compares the depletion time-scale due to the molecular outflow with
the depletion time-scale due to gas consumption by the star formation. The line indicates
the locus where the two time-scales are identical. In starburst-dominated galaxies the two time-scales
are similar, while in AGN-dominated galaxies the outflow depletion time-scale is much shorter
than the depletion time-scale due to star formation.

It is interesting to note that outflows can
expel the molecular gas on time-scales shorter than 100~Myr, which is the 
minimum quenching time-scale required to explain the $\alpha$-enhancement in massive spheroidal galaxies.
Therefore, if the same mechanism is in place at high redshift, it can actually help to explain the
enhancement of $\alpha$-elements, relative to iron, observed in local elliptical galaxies.
It is also worth noting that the outflow depletion
time-scale, especially in the powerful AGNs of our sample, can be even shorter than $\rm 10-20~Myr$, which is
the quenching time inferred by recent studies of post-starburst galaxies at intermediate and high redshift
(Christy Tremonti, Priv. Comm.).
Conversely, the depletion time-scales associated with the consumption of gas by star formation are too long and 
fail to meet these conditions in most objects.

\begin{figure}[tb]
\centering
   \includegraphics[width=.8\columnwidth,angle=0]{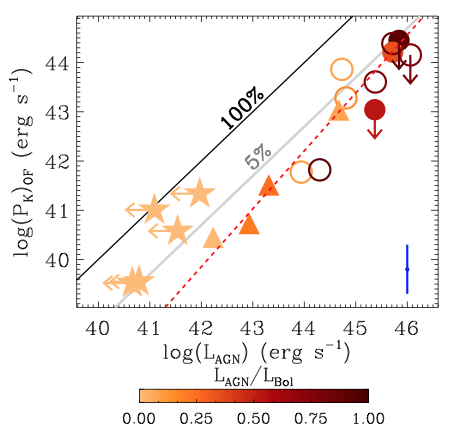}\\
   \caption{Correlation between the kinetic power of the outflow and the AGN bolometric luminosity. 
   		Symbols and colour-coding as in Fig. \ref{fig:plot1}. 
		The grey line represents the theoretical expectation of models of AGN feedback, 
		for which ${\rm P_{K, OF} = 5\% L_{AGN}}$. The red dashed line represents the linear fit to our data,		
		excluding the upper limits. The error-bar shown at the bottom-right of the plot corresponds to an average 
		error of $\pm$0.5 dex.}
   \label{fig:plot4}
\end{figure}

\begin{figure}[tb]
\centering
	\includegraphics[ width=.8\columnwidth,angle=0]{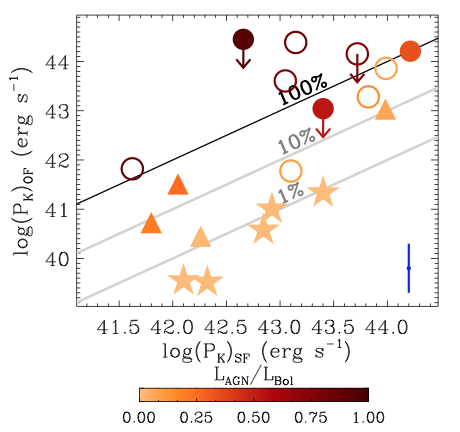}\\
   \caption{Kinetic power of the outflow plotted as a function of the kinetic power of a supernova-driven wind. 
   		Symbols and colour-coding as in Fig. \ref{fig:plot1}. Error bars as in Fig. \ref{fig:plot4}. 
		The black and grey lines mark the relations 
		${\rm P_{K, OF} = P_{K, SF}}$, ${\rm P_{K, OF} = 10\%\,P_{K, SF}}$ and ${\rm P_{K, OF} = 1\%\,P_{K, SF}}$. }
   \label{fig:plot5}
\end{figure}

\subsection{Kinetic Power of the Outflows}

Fig. \ref{fig:plot4} shows the kinetic power of the molecular outflow as a function of AGN luminosity.
Theoretical models of AGN feedback and cosmological simulations predict a coupling efficiency between
AGN-driven outflows and AGN power of about $\sim$5\%, for AGN accreting close to the Eddington limit (which
is likely the case for, at least, the most luminous AGNs in our sample). This is also the $\rm P_{kin}/L_{AGN}$
fraction needed to explain the $\rm M_{BH}-\sigma$ relation in local galaxies (e.g. \citealt{King10, Zubovas+12, Lapi+05}).
Our observations
of massive molecular outflows in AGN-host galaxies, overall, 
appear to confirm this prediction. The solid grey line in Fig. \ref{fig:plot4} indicates the locus
of points having an outflow kinetic power that is 5\% of the AGN luminosity, and galaxies hosting
powerful quasars are indeed located close to this value.

The best-fit to our data points by excluding the upper limits is:
\begin{equation}
{\rm log(P_{kin, OF})=(-9.6 \pm 6.1) + (1.18 \pm 0.14)\,log(L_{AGN})}
\end{equation}
and it is indicated by the red dashed line in Figure \ref{fig:plot4}.
It is interesting that low luminosity AGNs seem to show an efficiency lower than 5\%. Likely, these AGNs (especially
the LINERs, indicated with triangles) are accreting
at a rate lower than Eddington. One should also note that in some
of these low luminosity AGNs (e.g. IC~5063, see \cite{Morganti+13}, and possibly NGC~1266 and
NGC~6764, as suggested by \cite{Alatalo+11} and \cite{Leon+07}, respectively) 
a radio-jet is thought to contribute to the acceleration of the molecular
gas. Additional detailed observations are required to better understand the 
outflow driving mechanism in these objects.

Starburst galaxy upper limits are located above the 5\% line, indicating that in these objects
a different source of energy is required, most likely provided by SN ejecta and radiation pressure from the young stars.

The kinetic power of the outflow is compared in Figure \ref{fig:plot5} with the kinetic power injected by supernovae,
as inferred from the SFR, following \cite{Veilleux+05} (see Table \ref{table:res1} and relevant
explanation in Section 5).
Fig. \ref{fig:plot5} shows that the outflow kinetic power achieved in the ``pure''
starburst galaxies and in some of the starburst-dominated objects is compatible
with a supernova-driven wind, with a coupling efficiency of a few percent up to a few 10 percent.
Conversely, for most of the heavily AGN-dominated sources,
it is clear that the additional contribution of the AGN is needed to produce the observed
outflow energetics.


\subsection{Momentum Rate of the Outflows}

\begin{figure}[tb]
\centering
	\includegraphics[width=.8\columnwidth,angle=0]{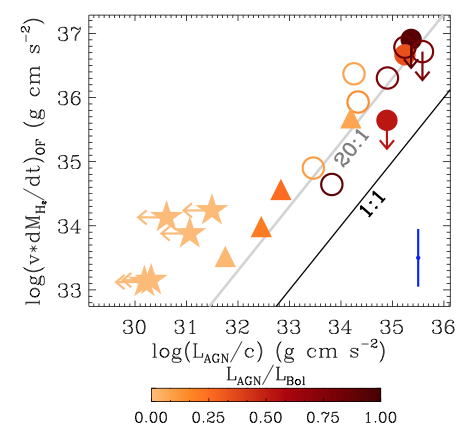}\\
   \caption{Outflow momentum rate (${\rm (v~\dot{M}_{H_2, OF})}$) 
   		versus photon momentum output of the AGN (${\rm L_{AGN}/c}$). 
   		Symbols and colour-coding as in Fig. \ref{fig:plot1}. 
		The grey line shows the prediction of models of AGN feedback, i.e. 
		${\rm \dot{M}_{H_2, OF}~v \sim 20~L_{AGN}/c}$. The error-bar shown at the bottom-right of 
		the plot corresponds to an average error of $\pm$0.45 dex.}
   \label{fig:plot6}
\end{figure}

\begin{figure}[tb]
\centering
	\includegraphics[width=.8\columnwidth,angle=0]{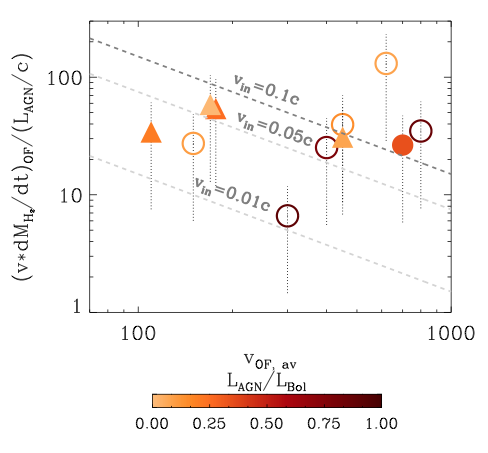}\\
   \caption{Outflow momentum ``boost" with respect to the AGN radiative momentum 
   		output, i.e. ${\rm (v~\dot{M}_{H_2, OF})/(L_{AGN}/c)}$,
   		as a function of the outflow velocity, following Fig. 5 of \cite{Faucher-Giguere+12}. 
   		Symbols and colour-coding as in Fig. \ref{fig:plot1}. 
		The uncertainties on the momentum boost are dominated by the errors on the 
  	         outflow momentum rates.
   		The upper limits on the outflow detection are excluded from
   		this plot, as well as the ``pure" starbursts.}
   \label{fig:plot8}
\end{figure}

The momentum rate provides an additional important indicator of the nature of the outflow and an important
test for models. In models in which the outflow is generated by a nuclear AGN-driven wind, the momentum rate
transferred by the AGN photons to the surrounding medium is given by the average number of scattering by each photon.
Some of these models predict, for AGNs accreting close to the Eddington limit,
momentum fluxes of the order of ${\rm \sim 20~L_{AGN}/c}$ \citep[e.g.][]{Zubovas+12, Faucher-Giguere+12}.

Very interestingly,
we find that the ``momentum boost'', i.e. the ratio of ${\rm v\,\dot{M}_{H_2, OF}}$ to the AGN radiative momentum output
${\rm L_{AGN}/c}$, ranges 
from $\sim$10 to $\sim$50 in the galaxies
where the AGN contributes more than 10\% to the total bolometric luminosity (Table \ref{table:res2}).
In particular,
most of our sources (except the ``pure'' starbursts) do follow, within the errors, 
the relation $\rm v\,\dot{M}_{H_2, OF} \sim 20~L_{AGN}/c$ (Fig. \ref{fig:plot6}). This finding both supports the AGN energy-driven
nature of these
outflows and the AGN feedback models that have been proposed so far, in which
a fast and highly ionised wind, arising from the nuclear regions of the AGN, creates a shock wave that propagates into the ISM of the galaxy
\citep{Zubovas+12, Faucher-Giguere+12}. We further
explore this hypothesis in Fig. \ref{fig:plot8}, by reproducing the diagram originally proposed by \cite{Faucher-Giguere+12} (Figure 5 in their paper), in
which the momentum ``boost'' of the outflows is plotted as a function of the outflow velocity. According to 
\cite{Faucher-Giguere+12}, under the assumption that
the observed molecular outflows are energy-conserving, the momentum boost ${\rm \dot{M}_{H_2, OF}/(L_{AGN}/c)}$ and the molecular outflow velocity ${\rm v_{OF}}$
are linked to the velocity of the initial nuclear and fast wind, ${\rm v_{in}}$, by the relationship:
\begin{equation}
\frac{\rm \dot{M}_{H_2, OF}}{(L_{AGN}/c)} \sim \frac{1}{2}\frac{\rm v_{in}}{\rm v_{OF}}.
\end{equation}
This relationship, computed for different values of ${\rm v_{in}}$, is represented by the grey dashed lines in Fig. \ref{fig:plot8}.
It is clear from Fig. \ref{fig:plot8} that we do not observe the correlation found by \cite{Faucher-Giguere+12}; however,
the trend shown in their paper is 
mainly driven by the two sources with outflow velocities of $\sim$5000 km s$^{-1}$, in which the powerful winds have been observed in ionised gas (FeLoBALs).
Due to the narrow range of velocities sampled by our molecular outflows, we probably cannot test the correlation of \cite{Faucher-Giguere+12}, but we
can still comment on the consistency of their model with our observations.  
Fig.\ref{fig:plot8} illustrates that nuclear ionised winds with ${\rm v_{in}} \in$ (0.01c, 0.1c) can in principle generate the molecular outflows observed in the AGN-dominated sources.

However, we should still keep in mind the bias affecting our sample, which has been selected to mostly include objects with known
(powerful outflows). Within this context it is useful to note that the upper limit given by I~Zw~1 points towards
a much lower ``momentum boost'' in the plot in Fig. \ref{fig:plot6}. If this object is representative
of a wider population (under-represented in our work because of our selection biases), then it may hint at other outflow acceleration mechanisms
at work in AGNs, such as direct radiation pressure onto the dusty clouds in the host galaxy (e.g. \citealt{Fabian12}).
Moreover, if the distribution of the clouds in the outflow is different with respect to what assumed by us in Sect. 3.2 (i.e. continuous versus
explosive scenarios) then the momentum rate may be lower by a factor of three, hence possibly favouring the latter models.


\subsection{The Driving Mechanism in the ``Pure'' Starbursts}

\begin{figure}[tb]
\centering
	\includegraphics[width=.8\columnwidth,angle=0]{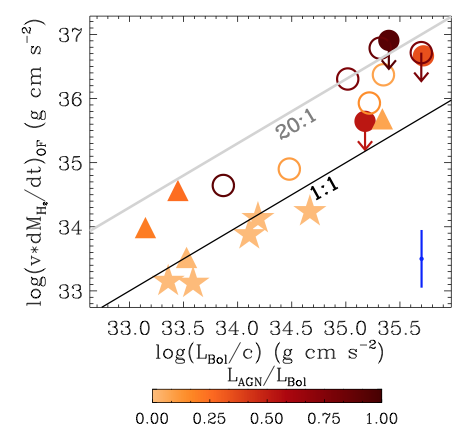}\\
   \caption{Outflow momentum rate versus total photon momentum output of the galaxy (${\rm L_{Bol}}$/c). 
   Symbols and colour-coding as in Fig. \ref{fig:plot1}. Error bars are as in Fig. \ref{fig:plot6}.
   The grey line shows the relation ${\rm \dot{M}_{H_2, OF}~v \sim 20~L_{Bol}/c}$. 
  }
   \label{fig:plot7}
\end{figure}

\begin{figure}[tb]
\centering
	\includegraphics[width=.8\columnwidth,angle=0]{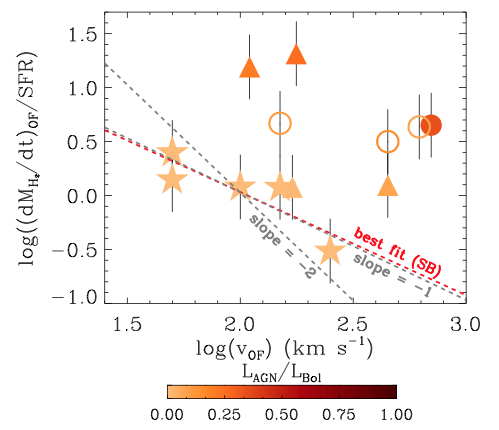}\\
   \caption{Outflow mass loading factor (${\rm \dot{M}_{H_2, OF}}$/SFR) plotted as a function of the outflow
   velocity. Only the objects with an AGN contribution to the total bolometric luminosity of $\leq$50\% are plotted.
   Symbols and colour-coding as in Fig. \ref{fig:plot1}.  Error bars as in Fig. \ref{fig:plot1}.
   The red dashed line represents the linear fit to the five ``pure" starbursts galaxies of our sample  (M82, NGC 3256,
   NGC 3628, NGC 253, NGC 2146).}
   \label{fig:plot_1_sbonly}
\end{figure}

In all of the previous sections, we have seen that ``pure'' starburst galaxies deviate from the outflow 
relations found for galaxies hosting AGNs,
indicating that in these systems the outflow driving mechanism is likely different. 
Two main outflow mechanisms have been proposed for
starburst galaxies: ``energy-driven'', commonly associated with the energy injected into the ISM by SNe explosions, and ``momentum
driven'', generally associated with 
radiation pressure from the young stellar population on the dusty clouds 
\citep{Murray+05, Murray+11, Thompson+05, Oppenheimer+Dave06, Oppenheimer+Dave08, Hopkins+12}.
In this final section we investigate the outflow driving
mechanism in the starburst galaxies included in our sample by exploiting our results. 

It has already been shown in Fig.~\ref{fig:plot5} 
that the kinetic power of the outflow in starburst galaxies corresponds to only about 1\% of the
kinetic power produced by the SNe associated with the starburst. Therefore, if supernovae power the outflows, the coupling efficiency between supernova shocks and ISM must be quite low.

Additional information is given by the investigation of the outflow momentum rates. In particular,
it is interesting to compare the outflow momentum rates with the \textit{total} radiative momentum output of these galaxies,
i.e. with ${\rm L_{Bol}/c}$. For the heavily AGN-dominated sources, for which ${\rm L_{Bol} \sim L_{AGN}}$, such a relation, shown in Fig. \ref{fig:plot7}, 
naturally resembles that of Fig. \ref{fig:plot6}, showing momentum ``boosts" of up to a factor of 20. However, for the other AGN hosts (with lower AGN
contribution), Fig. \ref{fig:plot7} demonstrates that, when comparing the outflow momentum rate
with ${\rm L_{Bol}/c}$,  i.e. when including the contribution of the starburst luminosity in the 
photon momentum computation, ``momentum boosts'' greater than $\sim$20 are no longer required to explain our observations. Instead,
a significant contribution to the momentum rates of these molecular outflows may come from momentum deposition into the ISM from star formation.
In short, in these objects probably the radiation pressure from the AGN and from starbursts synergically combine to drive the outflow.

For the five ``pure'' starburst galaxies, (along with the low luminosity AGN NGC 6764, which however
has an AGN contribution of only $\sim$2\%) Fig. \ref{fig:plot7} shows that the momentum rate is comparable to ${\rm L_{Bol}/c}$,
suggesting that (single-scattering) radiation pressure onto the dusty clouds of the ISM may contribute substantially to, or even dominate, the
driving mechanism of outflows in starbursts.

The latter scenario is further supported by Fig.~\ref{fig:plot_1_sbonly}, which shows the outflow mass loading factor
(${\rm \dot{M}_{H_2, OF}}$/SFR) as a function of outflow velocity. In energy-driven outflows the mass loading
factor should be proportional
to $\rm v^{-2}_{OF}$, while in momentum-driven outflows the mass loading factor should be proportional to $\rm v^{-1}_{OF}$ (e.g. \citealt{Murray+05}).
The two dashed gray lines in Fig.~\ref{fig:plot_1_sbonly}
show a fit to the starburst galaxies with the slopes fixed to --1 and --2. Clearly the fit with slope --2, which would be compliant with the energy-driven
scenario, is inconsistent with the starburst data. Instead, the fit with slope --1, expected from momentum-driven models, is in nice agreement with the
starburst data. The red dashed line shows the best fit to the starburst data obtained by leaving the slope free, which gives:

\begin{equation}
{\rm log(\dot{M}_{H_2, OF}/SFR) = (1.9 \pm 1.0) + (-1.0 \pm 0.5)~log(v_{OF})}
\end{equation}

As shown by Eq. (5), the best fit to the pure starbursts 
is fully consistent with the slope --1,
while it is only marginally consistent (2$\sigma$) with slope --2.
In summary, our results, although based on a very modest sample size, 
suggest that in starburst galaxies outflows are mostly momentum-driven, at least for 
the molecular component (which generally dominates the outflow mass budget).

\section{Summary and Conclusions}

We have investigated the properties of massive molecular outflows in a
sample of 19 nearby galaxies, spanning a wide range of AGN and starburst activities.
For seven of these sources, we have presented new PdBI observations of the CO(1-0) emission line; in four of them,
we have detected, for the first time, CO broad wings tracing molecular outflows extended on
kpc scales, which indicates that 
these outflows are affecting their host galaxies on large scales. In other two sources
we have found some marginal evidence for the presence of broad CO wings, but
additional observations are required to confirm the presence and the nature
of this high velocity gas.
We have measured very large outflow mass-loss rates of several times 100~${\rm M_{\odot}~yr^{-1}}$, 
similarly to the prototypical giant molecular wind discovered in Mrk 231 \citep{Feruglio+10, Fischer+10}.
With the aim to shed light on the possible correlations between outflow characteristics and
intrinsic properties of a larger sample of objects, we have included in our study
12 additional local galaxies, in which outflows of molecular gas have been observed.
For homogeneity, we have restricted our investigation to molecular outflows revealed by
observations of high velocity components of the CO(1-0) or CO(2-1) emission lines.
We emphasize that the sample employed for our analysis (both our new sample and the sample
taken from the literature) is clearly biased
towards galaxies for which a molecular outflow {\it has} been detected.
By comparing the intrinsic characteristics of this sample of nearby galaxies and AGN hosts
with their outflow properties and energetics, we have found that:
\begin{itemize}
\item Starbursts are effective at powering massive molecular outflows, but the presence of
a powerful AGN, when it dominates the energetic output of the galaxy, can significantly boost the outflow rates.
More specifically, while in starburst-dominated galaxies the outflow mass loading factor,
(${\rm \eta = \dot{M}_{H_2, OF}/SFR}$), is close to unity, with maximum values of 2--4, the mass loading factor
$\eta$ is found to increase with the fraction of total bolometric luminosity that is 
associated with the AGN, approaching values of $\sim$100 in the most powerful QSOs.
\item We have found a correlation 
between outflow rate and AGN luminosity for the
AGN host galaxies of our extended sample. However, this correlation may be tracing the upper envelope of a 
broader distribution. Indeed our sample is certainly biased towards objects whose presence 
of a molecular outflow was known a priori. A completely unbiased survey may deliver more 
sources with lower outflow rates, placed below the relation found by us. The presence of two 
objects, in our sample, hosting powerful AGNs and placed below the relation, hints at this scenario.
\item We have found that these AGN-driven massive molecular outflows are capable of
exhausting the cold gas reservoir of the host galaxy and, hence, of quenching star
formation, on time-scales as short as a few times 10$^6$ yrs, for the most luminous AGNs. 
In particular, we have found an anti-correlation 
between depletion time-scale and AGN power, 
confirming previous studies based on the
far-IR OH absorption lines \citep{Sturm+11}.
In AGN-hosts, the depletion time-scale due to the molecular outflow is much
shorter than the depletion time-scale of gas consumption due to star formation.
\item A comparison between kinetic power of the outflow and AGN power has
shown that these two quantities are correlated, and that their ratio, especially for the most powerful AGNs (likely
accreting close to the Eddington limit), is very close
to the value of 5\% predicted by theoretical models, commonly adopted in simulations of AGN negative
feedback, and needed to match the $\rm M_{BH}-\sigma$ relationship. Further confirmation to the hypothesis of AGN-driven outflows
is provided by the finding that
the energetics of the molecular outflows is totally inconsistent with supernovae-driven winds
in AGN-dominated sources.
\item We find that outflow momentum rates of ${\rm \sim 20~ L_{AGN}/c}$ are common
among the AGN-host galaxies of our extended sample. This likely indicates that the molecular outflows that we observe
in these sources are energy-conserving, and they may have been generated by fast nuclear
winds that subsequently
impact the interstellar medium of the host galaxy. However, other driving mechanisms (such as direct acceleration
of dusty clouds through radiation pressure),
may also contribute, especially for outflows with lower ``momentum boosts'', whose fraction in our sample
may be heavily under-represented due to selection biases. Moreover one should take into account that
the distribution of clouds in the outflow may be
different from what assumed by us, and this may reduce the inferred momentum rate by a factor of about three.
\item Interestingly,
in the starburst-dominated galaxies
(i.e. those where an AGN is present, but it only contributes less than 10\% to the total
bolometric luminosity of the galaxy), the power
source of the molecular outflow is ambiguous. On the one hand, the outflow rates 
are not much larger than their
SFRs. On the other hand, in terms of outflow energetics and momentum rates, many of them follow the same
relation as AGN-dominated galaxies, strongly suggesting that even in these SB-dominated objects
the AGN can be the main driving source of the outflow. This suggests that even if the AGN is not the main source 
of energy, it can be more effective than the starburst in driving outflows. Alternatively, a more powerful AGN may have
been present in the past and it may have been responsible for driving most of the outflow, which can continue for a few hundred million years since
the AGN has faded, as predicted by some models.
This would also explain the observations of high velocity (i.e. v$\sim$1000 km s$^{-1}$) 
molecular outflows in stacked spectra of galaxies which do not show a clear AGN contribution \citep{Chung+11},
as well as the detection of very fast atomic outflows (${\rm v<1000~km~s^{-1}}$) in galaxies at
z$\sim$0.6 showing no evidence of strong AGN activity \citep{Diamond-Stanic+12}: in these cases the AGN may
not dominate the total luminosity (and its detection may be made difficult by the dilution from the starburst in the host galaxy), but it may
still dominate the outflow mechanism. AGN variability can also account for the lack of AGN detection in these sources.
Moreover, a bias against detecting powerful AGNs, even based on mid-IR or X-ray diagnostics,
in outflows oriented perpendicular to our line of sight or during the early stages of outflow (when the AGN
is still mostly embedded in 
high column densities), may affect the classification of some objects.
\item For ``pure'' starburst galaxies (i.e. galaxies which do not host an AGN) 
our data suggest that outflows (at least their molecular phase)
are probably momentum-driven. This is supported by the finding that the outflow mass loading factor is proportional to $\rm v^{-1}$, 
as expected
by momentum-driven models, and by the finding that their momentum rate is comparable to $\rm L_{bol}/c$, suggesting that
(single scattering) radiation pressure onto dusty clouds is responsible for expelling these clouds. 
\end{itemize}
In conclusion, we have studied massive molecular outflows in galaxies spanning a wide
range of starburst and AGN activities. In most cases, these outflows are powerful enough
to quench the star formation in the host galaxy, by removing the cold molecular gas out
of which stars form. Our results suggest that such negative
feedback on star formation is most efficient and dramatic in powerful AGNs, whose presence
can significantly boost outflow rates, kinetic powers and momentum rates of these molecular winds
and reduce the depletion time-scale of molecular gas. Starbursts are also effective
in driving massive outflows (likely through radiation pressure), although not as much as AGNs.
However, one shall take into account that star formation likely lasts for a longer period than the AGN activity. As a consequence,
while AGN feedback has likely a dramatic effect in rapidly clearing massive galaxies of their gas content on short time-scales,
the integrated effect of starburst-driven winds may be more important on longer time-scales, especially for lower mass galaxies.
Further observations, carried on with the new generation of millimiter-wave interferometers (ALMA, NOEMA)
will allow us to extend this investigation to a much larger sample of 
local galaxies, and, most importantly, overcome the biases that are affecting the current sample.

\begin{acknowledgements}
 We thank the anonymous referee for constructive comments
that improved the paper. We also thank Andy Fabian,
Andrea Cattaneo and Claude-Andr\`{e} Faucher-Gigu\'{e}re for extremely
helpful discussions and suggestions.
 C.C. acknowledges support by the Isaac Newton Studentship.
This work is based on observations carried out with the IRAM Plateau de Bure Interferometer. 
IRAM is supported by INSU/CNRS (France), MPG (Germany), and IGN (Spain).
Basic research in IR astronomy at NRL is funded by the US ONR and J.F. acknowledges 
support from the NASA Herschel Science Center.
\end{acknowledgements}

\bibliography{ref}
\bibliographystyle{aa}

\appendix \section{Additional Information}

We report in this appendix additional information about the
seven galaxies that we observed with the PdBI
in their CO(1-0) transition. We further discuss
our PdBI CO(1-0) observations, by showing the results
of the fits to the {\it uv} plots of the CO wings, which provide
an estimate of the size of the molecular outflow (when detected).
We also present the integrated map, the corresponding {\it uv} plot
and the velocity field of the narrow core CO(1-0)
emission, which traces the bulk of the 
molecular gas in rotation in the galaxy disk/ring.

\subsection{IRAS F08572+3915}

\begin{figure}[h!]
\centering
{\includegraphics[width=.42\columnwidth, angle=270]{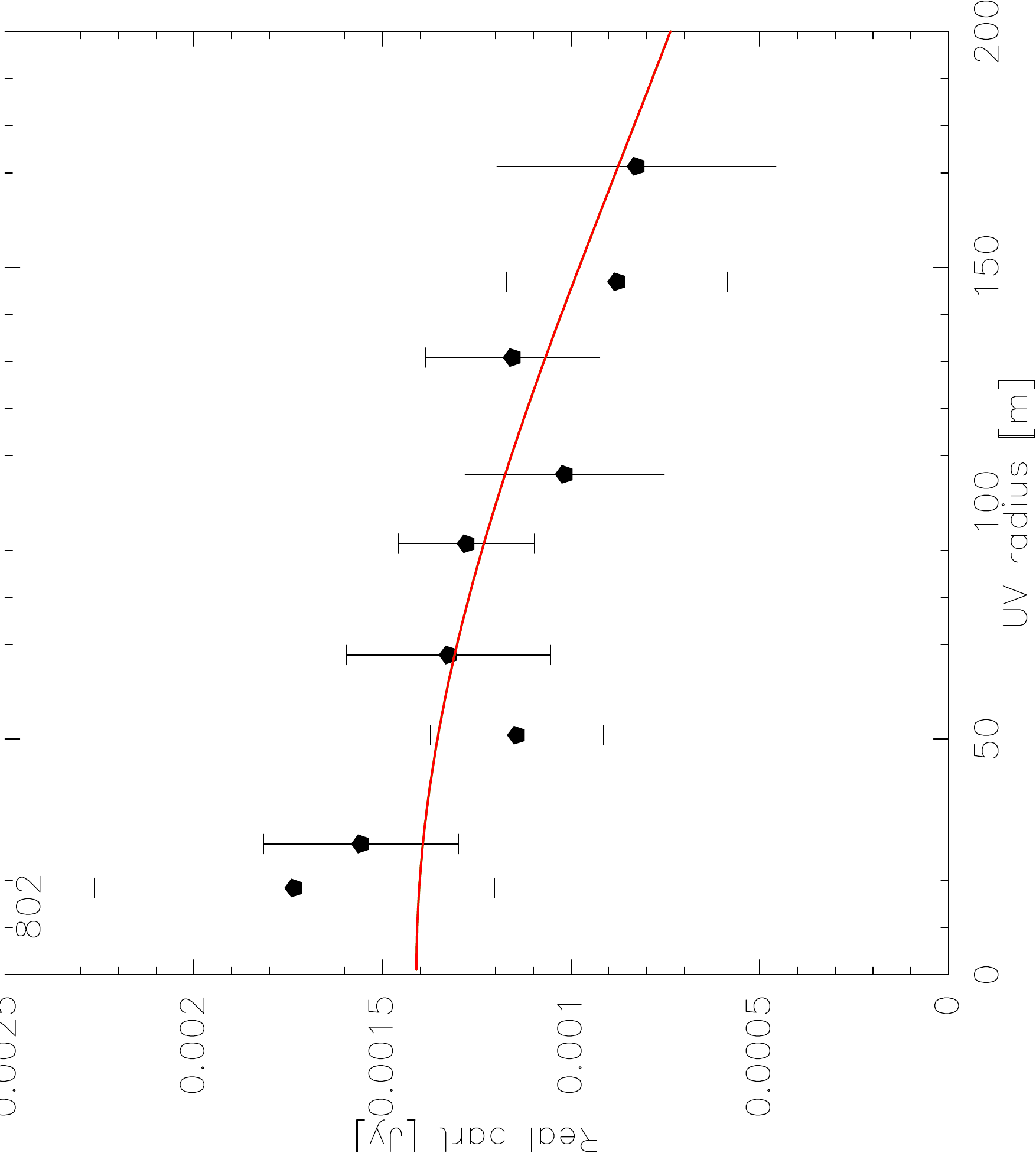}} \quad
{\includegraphics[width=.42\columnwidth, angle=270]{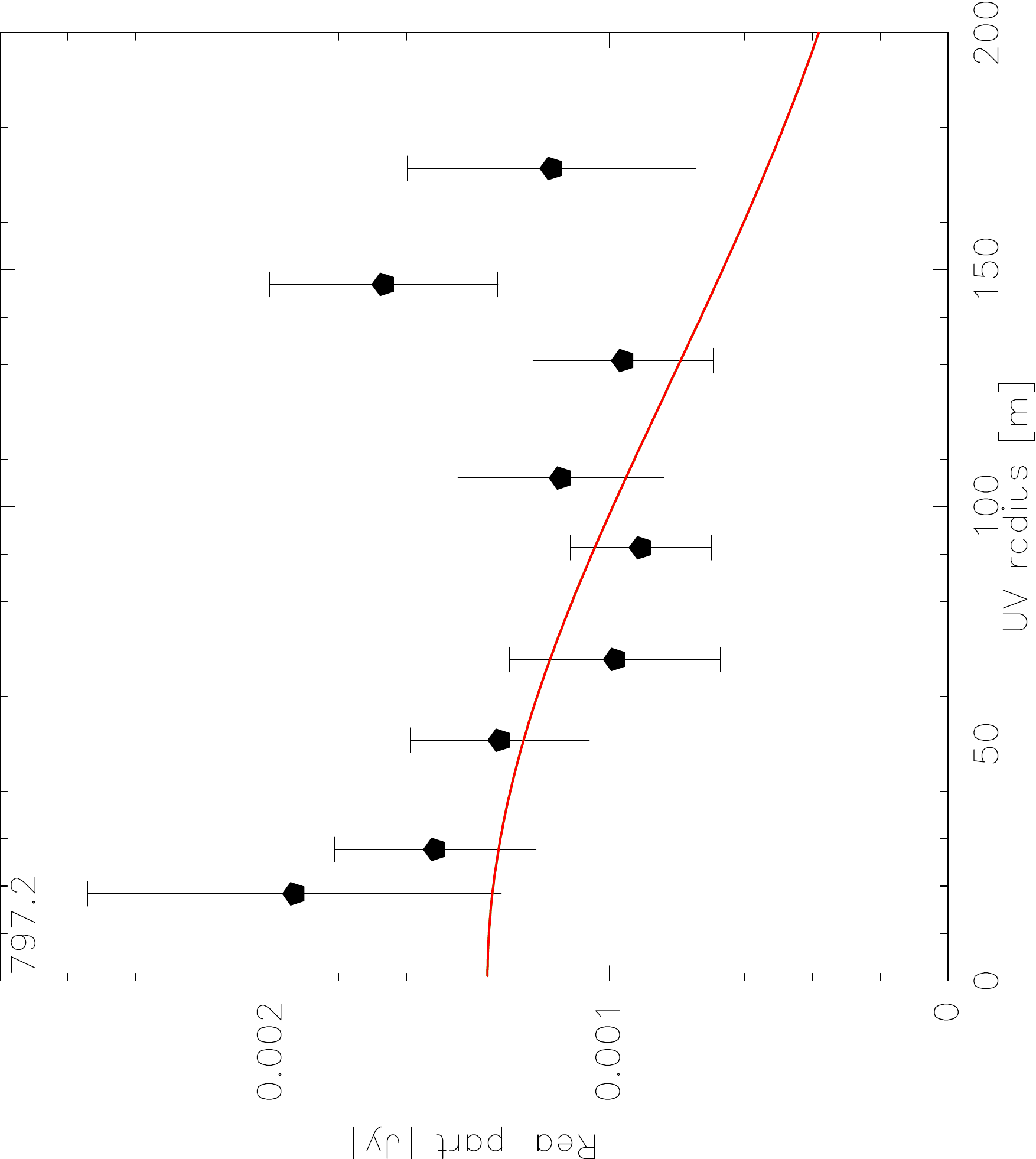}} \\
\caption{Real part of visibilities as a function of the {\it uv} distance for the blue (\emph{left})
and red (\emph{right}) CO(1-0) wings of IRAS F08572+3915, binned in baseline steps of 20m. 
The red curve is the best fit with a circular Gaussian model. Note that the {\it uv} tables
of the two CO(1-0) wings have been re-centred to their emission centroid positions (obtained from their maps
shown in Fig. \ref{fig:f0857}) prior to plotting the {\it uv} visibilities in order to correctly estimate their size.
} \label{fig:wings_uvplot_va26} \end{figure}

\begin{figure}[h!]
\centering
{\includegraphics[width=\columnwidth]{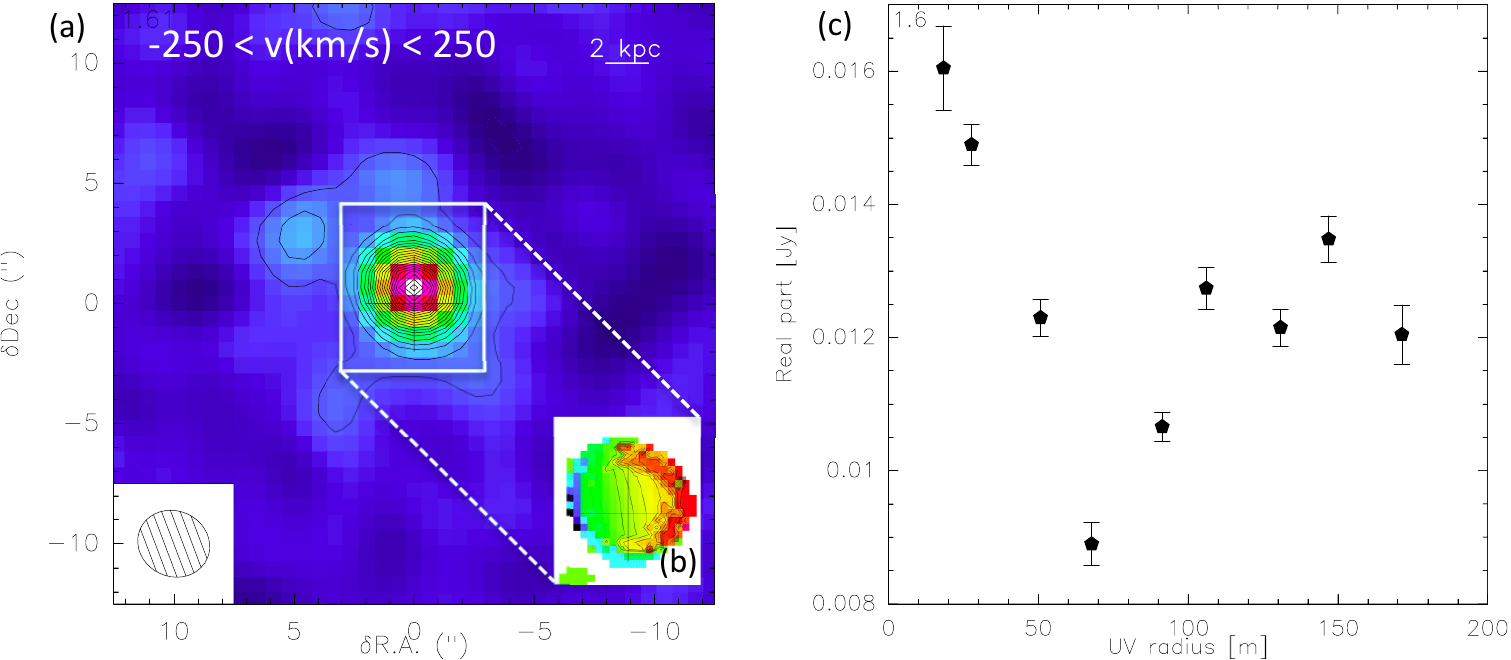}} \\
\caption{Panel (a): IRAM PdBI continuum-subtracted map and of the CO(1-0) narrow core emission in IRAS F08572+3915. Contour
levels correspond to 5$\sigma$. The inset (b) shows 
the first moment map obtained within v$\in$(-150, 150) km s$^{-1}$.
Contours correspond to 3 km s$^{-1}$. The size of the inset is 6$\times$7 arcsec.
Panel (c): \emph{uv} plot for the narrow core, with visibilities binned in {\it uv} radius steps of 20m.
} \label{fig:core_va26} \end{figure}

The Sy2-ULIRG IRAS F08572+3915  is an interacting system composed by two nuclei (NW and SE)
separated by a projected distance of 5.4 arcsec (6.1 kpc); only the north-west (NW) component,
which is the most luminous one and hosts a buried QSO, has been detected in CO(1-0) interferometric
maps \citep{Evans+02}. \cite{Papadopoulos+12a} measure a very large ${\rm L'_{CO(6-5)}/L'_{CO(1-0)}}$ ratio
in the NW nucleus, indicating very high ISM excitation, among the most extreme in their ULIRG sample.

The NW galaxy hosts powerful, multi-phase and large-scale outflows, characterised by very high gas velocities
in each component. The ionised and neutral winds have been resolved at $\sim$kpc scales by \cite{Rupke+Veilleux13a},
based on several optical emission lines and NaID absorption, and they appear to be partially overlapping. 
The ionised outflow shows blue-shifted velocities of up to $\sim$3350 km s$^{-1}$, while the maximum blue-shifted velocity detected in
NaID absorption is $\sim$1150 km s$^{-1}$.
Herschel--PACS observations of IRAS F08572+3915 revealed the presence of a powerful molecular outflow,
with OH 79 and 119$\mu$m blue-shifted velocities of up to $\sim$1100 km s$^{-1}$ (\citealt{Sturm+11, Veilleux+13}).
Interestingly, recent Keck--OSIRIS observations of the near-IR H$_2$ roto-vibrational transitions show that warm molecular gas is
also outflowing with velocities strikingly similar to those observed in the cold molecular phase \citep{Rupke+Veilleux13b}. 
This would suggest that molecular clouds may be affected by shocks during the acceleration process in this source.

In IRAS F08572+3915, both CO(1-0) wings are detected at high significance 
($>$ 10$\sigma$, see Fig. \ref{fig:f0857}), which allows us to investigate their spatial extent separately. 
For this purpose, we fit the visibility as a function of the {\it uv} radius (the plot of the real part of the visibility vs {\it uv} radius is shown in
Figure \ref{fig:wings_uvplot_va26}) with
a circular Gaussian model and we obtain a size (FWHM) of 1.36 $\pm$ 0.45 kpc
for the blue wing, and 1.91 $\pm$ 0.50 kpc for the red wing. The value reported in Table \ref{table:COwings} is
the average of the two. 

In Fig. \ref{fig:core_va26}  we present the map of the CO(1-0) narrow core integrated 
within $\pm$ 250 km s$^{-1}$, the corresponding {\it uv} plot and the velocity field of this central concentration of
molecular gas. The CO(1-0) narrow emission 
shows a non compact structure, which is also clearly appreciable in the {\it uv} visibility data. 
Multiple blobs of molecular gas at the systemic velocity are detected, one of which may be
part of the tidal bridge between the NW and the SE galaxy of the merger. The SE companion
is discernible in optical and near-IR images, but it has never been detected in molecular gas
\citep{Evans+02, Papadopoulos+12b}.
The first moment map shows that the molecular gas follows a regular rotation pattern
oriented east-west, with an approximate rotation major axis at a PA=-80 deg.
The {\it uv} plot
cannot be modelled with a single Gaussian but probably requires a more complex function.

\subsection{IRAS F10565+2448}

\begin{figure}[h!]
\centering
{\includegraphics[width=.42\columnwidth, angle=270]{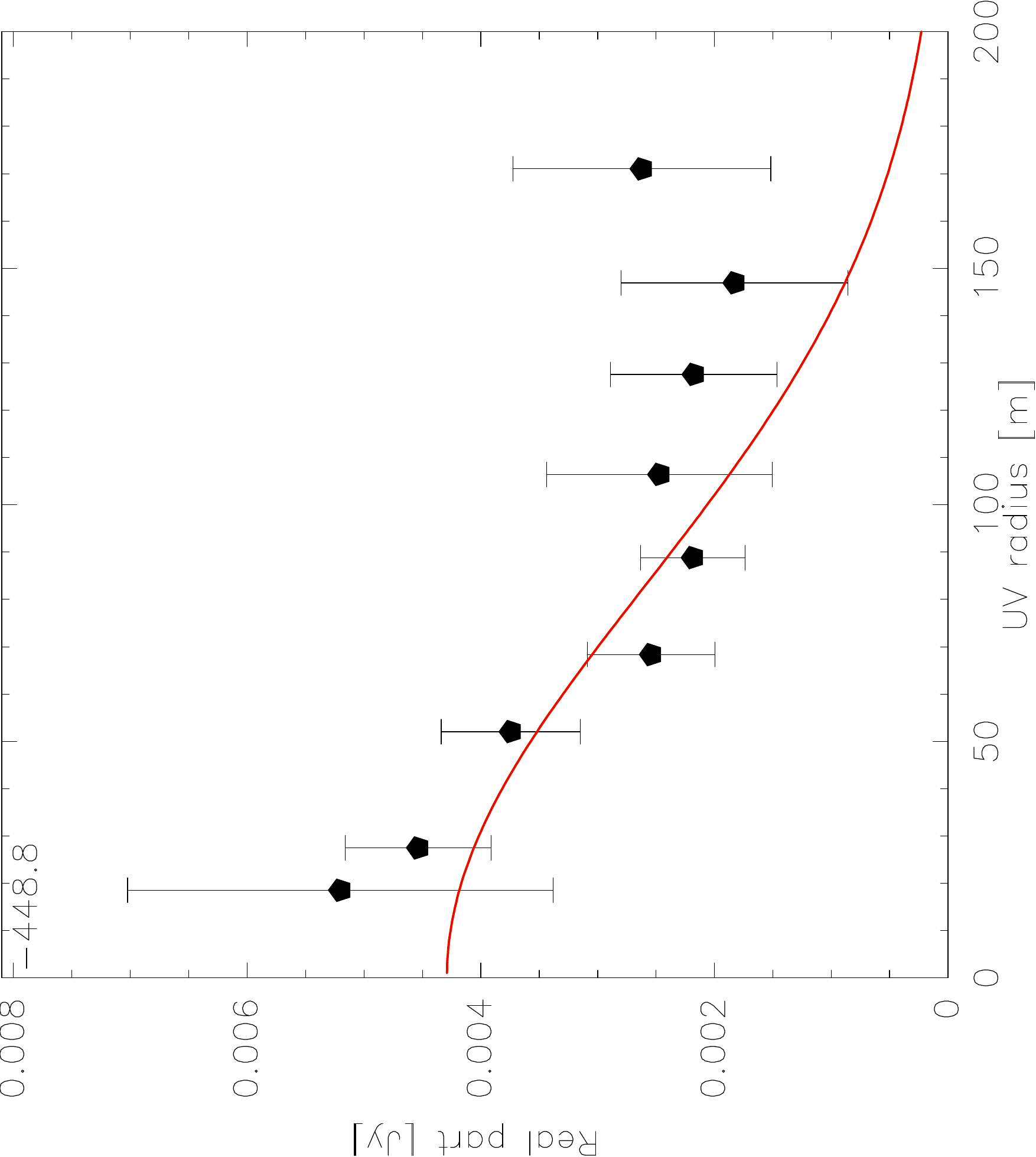}} \quad
{\includegraphics[width=.42\columnwidth, angle=270]{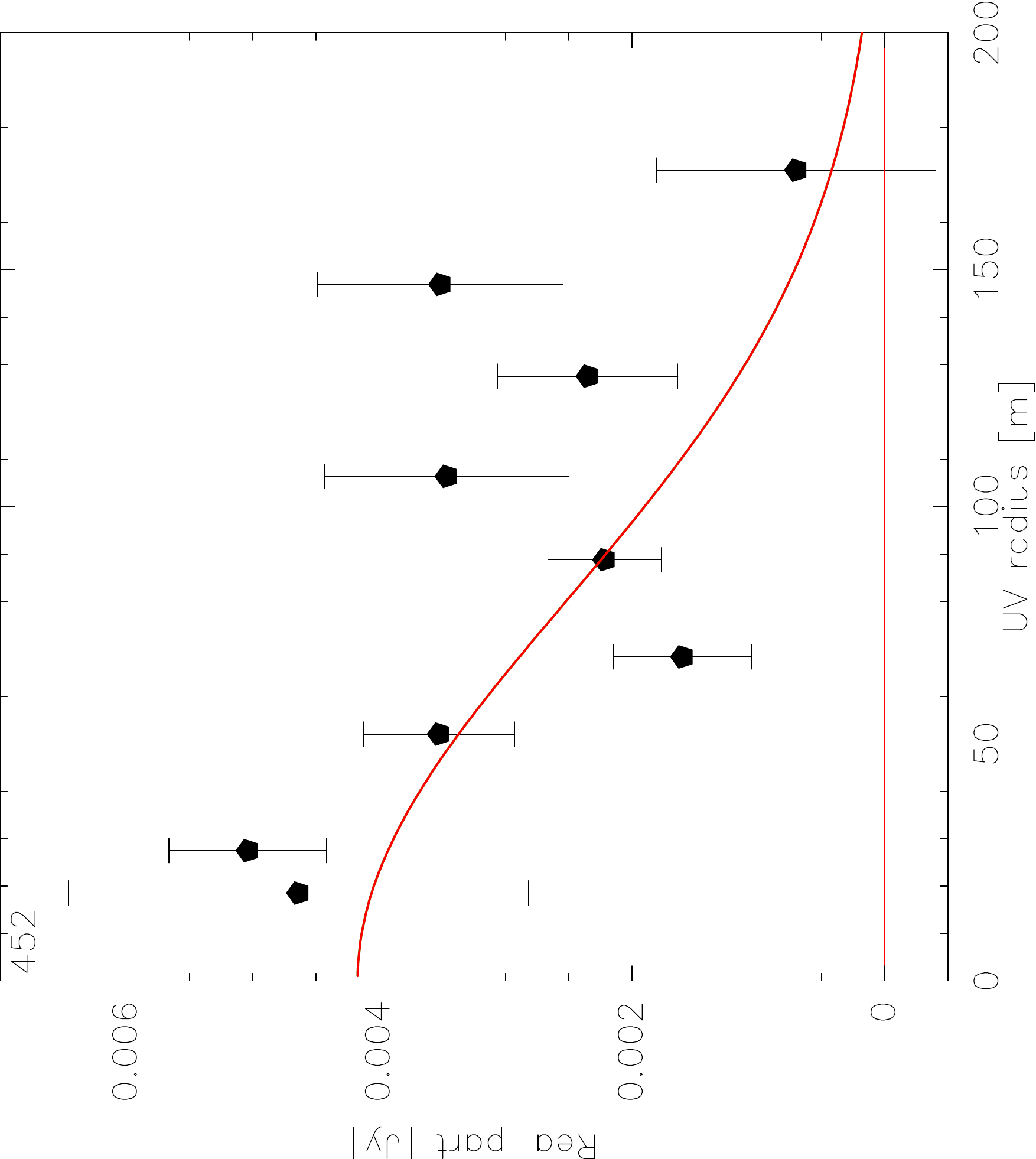}}\\
\caption{Real part of visibilities as a function of the {\it uv} distance for the blue (\emph{left})
and red (\emph{right}) CO(1-0) wings of IRAS F10565+2448, binned in baseline steps of 20m. 
The red curve is the best fit with a circular Gaussian model. Note that the {\it uv} tables
of the two CO(1-0) wings have been re-centered to their emission centroid positions (obtained from their maps
shown in Fig. \ref{fig:f1056}) prior to plotting the {\it uv} visibilities in order to correctly estimate their size.}
\label{fig:wings_uvplot_vb26} \end{figure}

\begin{figure}[h!]
\centering
{\includegraphics[width=\columnwidth]{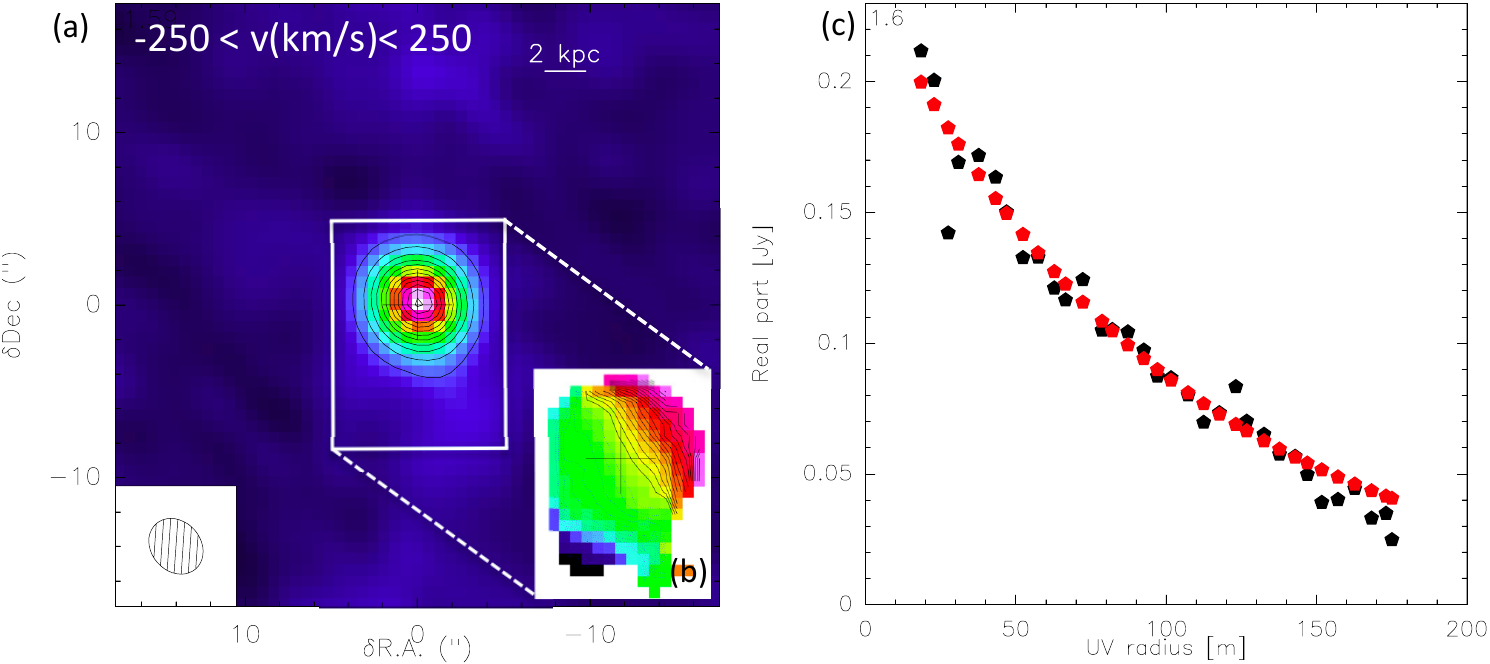}} \\
\caption{Panel (a): IRAM PdBI continuum-subtracted 
map of the CO(1-0) narrow core emission in IRAS F10565+2448. 
Contours correspond to 50$\sigma$. 
The inset (b) shows the first moment map obtained within v$\in$(-200, 200) km s$^{-1}$.
Contours correspond to 3 km s$^{-1}$. The size of this inset is 10 $\times$ 13 arcsec.
Panel (c): \emph{uv} plot for the narrow core, with visibilities binned in {\it uv} radius steps of 5m,
fitted with a power-law function model of the form ${\rm \propto r^{-3}}$ (red points).
} \label{fig:core_vb26} \end{figure}

The second Sy2-ULIRG of our sample, IRAS F10565+2448, is a possible triple
merger galaxy system, whose emission is largely dominated by the westernmost member. This is
also the only merger component that has been detected in CO(1-0) emission 
by \cite{Downes+Solomon98} with the PdBI and, later on,
by \cite{Chung+09} with the FCRAO 14m telescope. 
As for IRAS F08572+3915, the optical spectrum of IRAS F10565+2448 shows evidence of kpc-scale
ionised and neutral winds, although with outflowing velocities much more modest than IRAS F08572+3915
\citep{Rupke+Veilleux13a}.

The circular Gaussian fit to the {\it uv} plots reported in 
Fig. \ref{fig:wings_uvplot_vb26}, binned in baseline steps of 20m, provides
sizes (FWHM) of 2.15 $\pm$ 0.32 kpc and 2.22 $\pm$ 0.30 kpc for the blue and 
red-shifted CO(1-0) wings, respectively.

The emission from the CO(1-0) narrow core (i.e. within $\pm$ 250 km s$^{-1}$), 
reported in Fig. \ref{fig:core_vb26}a, appears more compact than the wings, 
although we find evidence, in the 
channel maps, for a plume of CO(1-0) emission extending to 
the south-west both at the systemic velocity and at slightly blue-shifted velocities of
about --150 km s$^{-1}$. Such plume is also clearly visible in the 
inset (b) of Fig. \ref{fig:core_vb26}, which shows the velocity field
within the velocity range of v$\in$(-200, 200) km s$^{-1}$. The major axis of the rotation of the
molecular gas is oriented with a PA$\simeq$-60 deg. 
The plot of the visibility as a function of the {\it uv} radius proves that the narrow component is very well
resolved in our data, and its emission can be fitted with a power law of the form ${\rm \propto r^{-3}}$,
resulting in a FWHM size of 0.781 $\pm$ 0.078 kpc. 

\subsection{IRAS 23365+3604}

\begin{figure}[h!]
\centering
{\includegraphics[width=.42\columnwidth, angle=270]{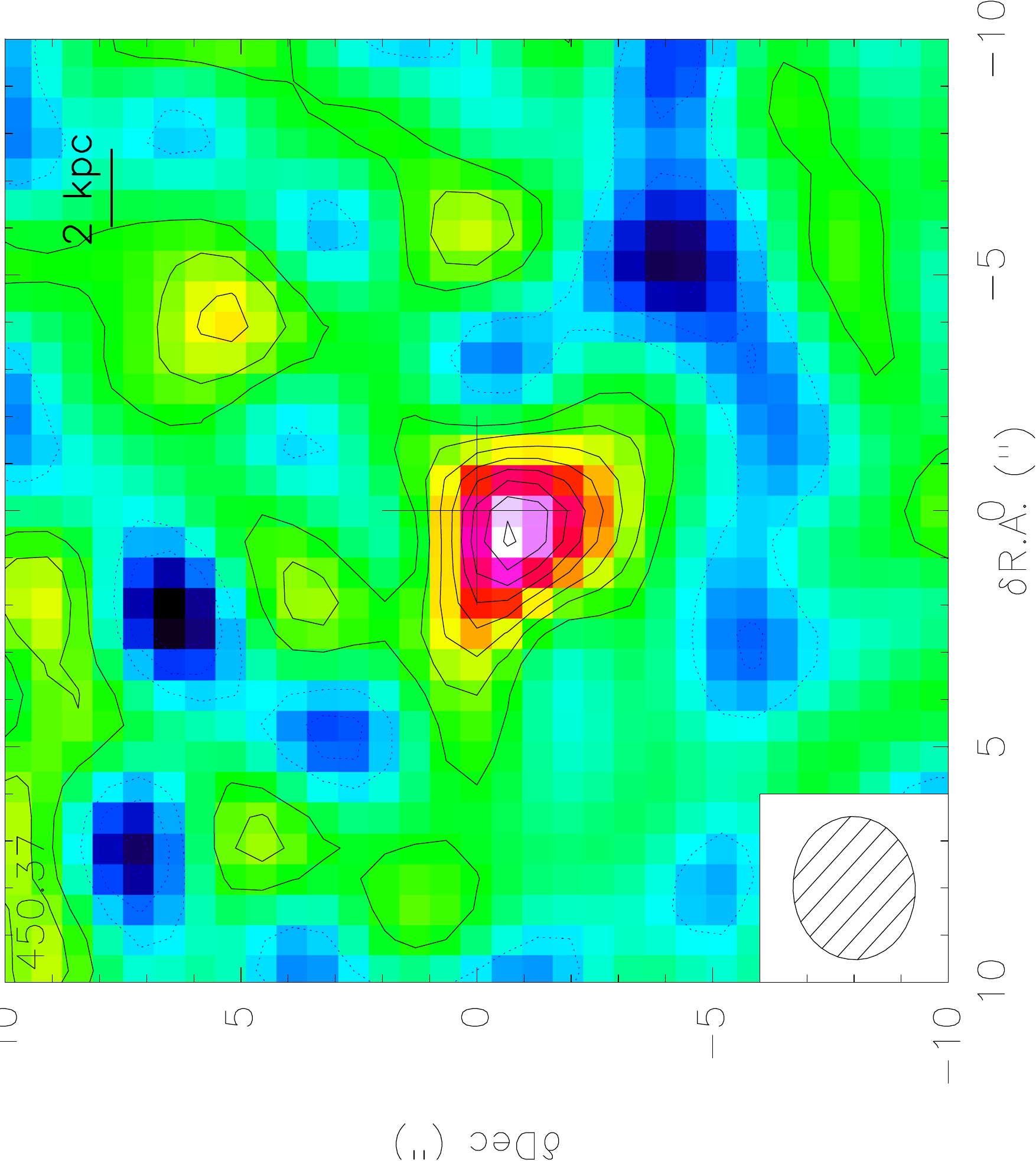}} \quad
{\includegraphics[width=.42\columnwidth, angle=270]{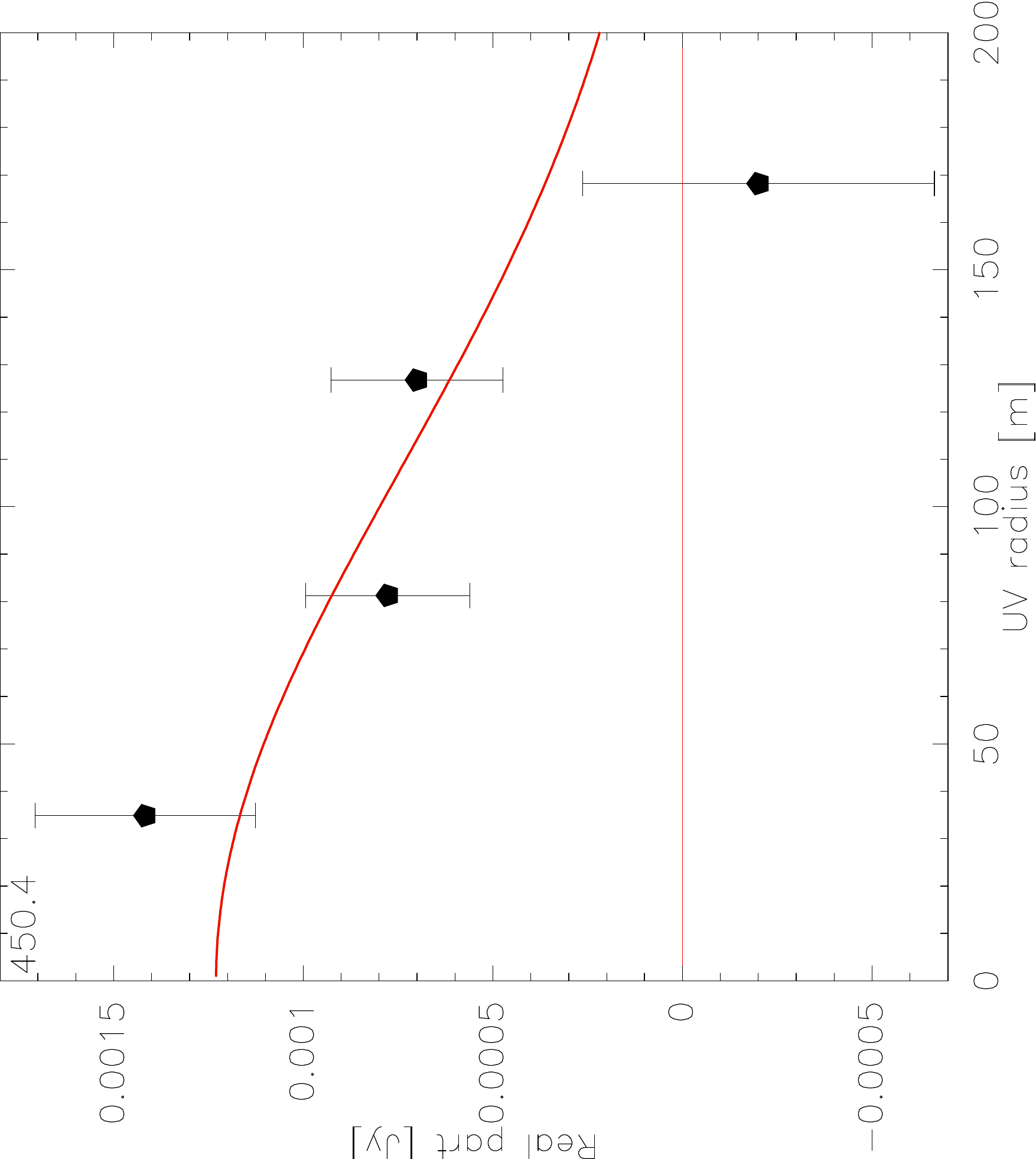}}\\
\caption{\emph{Left:} IRAM PdBI cleaned map of the CO(1-0) blue and red-shifted wings 
of IRAS 23365+3604, integrated within ($\pm$ 300, $\pm$ 600) km s$^{-1}$ and combined together.
Contour levels correspond to 1$\sigma$ ( 1$\sigma$ rms level is 0.15 mJy beam$^{-1}$).
\emph{Right:} Real part of visibilities as a function of the {\it uv} radius for the 
combined CO(1-0) wings; visibilities are binned in baseline steps of 50m. The red curve is the best fit
with a circular Gaussian model.}
 \label{fig:wings_vc26} \end{figure}

\begin{figure}[h!]
\centering
{\includegraphics[width=\columnwidth]{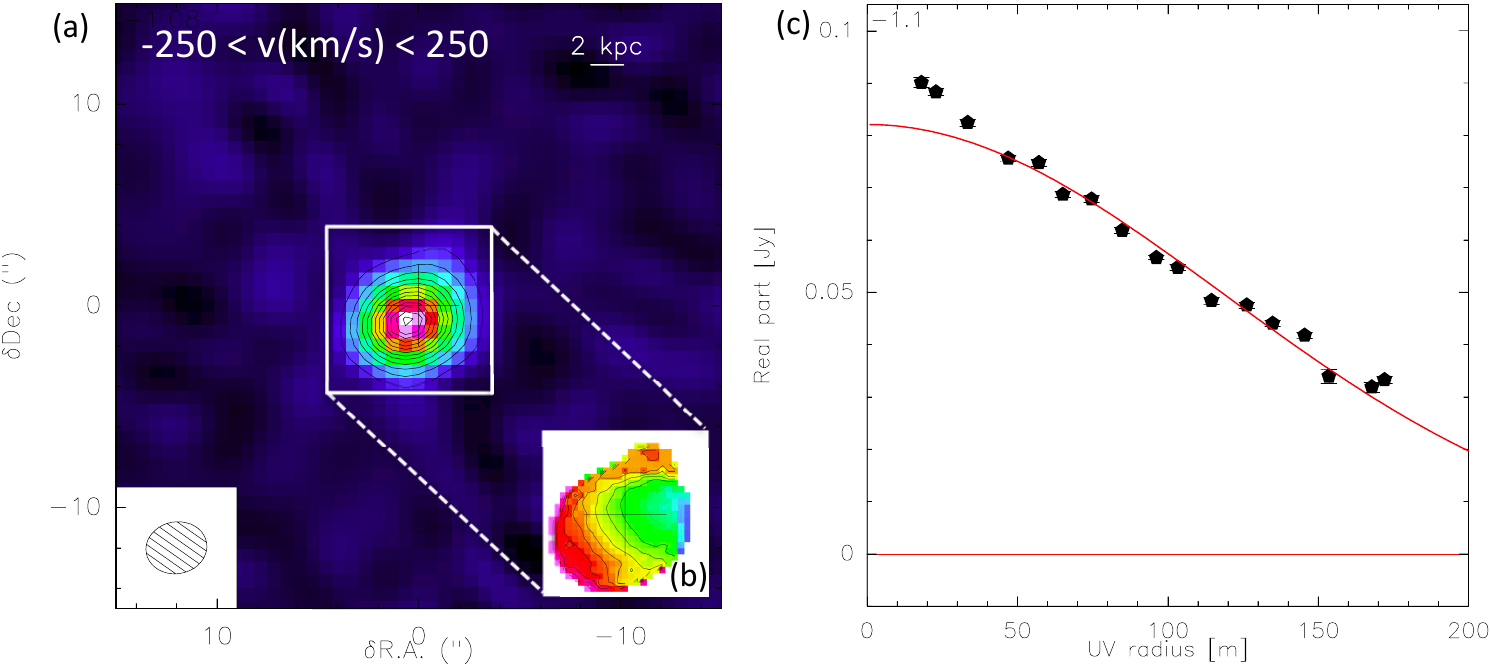}} \\
\caption{Panel (a): IRAM PdBI continuum-subtracted 
map of the CO(1-0) narrow core emission in IRAS 23365+3604. 
Contours correspond to 30$\sigma$ (1$\sigma$ rms level is 0.15 mJy beam$^{-1}$). 
The inset (b) shows the first moment map obtained within v$\in$(-200, 200) km s$^{-1}$. 
The size of the inset is 8 $\times$ 8 arcsec.
Panel (c): \emph{uv} plot for the narrow core, with visibilities binned in {\it uv} radius steps of 10m,
fitted with a circular Gaussian model.
} \label{fig:core_vc26} \end{figure}

Tidal tails revealed by near-IR imaging of the LINER--type ULIRG IRAS 23365+3604 indicate that this 
system is in a later merger stage, and no evidence for a double nucleus has been found so far
\citep{Downes+Solomon98}. Interestingly, based on a study of the physical conditions
of the dense molecular gas, \cite{Papadopoulos+12b} suggest an AGN influence on the ISM of this source.
IRAS 23365+3604 is one of the most spectacular cases of OH P-Cygni profiles
detected by Herschel--PACS, with blue-shifted velocities of up to 1300 km s$^{-1}$
\citep{Veilleux+13}.

In the case of IRAS 23365+3604, to improve the signal-to-noise 
in order to estimate the size of the outflow, we combine the 
two wings together: the corresponding map and {\it uv} plot
are shown in Figure \ref{fig:wings_vc26}. A circular Gaussian 
fit to the {\it uv} visibility data, binned in baseline steps of 50m, gives a
FWHM of 2.45 $\pm$ 0.70 kpc, which we use as an estimate of the size (diameter)
of the molecular outflow.

In Fig.~\ref{fig:core_vc26}a we also show the cleaned map of the CO(1-0) narrow
core, integrated within $\pm$ 250 km s$^{-1}$; with a
flux of ${\rm S_{CO, CORE}}$ = (45.24 $\pm$ 0.48) Jy km s$^{-1}$, this component
contributes more than 90\% to the total CO(1-0) emission of IRAS 23365+3604. Our observations resolve
reasonably well this core emission, and a circular Gaussian fit to the {\it uv} visibilities, binned in steps of
10m of {\it uv} distance, results in a FWHM of  3.174 $\pm$ 0.073 kpc.
We note in the channel maps the presence of a structure extended to the
north, which is also evident in the velocity map (Fig. \ref{fig:core_vc26}b), at 
slightly red-shifted velocities of 50-80 km s$^{-1}$. The narrow core
of the CO(1-0) emission in IRAS 23365+3604 traces a disk of molecular gas with 
a major rotation axis oriented at a PA $\simeq$ -55 deg.

\subsection{Mrk 273}

\begin{figure}[h!]
\centering
{\includegraphics[width=.42\columnwidth, angle=270]{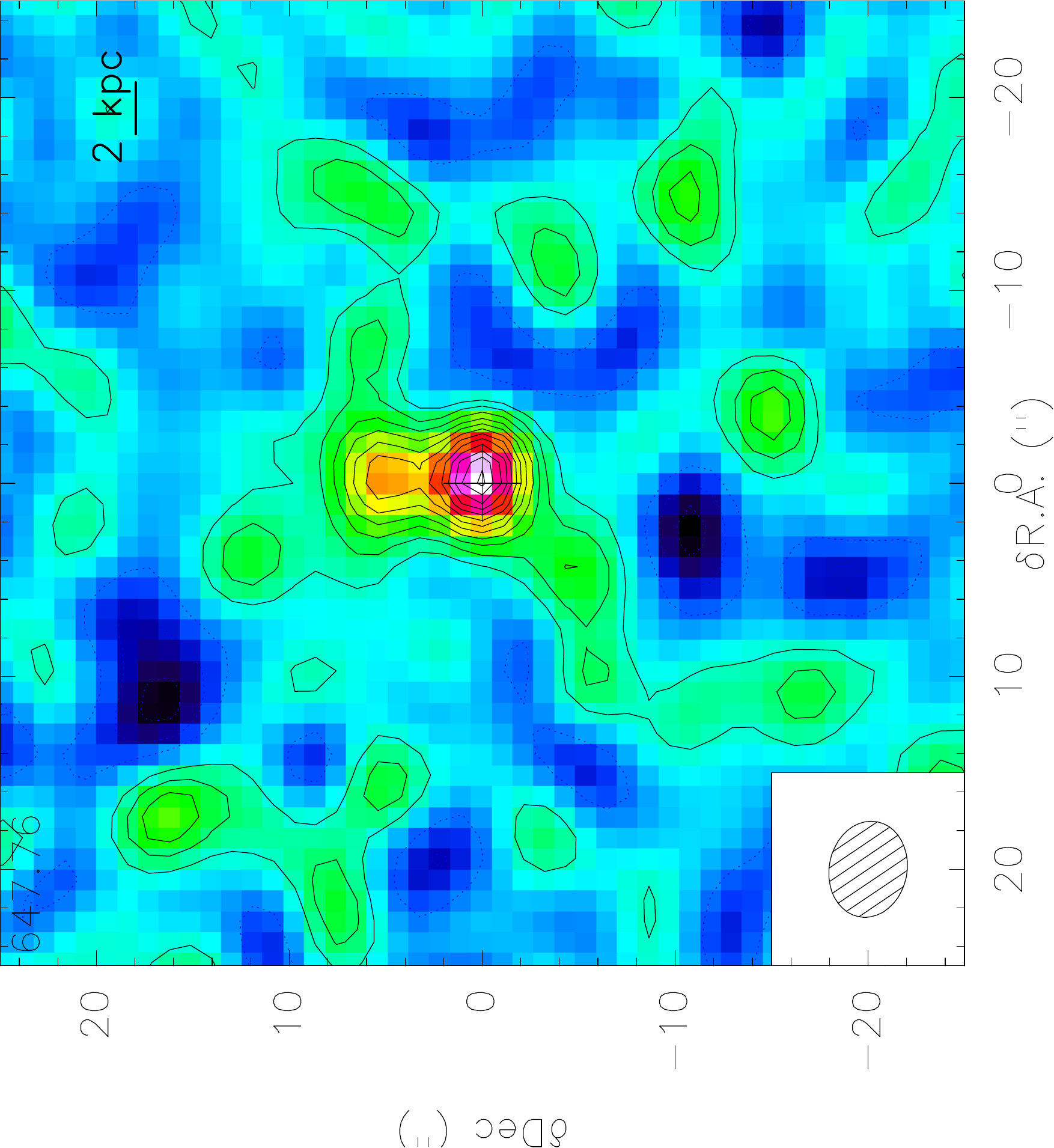}} \quad
{\includegraphics[width=.42\columnwidth, angle=270]{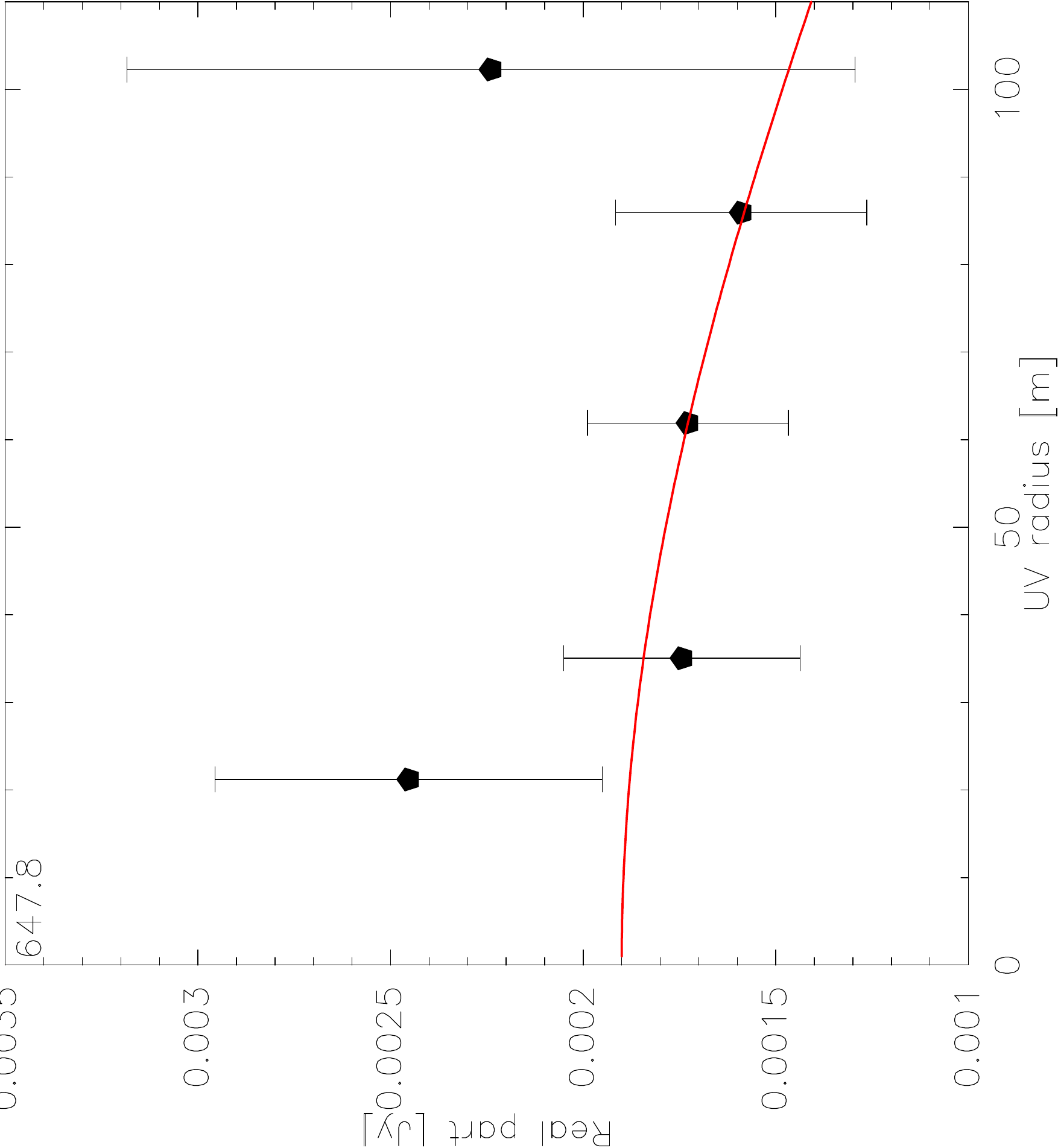}} \\
\caption{\emph{Left:} IRAM PdBI cleaned map of the CO(1-0) blue
and red-shifted wings of Mrk 273, integrated within 
v$\in$ (-800, -400) km s$^{-1}$ and v$\in$ (400, 900) km s$^{-1}$ and combined together.
Contour levels correspond to 1$\sigma$ (1$\sigma$ rms level is 0.2 mJy beam$^{-1}$).
\emph{Right:} Real part of {\it uv} visibilities as a function of the {\it uv} radius for the combined CO(1-0) 
wings, with visibilities binned in baseline steps of 25m. The red curve is the best fit 
with a circular Gaussian model.
} \label{fig:wings_wa26} \end{figure}

\begin{figure}[h!]
\centering
{\includegraphics[width=\columnwidth]{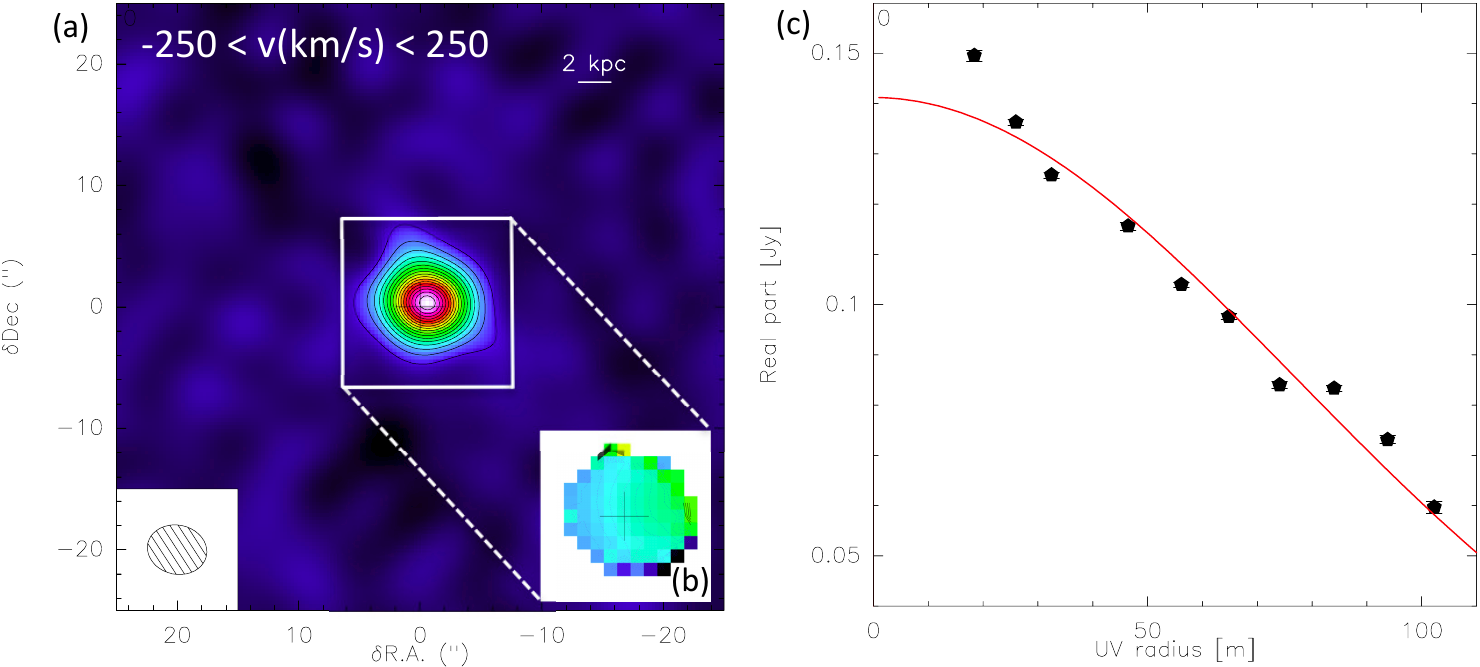}} \\
\caption{Panel (a): IRAM PdBI continuum-subtracted 
map of the CO(1-0) narrow core emission in Mrk~273. 
Contours correspond to 30$\sigma$ (1$\sigma$ rms level is 0.2 mJy beam$^{-1}$). 
The inset (b) shows the first moment map obtained within v$\in$(-100, 100) km s$^{-1}$
by applying a flux density threshold of 8 mJy. 
The size of the inset is 14 $\times$ 14 arcsec.
Panel (c): \emph{uv} plot for the narrow core, with visibilities binned in {\it uv} radius steps of 10m,
fitted with a circular Gaussian model.
} \label{fig:core_wa26} \end{figure}

Mrk 273 is a Sy2-ULIRG with a double nucleus in near-IR images and a 44 kpc long tidal tail,
which are signatures of merging. The physical conditions of the molecular gas in this
galaxy suggest the presence of prodigious starburst activity, fed by large amounts of
cold molecular gas \citep{Papadopoulos+12b}. The ionised and neutral gas kinematics is complex in this object,
characterised by kpc-scale outflows. In particular, \cite{Rupke+Veilleux13a}
detect two super-bubbles of outflowing ionised gas, aligned in the north-south direction,
which, due to the large velocities, are likely generated by an AGN.

The map of the blue and red-shifted CO(1-0) wings, integrated respectively
within the velocity ranges v$\in$(-800, -400) km s$^{-1}$ and v$\in$(400, 900) km s$^{-1}$
and combined together, is reported in the left panel of Fig. \ref{fig:wings_wa26}. 
We note that a significant fraction of the outflow (especially the red-shifted wing, see also Fig. \ref{fig:mrk273}d), 
is elongated towards north. Notably, despite the narrow bandwidth
available at that time, the CO(1-0) red-shifted component, peaking 5 arcsec to the north,
had already been observed by \cite{Downes+Solomon98}, although only up to
velocities of $\sim$350 km s$^{-1}$. These authors stressed
that the velocity and orientation of this structure are inconsistent with 
the nuclear disk and its east-west velocity gradient. 
This consideration by \cite{Downes+Solomon98} supports
our hypothesis (also discussed in Section 3) that the asymmetry of the CO(1-0) 
line at v$\in$(200, 400) km s$^{-1}$
traces outflowing gas. Indeed, the data reported by these authors
only recovered velocities ${\rm v\leq 350 km~s^{-1}}$, and, consequently,
they could not be observing the high velocity wings at ${\rm v>400 km~s^{-1}}$ which are
instead revealed, for the first time, by our new PdBI observations.
However, we emphasise that in our analysis we have chosen to be conservative 
and to define the red-shifted wing only from ${\rm v \geq 400~km~s^{-1}}$.

As for IRAS 23365+3604, we estimate the size of the outflow using
the {\it uv} plot of the two wings combined together, which is shown in the right panel 
of Fig. \ref{fig:wings_wa26}.
The partially decreasing trend of the 
{\it uv} plot is inconsistent with a point source, indicating that the outflow
is marginally resolved, despite the low spatial resolution of our D-conf data. 
The fit with a simple circular Gaussian (red curve in the plot) provides an approximate
size of (FWHM) of 1.1 $\pm$ 0.5 kpc.

For comparison, we show in Fig.~\ref{fig:core_wa26} the emission from 
the narrow core of the CO(1-0) line, integrated within ${\rm \pm 250~km~s^{-1}}$.
This component traces a central concentration of gas, which is in regular 
rotation, with a major axis oriented with a PA of about -75 deg. (panel b of Fig. \ref{fig:core_wa26}).
The {\it uv} plot of the narrow core emission can be 
modelled with a circular Gaussian with a FWHM of 
2.026 $\pm$ 0.046 kpc (panel (c) of Fig. \ref{fig:core_wa26}).

\subsection{IRAS F23060+0505}

\begin{figure}[h!]
\centering
{\includegraphics[width=.42\columnwidth, angle=270]{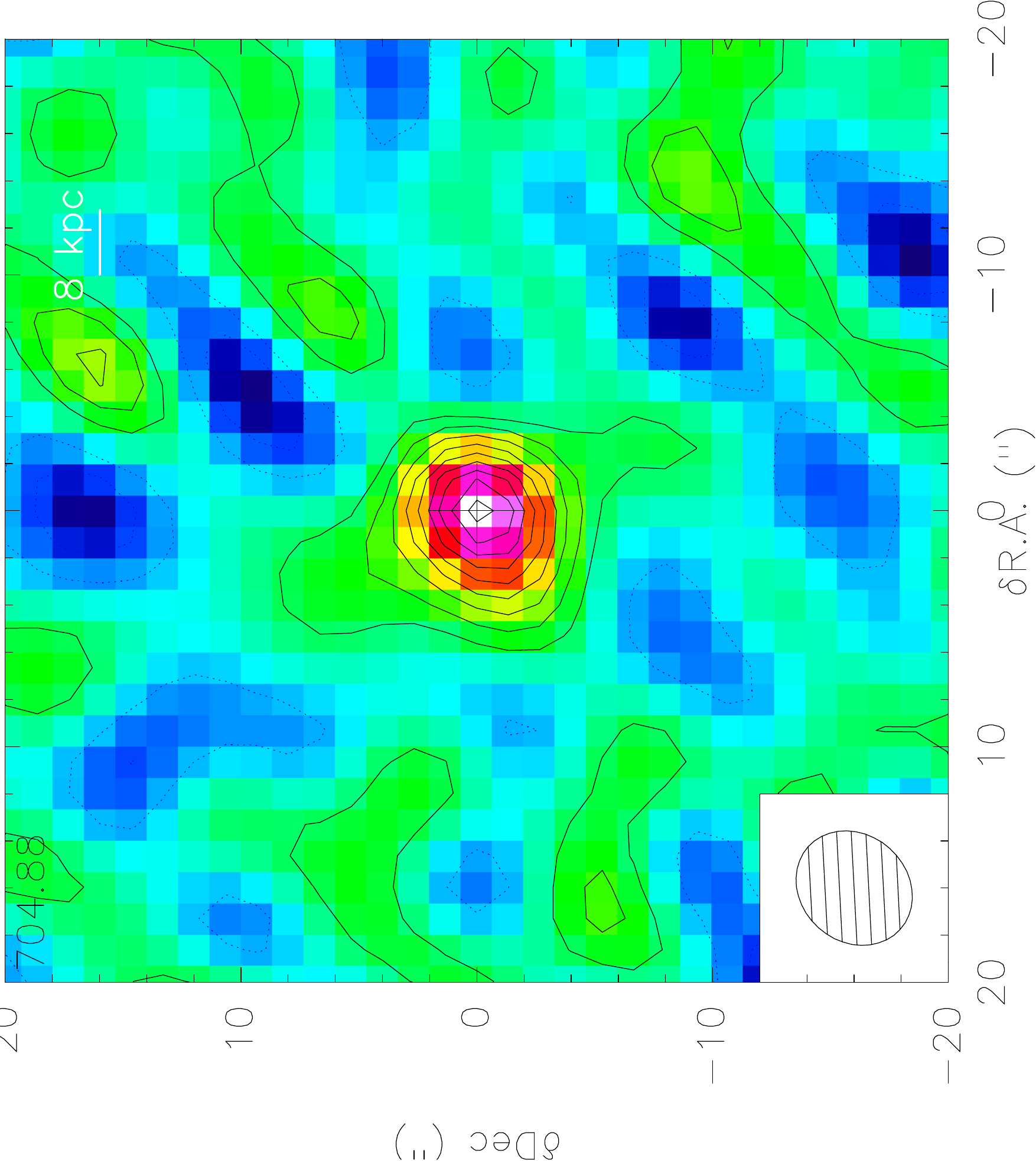}} \quad
{\includegraphics[width=.42\columnwidth, angle=270]{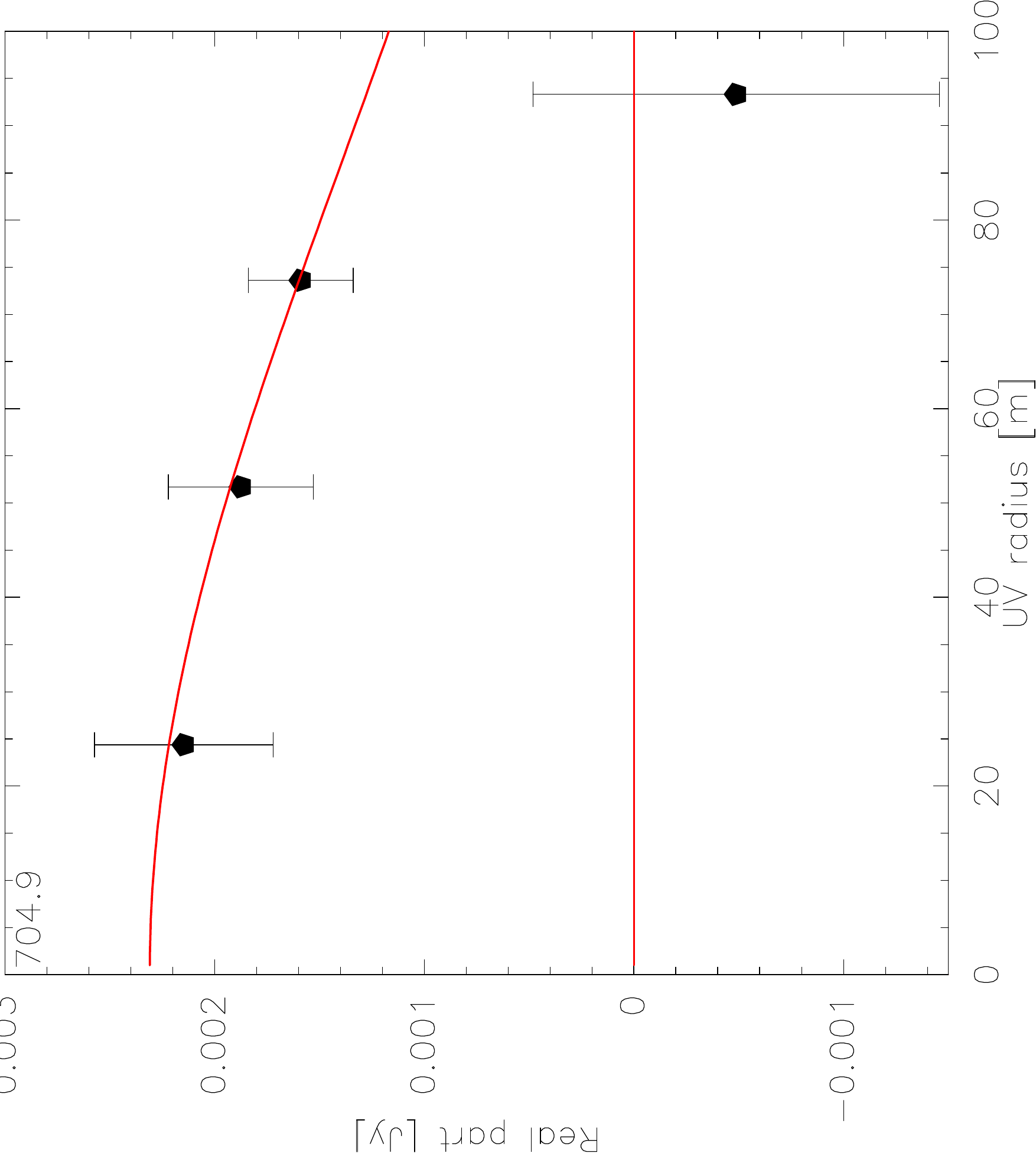}} \\
\caption{\emph{Left:} IRAM PdBI cleaned map of the CO(1-0) blue-shifted 
and red-shifted wings of IRAS F23060+0505, integrated in the velocity ranges
v$\in$ (-500, -300) km s$^{-1}$ and v$\in$ (300, 1100) km s$^{-1}$, and combined together.
Contours correspond to 1$\sigma$ ( 1$\sigma$ rms noise is 0.2 mJy beam$^{-1}$).
\emph{Right:} Real part of visibilities plotted as a function of the {\it uv} radius for the 
CO(1-0) broad wings. Visibilities are binned in baseline steps of 30m. The red curve is the result of the
best fit with a circular Gaussian model.}
 \label{fig:wings_uf2a} \end{figure}

 \begin{figure}[h!]
\centering
{\includegraphics[width=\columnwidth]{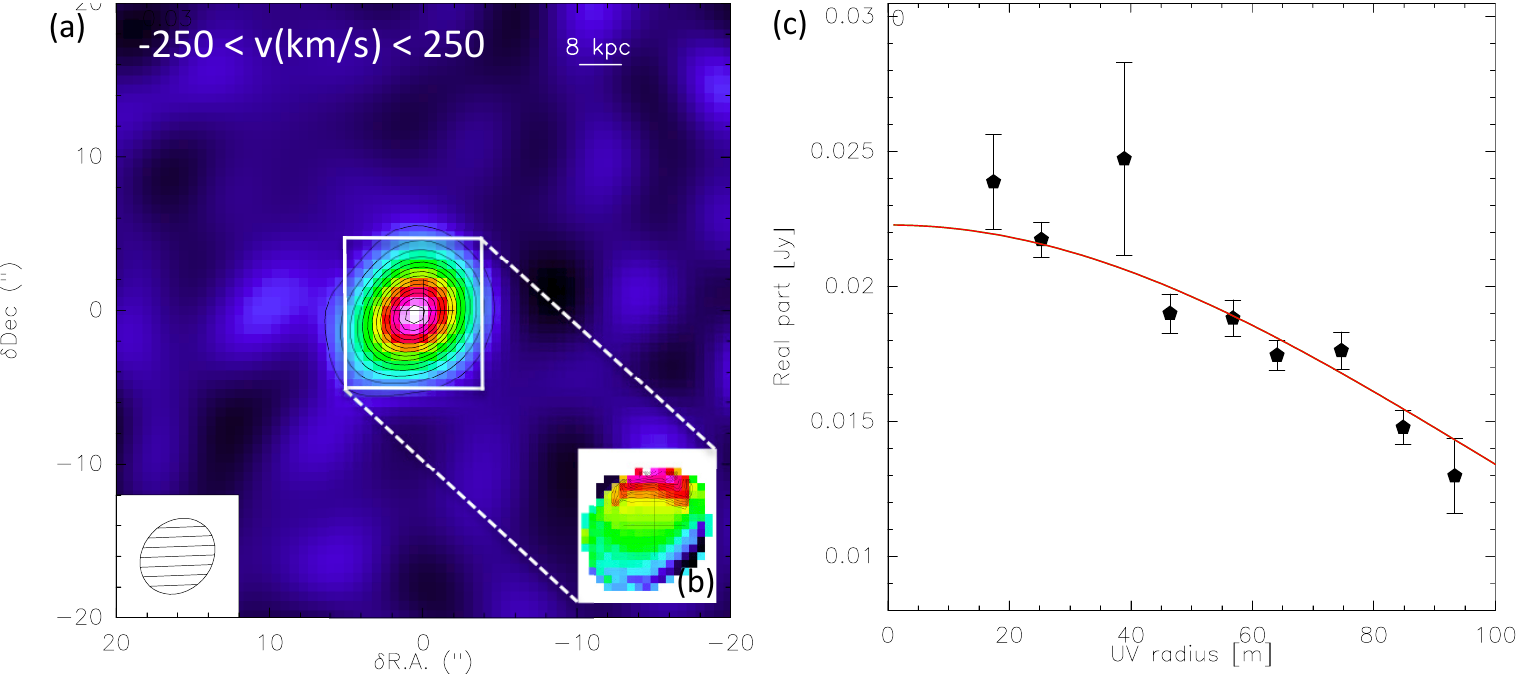}} \\
\caption{Panel (a): IRAM PdBI continuum-subtracted 
map of the CO(1-0) narrow core emission in IRAS F23060+0505. 
Contours correspond to 5$\sigma$ (1$\sigma$ rms level is 0.3 mJy beam$^{-1}$). 
Panel (b) shows the first moment map obtained within v$\in$(-200, 200) km s$^{-1}$
by applying a flux density threshold of 8 mJy. 
The size of the inset is 9 $\times$ 10 arcsec.
Panel (c): \emph{uv} plot for the narrow core, with visibilities binned in {\it uv} radius steps of 10m,
fitted with a circular Gaussian model.
} \label{fig:core_uf2a} \end{figure}

The ULIRG IRAS F23060+0505 hosts a heavily obscured AGN (type 2 QSO). 
Previous studies of the morphology and kinematics of its extended narrow line region
indicated the presence of outflows of ionised gas through blue-shifted asymmetries of the emission
line profiles \citep{Wilman+99}.

In IRAS F23060+0505, we define the blue and red-shifted wings of the CO(1-0) line within 
the velocity ranges (-500, -300) km s$^{-1}$ and (300, 1100) km s$^{-1}$; because
the red wing is detected at a significance below 8$\sigma$ (Fig. \ref{fig:i2306}(c-d)), we combine them together
to estimate their size; the map of the two wings merged together and the corresponding {\it uv} plot are
shown in Fig. \ref{fig:wings_uf2a}.
From the latter, by performing a fit 
with a simple circular Gaussian function, we estimate 
their spatial extension to be {\rm 8.1 $\pm$ 2.9 kpc (FWHM)}.

The cleaned map, the first momentum map and  the corresponding {\it uv} plot for the narrow core of the CO(1-0) 
emission, integrated
within (-250, 250) km s$^{-1}$, are shown in Fig. \ref{fig:core_uf2a}(a-c).
The inferred flux of this CO(1-0) narrow component is ${\rm S_{CO, CORE}}$ = (12.06 $\pm$ 0.85) Jy km s$^{-1}$.
The inset (b) of Fig. \ref{fig:core_uf2a} shows that the narrow core of the CO(1-0) emission traces molecular gas in regular rotation
in a disk, with a major axis of rotation oriented with a PA $\sim$ 0 deg.
A fit to the {\it uv} plot with a circular Gaussian function provides an estimate of the size of this narrow
component of about 6.96 $\pm$ 0.52, which, within the errors, is
comparable to the extension of the broad CO wings in this source.

\subsection{Mrk 876 (PG 1613+658)}

\begin{figure}[h!]
\centering
{\includegraphics[width=.42\columnwidth, angle=270]{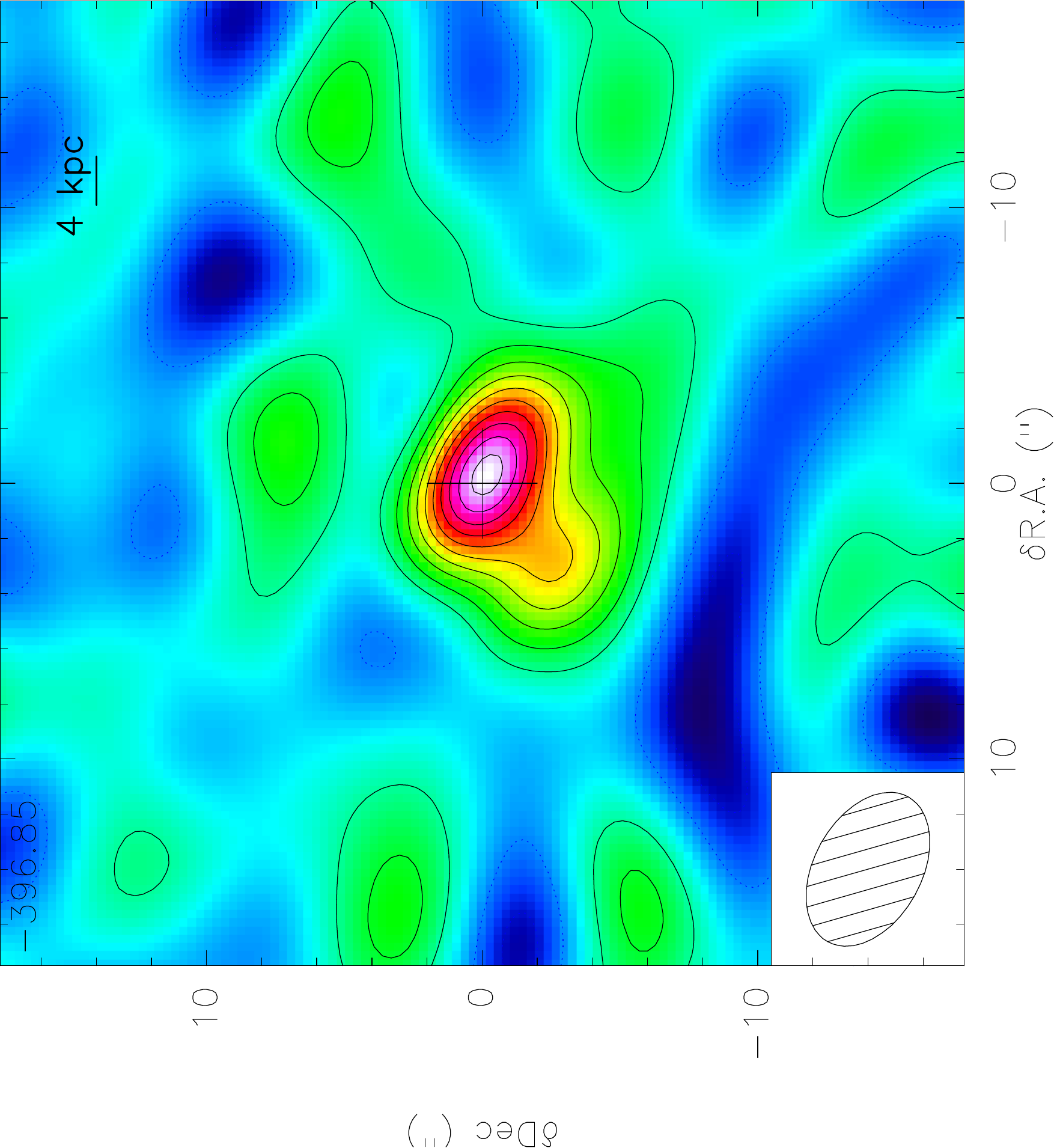}} \quad
{\includegraphics[width=.42\columnwidth, angle=270]{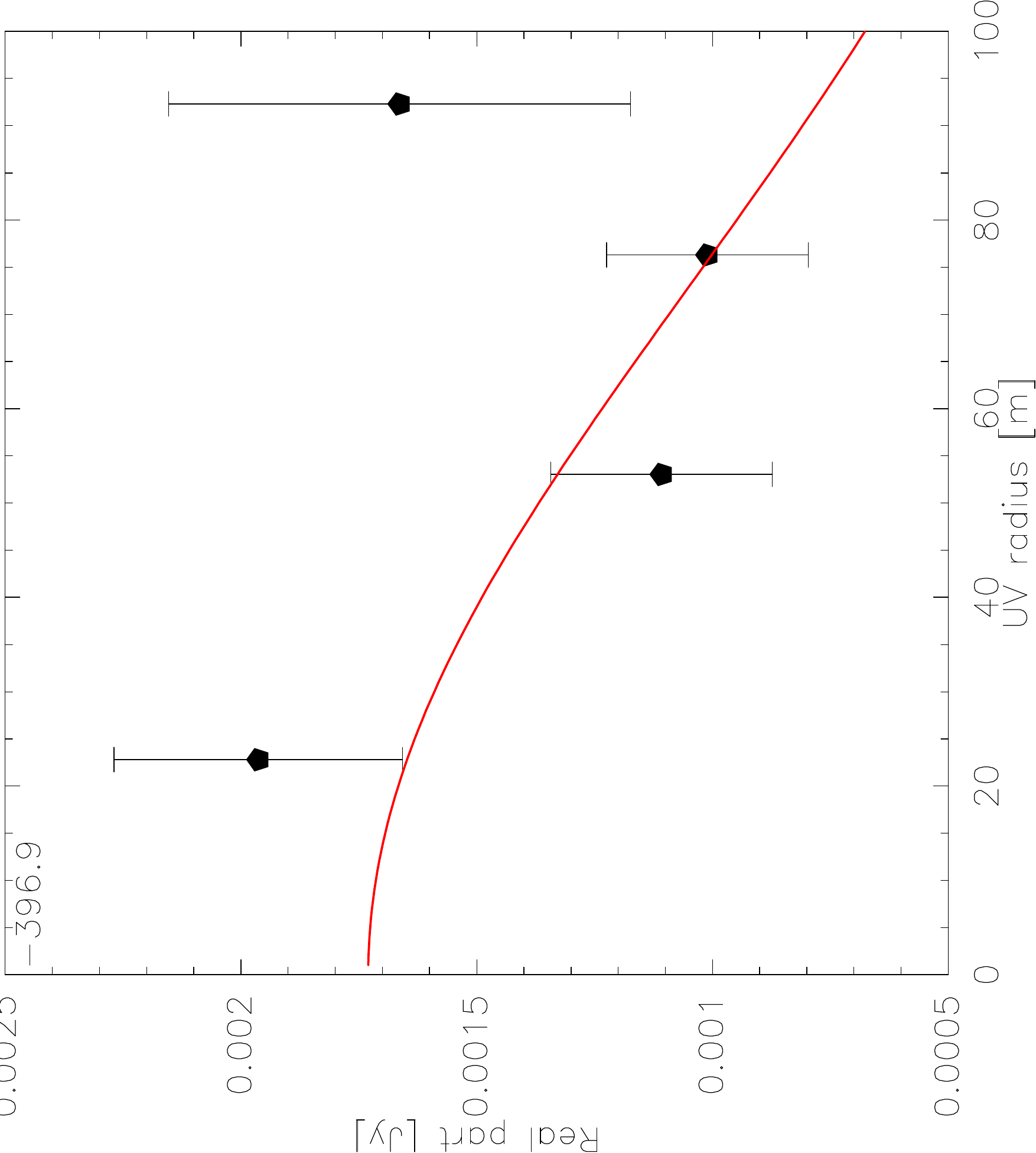}} \\
\caption{\emph{Left:} IRAM PdBI cleaned map of the CO(1-0) blue and red-shifted wings 
of Mrk~876, integrated in the velocity ranges
v$\in$ (-500, -300) km s$^{-1}$ and v$\in$ (400, 1700) km s$^{-1}$, and merged together. 
Contour levels correspond to 1$\sigma$ 
(1$\sigma$ rms level is 0.2  mJy beam$^{-1}$).
\emph{Right:} Real part of {\it uv} visibilities as a function of the {\it uv} radius for the 
combined CO(1-0) wings. Visibilities are binned in baseline steps of 30m. The red curve 
is the result of the best fit with a circular Gaussian model.
} \label{fig:wings_uc2a} \end{figure}

\begin{figure}[h!]
\centering
{\includegraphics[width=\columnwidth]{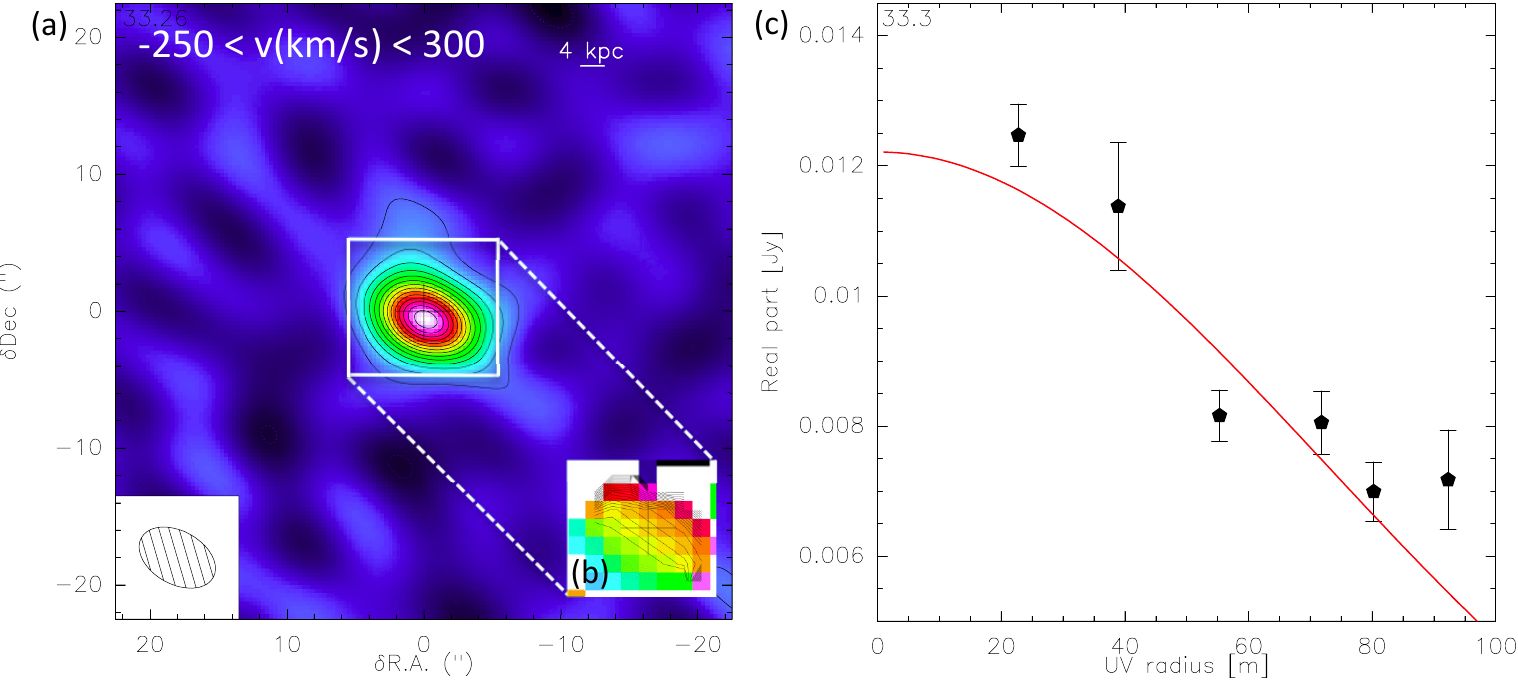}} \\
\caption{Panel (a): IRAM PdBI continuum-subtracted 
map of the CO(1-0) narrow core emission in Mrk~876. 
Contours correspond to 5$\sigma$ (1$\sigma$ rms level is 0.2 mJy beam$^{-1}$). 
Panel (b) shows the first moment map obtained within v$\in$(-200, 200) km s$^{-1}$
The size of the inset is 11 $\times$ 10 arcsec.
Panel (c): \emph{uv} plot for the narrow core, with visibilities binned in {\it uv} radius steps of 15m,
fitted with a circular Gaussian model.
} \label{fig:core_uc2a} \end{figure}

Despite being classified as a LIRG, the total IR luminosity 
({$\rm log(L_{IR}) = 11.97$}, \citealt{Veilleux+09a}) of the powerful QSO Mrk 876
(PG 1613+658) makes it very close to being a ULIRG.
The detection of a molecular outflow is unclear in Mrk~876, and the quality of our 
data only allows us to set an upper limit on the mass and outflow rate 
in this source. Although the PV diagram along the major
rotation axis of this source does not present any clear evidence for high-velocity outflowing gas 
(Fig. \ref{fig:mrk876}c), the map of the blue and red-shifted
CO wings combined together, reported in Fig. \ref{fig:wings_uc2a}, shows some emission
detected at 10$\sigma$. A Gaussian fit to the {\it uv} visibilities, binned in steps
of 30m, provides a size (FWHM) of this broad CO component of 7.1 $\pm$ 1.4 kpc. 

The narrow core of the CO(1-0) emission, although detected at high significance, is not completely resolved
by our observations, as shown by the cleaned map and corresponding {\it uv} plot shown in Fig. \ref{fig:core_uc2a}.
The inset (b) of Fig. \ref{fig:core_uc2a} shows that the central concentration of molecular gas
in Mrk~876, responsible for the bulk of its emission, is likely rotating in a regular disk or ring, 
with a rotation major axis oriented at a PA = -37 deg.
We infer for this component, within (-250, 300) km s$^{-1}$ (i.e. the maximum velocity extent of the rotation as evidenced
by the PV diagram in Fig. \ref{fig:mrk876}c), a physical size of 7.18 $\pm$ 0.46 kpc, which is perfectly
consistent with our estimate of the size of the putative broad CO wings.

\subsection{I~Zw~1 (PG 0050+124)}

\begin{figure}[h!]
\centering
{\includegraphics[width=\columnwidth]{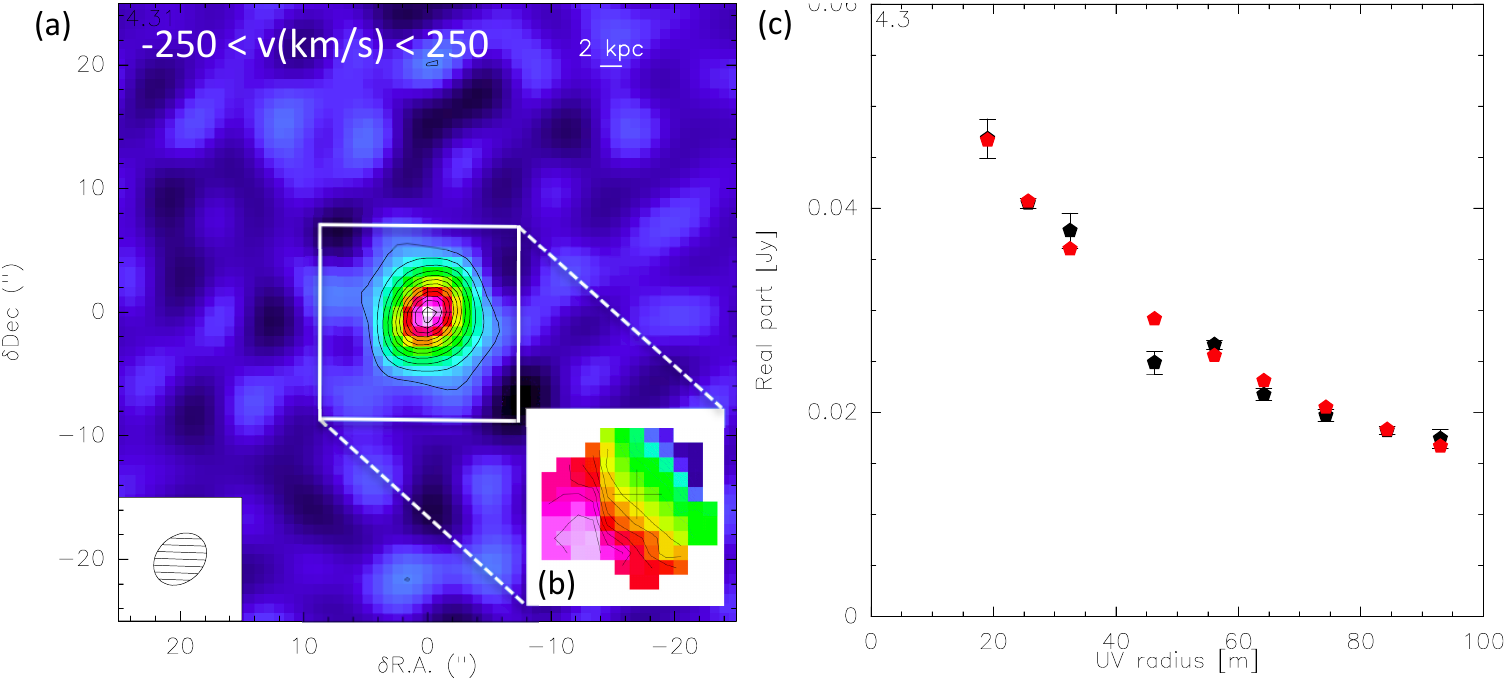}} \\
\caption{Panel (a): IRAM PdBI continuum-subtracted 
map of the CO(1-0) core emission in I~Zw~1. 
Contours correspond to 10$\sigma$ (1$\sigma$ rms level is 0.2 mJy beam$^{-1}$). 
Panel (b) shows the first moment map obtained within v$\in$(-200, 200) km s$^{-1}$,
by applying a flux threshold of 3 mJy.
The size of the inset is 16 $\times$ 16 arcsec.
Panel (c): \emph{uv} plot for the narrow core, with visibilities binned in {\it uv} radius steps of 5m,
fitted with a power law function model of the form ${\rm \propto r^{-2}}$.}
 \label{fig:core_ub2a} \end{figure}

In the LIRG and luminous QSO host I~Zw~1 (PG~0050+124) we
do not detect any CO(1-0) broad emission component, despite the fact 
that its UV and X-ray spectra show some evidence for
high velocity ($\sim$2000 km s$^{-1}$) ionised outflows \citep{Costantini+07}.
Figure \ref{fig:core_ub2a} shows the map and ${\it uv}$ plot of the narrow
CO(1-0) line of I~Zw~1, integrated within v = $\pm$ 250 km s$^{-1}$. The CO line emission appears well
resolved by our observations, and the {\it uv} visibilities, binned in baseline steps of 5m, can be 
well fitted with a power law model of the form ${\rm \propto r^{-2}}$, which gives a FWHM 
of 1.258 $\pm$ 0.041 kpc. The molecular gas traced by this narrow CO(1-0) component is mostly
rotating in a central disk or ring, with a major axis of 
rotation oriented with a PA = -50 deg (inset (b) of Fig. \ref{fig:core_ub2a}).

\end{document}